\newif\ifmnras
\mnrasfalse

\newif\ifaa
\aatrue
\newif\ifaadraft
\aadraftfalse

\newif\ifspie
\spiefalse

\ifmnras
  \documentclass[useAMS,usenatbib,a4paper]{mn2e}
\fi
\ifspie
  \documentclass[a4paper]{spie}
\fi
\ifaa
  \ifaadraft
    \documentclass[traditabstract,onecolumn]{aa}
  \else
    \documentclass[traditabstract]{aa}
  \fi
\fi

\usepackage{natbib}
\usepackage{amssymb}
\usepackage{amsmath}
\usepackage{amsfonts}    
\usepackage{graphicx}
\usepackage{subfigure}
\usepackage{booktabs}
\usepackage{txfonts}
\usepackage{ifpdf}
\ifpdf
  \usepackage[pdftex]{hyperref}
\fi\ifspie
  \usepackage[ps2pdf,colorlinks=true]{hyperref}
\fi

\newif\ifbw
\bwfalse

\newcommand{\eqn}[1]{(#1)}

\newcommand{\tbl}[1]{Table~#1}

\newcommand{\fig}[1]{Fig.~#1}

\newcommand{\sectn}[1]{Sec.~#1}

\newcommand{\eg}{\mbox{\it e.g.}}
\newcommand{\ie}{\mbox{\it i.e.}}

\newcommand{\cmb}{{CMB}}
\newcommand{\cmbtext}{{cosmic microwave background}}

\newcommand{\wmap}{{WMAP}}
\newcommand{\wmaptext}{{Wilkinson Microwave Anisotropy Probe}}

\newcommand{\healpix}{{HEALPix}}
\newcommand{\healpixtext}{{Hierarchical Equal Area isoLatitude Pixelisation}}

\newcommand{\spcend}{\ensuremath{\:}}

\newcommand{\cconj}{\ensuremath{\ast}}

\newcommand{\ltwo}{\ensuremath{\mathrm{L}^2}}
\newcommand{\sphere}{\ensuremath{{\mathrm{S}^2}}}

\newcommand{\dx}{\ensuremath{\mathrm{\,d}}}

\newcommand{\dmun}{\ensuremath{\dx \Omega}}

\newcommand{\sa}{\ensuremath{\omega}}
\newcommand{\saa}{\ensuremath{\theta}}
\newcommand{\sab}{\ensuremath{\varphi}}
\newcommand{\sas}{\ensuremath{\saa, \sab}}

\newcommand{\el}{\ensuremath{\ell}}

\newcommand{\kron}[2]{\ensuremath{\delta_{{#1}{#2}}}}

\newcommand{\wav}{\ensuremath{\psi}}

\newcommand{\nside}{\ensuremath{{N_{\rm{side}}}}}
\newcommand{\npix}{\ensuremath{{N_{\rm{pix}}}}}

\newcommand{\order}{\ensuremath{\mathcal{O}}}

\newcommand{\precision}{\ensuremath{p}}
\newcommand{\scalefun}{\ensuremath{\phi}}
\newcommand{\scal}{\ensuremath{j}}
\newcommand{\scalmax}{\ensuremath{J}}
\newcommand{\scalmin}{\ensuremath{J_0}}
\newcommand{\locat}{\ensuremath{k}}
\newcommand{\approxspace}{\ensuremath{V}}
\newcommand{\wavspace}{\ensuremath{W}}
\newcommand{\area}{\ensuremath{A}}
\renewcommand{\npix}{\ensuremath{N}}
\newcommand{\pixel}{\ensuremath{P}}
\newcommand{\wavtype}{\ensuremath{m}}
\newcommand{\acoeff}{\ensuremath{\lambda}}
\newcommand{\dcoeff}{\ensuremath{\gamma}}
\newcommand{\update}{}

\ifmnras

  \title[Data compression on the sphere]
    {Data compression on the sphere}

  \author[McEwen \& Eyers]
    {J.~D.~McEwen$^{1}$\thanks{E-mail: mcewen@mrao.cam.ac.uk} and D.~M.~Eyers$^{2}$\\ 
    $^1$Astrophysics Group, 
        Cavendish Laboratory,  J.~J.~Thomson Avenue,
        Cambridge CB3 0HE, UK\\
    $^2$University of Cambridge Computer Laboratory, J.~J.~Thomson Avenue,
        Cambridge CB3 0FD, UK\\
    }

  \date{Accepted ---. Received ---; in original form ---}
  \pagerange{\pageref{sec:intro}--\pageref{lastpage}} 
  \pubyear{2008}

  \def\LaTeX{L\kern-.36em\raise.3ex\hbox{a}\kern-.15em
      T\kern-.1667em\lower.7ex\hbox{E}\kern-.125emX}

\fi
\ifspie

  \title{Data compression on the sphere}

  \author{J.~D.~McEwen$^1$ and D.~M.~Eyers$^2$
    \skiplinehalf \small
    $^1$\emph{Astrophysics Group, Cavendish Laboratory, J.~J.~Thomson
    Avenue, Cambridge CB3 0HE, UK}\\
    $^2$\emph{University of Cambridge Computer Laboratory, J.~J.~Thomson Avenue,
    Cambridge CB3 0FD, UK}\
  }

  \authorinfo{Further author information\\ 
    E-mail: mcewen@mrao.cam.ac.uk \quad URL:
    \url{www.mrao.cam.ac.uk/~jdm57/} \quad (JDM)\\
    E-mail: david.eyers@cl.cam.ac.uk \quad URL:
    \url{www.cl.cam.ac.uk/~dme26/} \quad (DME)}

\fi

\begin{document}
\label{firstpage}

\ifaa

  \title{Data compression on the sphere}

   \author
    {J.~D.~McEwen\inst{1,2},
      Y.~Wiaux\inst{1,3,4} \and
      D.~M.~Eyers\inst{5,6}}
  \institute{Institute of Electrical Engineering, Ecole Polytechnique
               F{\'e}d{\'e}rale de Lausanne (EPFL), CH-1015 Lausanne, Switzerland           
         \and
             Astrophysics Group, Cavendish Laboratory, J.~J.~Thomson Avenue\, 
             Cambridge CB3 0HE, UK
         \and
             Institute of Bioengineering, Ecole Polytechnique 
             F{\'e}d{\'e}rale de Lausanne (EPFL), CH-1015 Lausanne, Switzerland
         \and          
             Department of Radiology and Medical Informatics, University of Geneva (UniGE), 
             CH-1211 Geneva, Switzerland
         \and
             Department of Computer Science, University of Otago, 
             Dunedin 9016, New Zealand
         \and
            University of Cambridge Computer Laboratory, J.~J.~Thomson Avenue,
             Cambridge CB3 0FD, UK\\
         \email{mcewen@mrao.cam.ac.uk}
       }

  \date{Received 9 September 2010 / Accepted 29 April 2011}
 
  \abstract
  {Large data-sets defined on the sphere arise in many fields.  In
  particular, recent and forthcoming observations of the anisotropies
  of the \cmbtext\ (\cmb) made on the celestial sphere contain
  approximately three and fifty mega-pixels respectively.  The
  compression of such data is therefore becoming increasingly
  important.  We develop algorithms to compress data defined on the
  sphere.  A Haar wavelet transform on the sphere is used as an energy
  compression stage to reduce the entropy of the data, followed by
  Huffman and run-length encoding stages.  Lossless and lossy
  compression algorithms are developed.  We evaluate compression
  performance on simulated \cmb\ data, Earth topography data and
  environmental illumination maps used in computer graphics.  The
  \cmb\ data can be compressed to approximately 40\% of its original
  size for essentially no loss to the cosmological information content
  of the data, and to approximately 20\% if a small cosmological
  information loss is tolerated.  For the topographic and illumination
  data compression ratios of approximately 40:1 can be achieved when a
  small degradation in quality is allowed.   \update{We make our {SZIP}
  program that implements these compression algorithms available publicly.}}


\fi

\newcommand{\subfigcapsize}{ }
\ifspie
  \renewcommand{\subfigcapsize}{\scriptsize}
\fi

\maketitle


\ifaa\else
\begin{abstract}
  Large data-sets defined on the sphere arise in many fields.  In
  particular, recent and forthcoming observations of the anisotropies
  of the \cmbtext\ (\cmb) made on the celestial sphere contain
  approximately three and fifty mega-pixels respectively.  The
  compression of such data is therefore becoming increasingly
  important.  We develop algorithms to compress data defined on the
  sphere.  A Haar wavelet transform on the sphere is used as an energy
  compression stage to reduce the entropy of the data, followed by
  Huffman and run-length encoding stages.  Lossless and lossy
  compression algorithms are developed.  We evaluate compression
  performance on simulated \cmb\ data, Earth topography data and
  environmental illumination maps used in computer graphics.  The
  \cmb\ data can be compressed to approximately 40\% of its original
  size for essentially no loss to the cosmological information content
  of the data, and to approximately 20\% if a small cosmological
  information loss is tolerated.  For the topographic and illumination
  data compression ratios of approximately 40:1 can be achieved when a
  small degradation in quality is allowed.  We make our {SZIP}
  program that implements these compression algorithms available publicly.
\end{abstract}
\fi

\ifmnras
\fi\ifspie
\fi

\section{Introduction}
\label{sec:intro}

Large data-sets that are measured or defined inherently on the sphere
arise in a range of applications.  Examples include environmental
illumination maps and reflectance functions used in computer graphics
(\eg\ \citealt{ramamoorthi:2004}), astronomical observations made on
the celestial sphere, such as the \cmbtext\ (\cmb) (\eg\
\citealt{bennett:1996,bennett:2003a}), and applications in many other
fields, such as planetary science (\eg\
\citealt{wieczorek:2006,wieczorek:1998,turcotte:1981}), geophysics
(\eg\ \citealt{whaler:1994,swenson:2002,simons:2006}) and quantum
chemistry (\eg\ \citealt{choi:1999,ritchie:1999}).  Technological
advances in observational instrumentation and improvements in
computing power are resulting in significant increases in the size of
data-sets defined on the sphere (hereafter we refer to a data-set
defined on the sphere as a \emph{data-sphere}).  For example, current
and forthcoming observations of the anisotropies of the \cmb\ are of
considerable size.  Recent observations made by the \wmaptext\ (\wmap)
satellite \citep{bennett:1996,bennett:2003a} contain approximately
three mega-pixels, while the forthcoming Planck mission
\citep{planck:bluebook} will generate data-spheres with approximately
fifty mega-pixels.  Furthermore, cosmological analyses of these data
often require the use of Monte Carlo simulations, which generate in
the order of a thousand-fold increase in data size.  The efficient and
accurate compression of data-spheres is therefore becoming
increasingly important for both the dissemination and storage of data.

In general, data compression algorithms usually consist of an energy
compression stage (often a transform or filtering process), followed
by quantisation and entropy encoding stages.  For example, JPEG
(ISO/IEC IS 10918-1) uses a discrete cosine transform for the energy
compression stage, whereas JPEG2000 (ISO/IEC 15444-1:2004)
uses a discrete wavelet transform.  Due to the simultaneous
localisation of signal content in scale and space afforded by a
wavelet transform, one would expect wavelet-based energy compression
to perform well relative to other methods.  Wavelet theory in
Euclidean space is well established (see \citet{daubechies:1992} for a
detailed introduction), however the same cannot yet be said for wavelet
theory on the sphere.
A number of attempts have been made to extend wavelets to the sphere.
Discrete second generation wavelets on the sphere that are based on a
multiresolution analysis have been developed
\citep{schroder:1995,sweldens:1996}.  Haar wavelets on the sphere for
particular pixelisation schemes have also been developed
\citep{tenorio:1999,barreiro:2000}.  These discrete constructions
allow for the exact reconstruction of a signal from its wavelet
coefficients but they may not necessarily lead to a stable basis (see
\citet{sweldens:1997} and references therein).  Other authors have
focused on continuous wavelet methodologies on the sphere
\citep{freeden:1997a,freeden:1997b,holschneider:1996,torresani:1995,dahlke:1996,antoine:1998,antoine:1999,antoine:2002,antoine:2004,demanet:2003,wiaux:2005,sanz:2006,mcewen:2006:cswt2}.
Although signals can be reconstructed exactly from their wavelet
coefficients in these continuous methodologies in theory, the absence
of an infinite range of dilations precludes exact reconstruction in
practice.  Approximate reconstruction formula may be developed by
building discrete wavelet frames that are based on the continuous
methodology (\eg\ \citealt{bogdanova:2004}).  More recently, filter
bank wavelet methodologies that are essentially based on a continuous
wavelet framework have been developed for the axi-symmetric
\citep{starck:2006} and directional \citep{wiaux:2007:sdw} cases.
These methodologies allow the exact reconstruction of a signal from
its wavelet coefficients in theory and in practice.  Compression
applications require a wavelet transform on the sphere that allows
exact reconstruction, thus the methodologies of \citet{schroder:1995},
\citet{tenorio:1999}, \citet{barreiro:2000}, \citet{starck:2006} and
\citet{wiaux:2007:sdw} are candidates.

Data compression algorithms on the sphere that use wavelet or
alternative transforms have been developed already.  Compression on
the sphere was considered first, to our knowledge, in the pioneering
work of \citet{schroder:1995}.  The lifting scheme was used here to
define a discrete wavelet transform on the sphere, however compression
was analysed only in terms of the number of wavelet coefficients
required to represent a data-sphere in a lossy manner and no encoding
stage was performed.  The addition of an encoding stage to this
algorithm was performed by \citet{kolarov:1997} using zero-tree coding
methods.  An alternative compression algorithm based on a Faber
decomposition has been proposed by \citet{assaf:1999}, however no
encoding stage is included and performance is again analysed only in
terms of the number of coefficients required to recover a lossy
representation of the data-sphere.  The data-sphere compression
algorithm devised by \citet{schroder:1995} and \citet{kolarov:1997}
therefore constitutes the current state-of-the-art.  This algorithm
relies on an icosahedron pixelisation of the sphere that is based on
triangular subdivisions.  The corresponding pixelisation of the sphere
precludes pixel centres located on rings of constant latitude.
Constant latitude pixelisations of the sphere are of considerable
practical use since this property allows the development of many fast
algorithms on pixelised spheres, such as fast spherical harmonic
transforms.  For example, the following constant latitude
pixelisations of the sphere have been used extensively in astronomical
applications and beyond: the equi-angular pixelisation
\citep{driscoll:1994}; the
\healpixtext\footnote{\url{http://healpix.jpl.nasa.gov/}} (\healpix)
\citep{gorski:2005}; the
IGLOO\footnote{\url{http://www.cita.utoronto.ca/~crittend/pixel.html}}
pixelisation \citep{crittenden:1998}; and the
GLESP\footnote{\url{http://www.glesp.nbi.dk/}} pixelisation
\citep{doroshkevich:2005}.  Furthermore, at present no
data-sphere compression tool is available publicly.

Motivated by the requirement for a data-sphere compression algorithm
defined on a constant latitude pixelisation of the sphere, and a
publicly available tool to compress such data, we develop
wavelet-based compression algorithms for data defined on the \healpix\
pixelisation scheme and make our implementation of these algorithms
available publicly.  We are driven primarily by the need to compress
\cmb\ data, hence the adoption of the \healpix\ scheme (the
pixelisation scheme used currently to store and distribute these data).
Wavelet transforms are expected to perform well in the energy
compression stage of the compression algorithm, thus we adopt a Haar
wavelet transform defined on the sphere for this stage (following a
similar framework to that outlined by \citealt{barreiro:2000}).  We
could have chosen a filter bank based wavelet framework, such as those
developed by \citet{starck:2006} and \citet{wiaux:2007:sdw}, however,
for now, we adopt discrete Haar wavelets due to their simplicity and
computational efficiency.

The remainder of this paper is organised as follows.  In
\sectn{\ref{sec:algorithm}} we describe the compression algorithms
developed, first discussing Haar wavelets on the sphere, before
explaining the encoding adopted in our lossless and lossy compression
algorithms.  The performance of our compression algorithms is then
evaluated in \sectn{\ref{sec:applications}}.  We first examine
compression performance for \cmb\ data and study the implications of
any errors on cosmological inferences drawn from the data.  We then
examine compression performance for topographical data and
environmental illumination maps.  Concluding remarks are made in
\sectn{\ref{sec:conclusions}}.

\section{Compression algorithms}
\label{sec:algorithm}

The wavelet-based compression algorithms that we develop to compress
data-spheres consist of a number of stages.  Firstly, a Haar wavelet
transform is performed to reduce the entropy of the data, followed by
quantisation and encoding stages.  The resulting algorithm is lossless
to numerical precision.  We then develop a lossy compression algorithm
by introducing an additional thresholding stage, after the wavelet
transform, in order to reduce the entropy of the data further.
Allowing a small degradation in the quality of decompressed data in
this manner improves the compression ratios that may be attained. In
this section we first discuss the Haar wavelet transform on the sphere
that we adopt, before outlining the subsequent stages of the lossless
and lossy compression algorithms.  \update{We make our SZIP program that
implements these algorithms available
publicly.\footnote{\url{http://www.szip.org.uk}}
Furthermore, we also provide an SZIP user manual \citep{mcewen:szip_manual}, which discusses
installation, usage (including a description of all compression
options and parameters), and examples.}

\subsection{Haar wavelets on the sphere}
\label{sec:algorithm_haar}

The description of wavelets on the sphere given here is based largely
on the generic lifting scheme proposed by \citet{schroder:1995} and
also on the specific definition of Haar wavelets on a \healpix\
pixelised sphere proposed by \citet{barreiro:2000}.  However, our
discussion and definitions contain a number of notable differences to
those given by \citet{barreiro:2000} since we construct an orthonormal
Haar basis on the sphere and describe this in a multiresolution
setting.

We begin by defining a nested hierarchy of spaces as required for a
multiresolution analysis (see \citet{daubechies:1992} for a more
detailed discussion of multiresolution analysis).  Firstly, consider
the approximation space $\approxspace_\scal$ on the sphere \sphere,
which is a subset of the space of square integrable functions on the
sphere, \ie\ $\approxspace_\scal\subset\ltwo(\sphere)$.  One may think
of $\approxspace_\scal$ as the space of piecewise constant functions
on the sphere, where the index $\scal$ corresponds to the size of the
piecewise constant regions.  As the resolution index $\scal$
increases, the size of the piecewise constant regions shrink, until in
the limit we recover $\ltwo(\sphere)$ as $\scal\rightarrow\infty$.  If
the piecewise constants regions of $\sphere$ are arranged
hierarchically as $\scal$ increases, then one can construct the nested
hierarchy of approximation spaces
\begin{equation}
\approxspace_1 \subset \approxspace_2 \subset \cdots 
\subset \approxspace_\scalmax
\subset \ltwo(\sphere)
\spcend ,
\label{eqn:space_hierarchy}
\end{equation}
where coarser (finer) approximation spaces correspond to a lower
(higher) resolution level $\scal$.  For each space
$\approxspace_\scal$ we define a basis with basis elements given by
the \emph{scaling functions}
$\scalefun_{\scal,\locat}\in\approxspace_\scal$, where the $\locat$ index
corresponds to a translation on the sphere.  Now, let us define
$\wavspace_\scal$ to be the orthogonal complement of $\approxspace_\scal$
in $\approxspace_{\scal+1}$, where the inner product of two square
integrable functions on the sphere $f, g \in \ltwo(\sphere)$ is
defined by
\begin{equation*}
\langle f | g \rangle \equiv \int_\sphere f(\sa) g^{\cconj}(\sa) \dmun
\spcend ,
\end{equation*}
where $\sa=(\sas)$ denotes spherical coordinates with colatitude
$\saa\in[0,\pi]$ and longitude $\sab\in[0,2\pi)$, $\cconj$ denotes
complex conjugation and $\dmun=\sin\saa \dx \saa \dx \sab$ is the
usual rotation invariant measure on the sphere.  $\wavspace_\scal$
essentially provides a space for the representation of the components
of a function in $\approxspace_{\scal+1}$ that cannot be represented
in $\approxspace_\scal$, \ie\ $\approxspace_{\scal+1} =
\approxspace_\scal \oplus \wavspace_\scal$.  For each space
$\wavspace_\scal$ we define a basis with basis elements given by the
\emph{wavelets} $\wav_{\scal,\locat}\in\wavspace_\scal$.  The wavelet
space $\wavspace_\scal$ encodes the difference (or details) between
two successive approximation spaces $\approxspace_{\scal}$ and
$\approxspace_{\scal+1}$.  By expanding the hierarchy of approximation
spaces, the highest level (finest) space $\scal=\scalmax$, can then be
represented by the lowest level (coarsest) space $\scal=1$ and the
differences between the approximation spaces that are encoded by the
wavelet spaces:
\begin{equation}
\label{eqn:multires}
V_\scalmax = V_1 \oplus \bigoplus_{\scal=1}^{\scalmax-1} W_j
\spcend .
\end{equation}

Let us now relate the generic description of multiresolution spaces
given above to the \healpix\ pixelisation.  The \healpix\ scheme
provides a hierarchical pixelisation of the sphere and hence may be
used to define the nested hierarchy of approximation spaces
explicitly.  The piecewise constant regions of the function spaces
$\approxspace_\scal$ discussed above now correspond to the pixels of
the \healpix\ pixelisation at the resolution associated with
$\approxspace_\scal$.  To make the association explicit, let
$\approxspace_\scal$ correspond to a \healpix\ pixelised sphere with
resolution parameter $\nside=2^{\scal-1}$ (\healpix\ data-spheres are
represented by the resolution parameter \nside, which is related to
the number of pixels in the pixelisation by $\npix=12\nside^2$).  In
the \healpix\ scheme, each pixel at level $\scal$ is subdivided into
four pixels at level $\scal+1$, and the nested hierarchy given by
\eqn{\ref{eqn:space_hierarchy}} is satisfied.  The number of pixels
associated with each space $\approxspace_\scal$ is given by
$\npix_\scal=12\times4^{\scal-1}$, where the area of each pixel is
given by $\area_\scal=4\pi/\npix_\scal=\pi / (3 \times 4^{\scal-1})$
(note that all pixels in a \healpix\ data-sphere at resolution $\scal$ have
equal area).  It is also useful to note that the number and area of
pixels at one level relates to adjacent levels through
$\npix_{\scal+1}=4\npix_{\scal}$ and $\area_{\scal+1}=\area_\scal / 4$
respectively.

\begin{figure*}
\centering
\includegraphics[height=100mm]{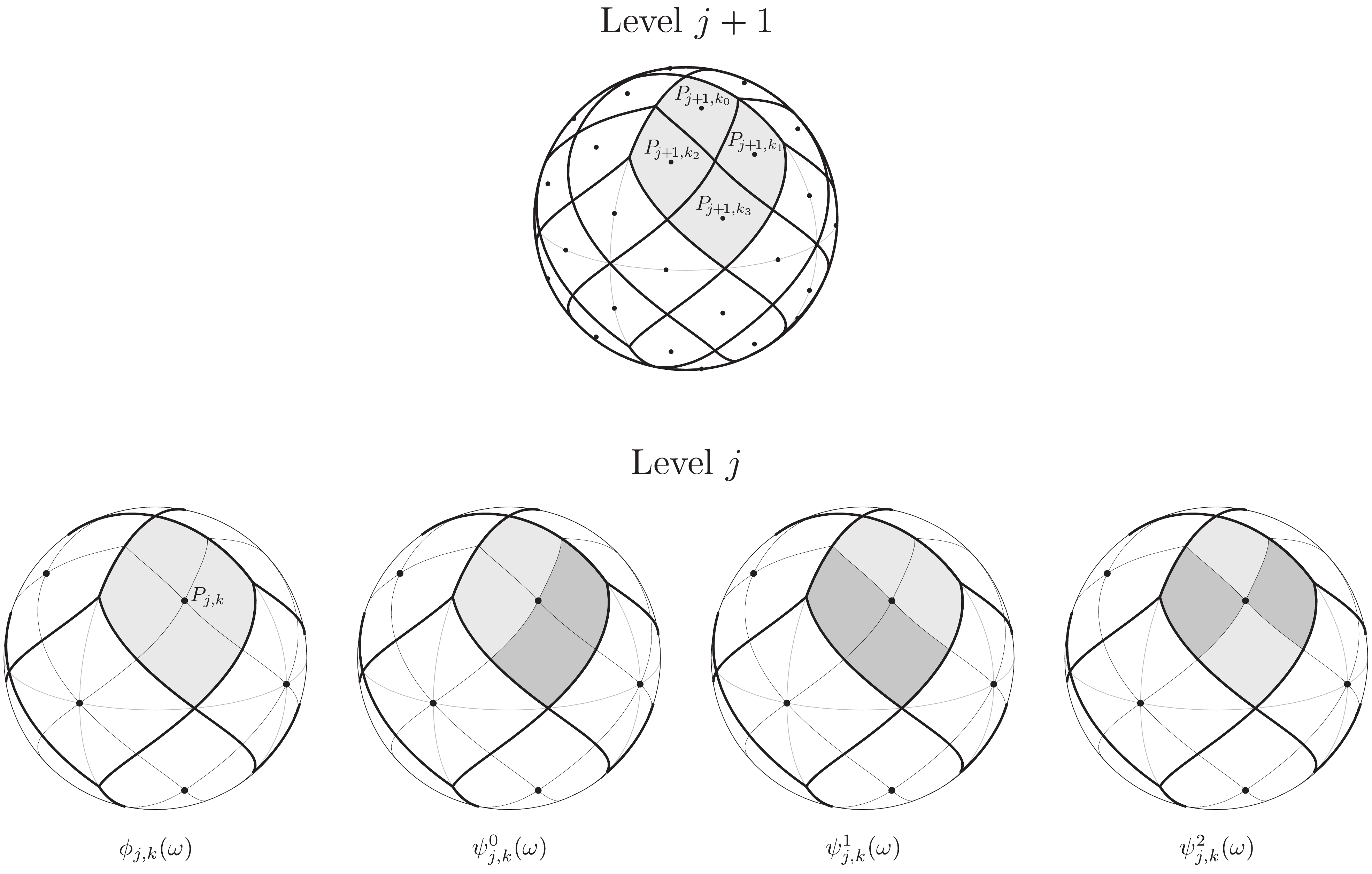}
\caption{Haar scaling function $ \scalefun_{\scal,\locat}(\sa)$ and
  wavelets $\wav_{\scal,\locat}^\wavtype(\sa)$.  Dark shaded regions
  correspond to negative constant values, light shaded regions
  correspond to positive constant values and unshaded regions
  correspond to zero.  The scaling function and wavelets at level
  \scal\ and position \locat\ are non-zero on pixel
  $\pixel_{\scal,\locat}$ only.  Pixel $\pixel_{\scal,\locat}$ at
  level $\scal$ is subdivided into four pixels at level $\scal+1$,
  which we label $\pixel_{\scal+1,\locat_0}$,
  $\pixel_{\scal+1,\locat_1}$, $\pixel_{\scal+1,\locat_2}$ and
  $\pixel_{\scal+1,\locat_3}$ as defined above.}
\label{fig:wavelets}
\end{figure*}

We are now in a position to define the scaling functions and wavelets
explicitly for the Haar basis on the nested hierarchy of \healpix\
spheres.  In this setting the index $\locat$ corresponds to the
position of pixels on the sphere, \ie\ for $\approxspace_\scal$ we get
the range of values $\locat=0,\cdots,\npix_\scal - 1$, and we let
$\pixel_{\scal,\locat}$ represent the region of the $\locat$th pixel of
a \healpix\ data-sphere at resolution $\scal$.  For the Haar basis, we
define the scaling function $\scalefun_{\scal,\locat}$ at level $\scal$
to be constant for pixel $\locat$ and zero elsewhere:
\begin{equation*}
\scalefun_{\scal,\locat}(\sa) \equiv
\begin{cases}
1/\sqrt{\area_\scal} & \sa \in \pixel_{\scal,\locat} \\
0                   & \text{elsewhere .}
\end{cases}
\end{equation*}
The non-zero value of the scaling function $1/\sqrt{\area_\scal}$ is
chosen to ensure that the scaling functions $\scalefun_{\scal,\locat}$
for $k=0,\cdots,\npix_\scal - 1$ do indeed define an orthonormal basis
for $\approxspace_\scal$.  Before defining the wavelets explicitly, we
fix some additional notation.  Pixel $\pixel_{\scal,\locat}$ at level $\scal$ is
subdivided into four pixels at level $\scal+1$, which we label
$\pixel_{\scal+1,\locat_0}$, $\pixel_{\scal+1,\locat_1}$,
$\pixel_{\scal+1,\locat_2}$ and $\pixel_{\scal+1,\locat_3}$, as
illustrated in \fig{\ref{fig:wavelets}}.  An orthonormal basis for the wavelet
space $\wavspace_\scal$, the orthogonal complement of
$\approxspace_\scal$, is then given by the following wavelets of type
$\wavtype=\{0,1,2\}$:
\begin{equation*}
\wav_{\scal,\locat}^0(\sa) \equiv \bigl [
\scalefun_{\scal+1,\locat_0}(\sa) -
\scalefun_{\scal+1,\locat_1}(\sa) +
\scalefun_{\scal+1,\locat_2}(\sa) -
\scalefun_{\scal+1,\locat_3}(\sa) \bigr ] / 2
\spcend ;
\end{equation*}
\begin{equation*}
\wav_{\scal,\locat}^1(\sa) \equiv \bigl [
\scalefun_{\scal+1,\locat_0}(\sa) +
\scalefun_{\scal+1,\locat_1}(\sa) -
\scalefun_{\scal+1,\locat_2}(\sa) -
\scalefun_{\scal+1,\locat_3}(\sa) \bigr ] / 2
\spcend ;
\end{equation*}
\begin{equation*}
\wav_{\scal,\locat}^2(\sa) \equiv \bigl [
\scalefun_{\scal+1,\locat_0}(\sa) -
\scalefun_{\scal+1,\locat_1}(\sa) -
\scalefun_{\scal+1,\locat_2}(\sa) +
\scalefun_{\scal+1,\locat_3}(\sa) \bigr ] / 2
\spcend .
\end{equation*}
We require three independent wavelet types to construct a complete
basis for $\wavspace_\scal$ since the dimension of
$\approxspace_{\scal+1}$ (given by $\npix_{\scal+1}$) is four times
larger than the dimension of $\approxspace_\scal$ (the approximation
function provides the fourth component).  The Haar scaling functions
and wavelets defined on the sphere above are illustrated in
\fig{\ref{fig:wavelets}}.

Let us check that the scaling functions and wavelets satisfy the
requirements for an orthonormal multiresolution analysis as outlined
previously.  We require $\wavspace_\scal$ to be orthogonal to
$\approxspace_\scal$, \ie\ we require
\begin{equation*}
\int_\sphere 
\scalefun_{\scal,\locat}(\sa) \wav_{\scal,\locat^\prime}^\wavtype(\sa)
\dmun = 0
\spcend.
\end{equation*}
This is always satisfied since for $\locat^\prime\neq\locat$ the
scaling function and wavelet do not overlap and so the integrand is
zero always, and for $\locat^\prime=\locat$ we find
\begin{equation*}
\int_\sphere 
\scalefun_{\scal,\locat}(\sa) \wav_{\scal,\locat}^\wavtype(\sa)
\dmun 
\propto
\int_\sphere 
\wav_{\scal,\locat}^\wavtype(\sa)
\dmun 
= 0
\spcend .
\end{equation*}
We also require $\wavspace_\scal$ to be orthogonal to
$\wavspace_{\scal^\prime}$ for all $\scal$ and $\scal^\prime$.  Again,
if the basis functions do not overlap (\ie\ $\locat\neq\locat^\prime$)
then this requirement is satisfied automatically, and if they do (\ie\
$\locat=\locat^\prime$) then the wavelet at the finer level
$\scal^\prime>\scal$ will always lie within a region of the wavelet at
level $\scal$ with constant value, and consequently
\begin{equation*}
\int_\sphere 
\wav_{\scal,\locat}^\wavtype(\sa) \wav_{{\scal^\prime},{\locat^\prime}}^{\wavtype^\prime}(\sa)
\dmun 
\propto
\int_\sphere 
\wav_{\scal^\prime,\locat^\prime}^{\wavtype^\prime}(\sa)
\dmun 
= 0
\spcend .
\end{equation*}
Finally, to ensure that we have constructed an orthonormal wavelet
basis for $\wavspace_\scal$, we check the orthogonality of all
wavelets at level $\scal$:
\begin{equation*}
\int_\sphere 
\wav_{\scal,\locat}^\wavtype(\sa) \wav_{{\scal},{\locat^\prime}}^{\wavtype^\prime}(\sa)
\dmun 
= \kron{\wavtype}{\wavtype^\prime}
\kron{\locat}{\locat^\prime}
\Biggl(\frac{1}{2\sqrt{\area_{\scal+1}}} \Biggr)^2 \area_\scal
=\kron{\wavtype}{\wavtype^\prime}
\kron{\locat}{\locat^\prime}
\spcend ,
\end{equation*}
where for $\wavtype\neq\wavtype^\prime$ the positive and negative
regions of the integrand cancel exactly and for
$\locat\neq\locat^\prime$ the wavelets do not overlap and so the
integrand is zero always.  Note that in the previous expression the
final $\area_\scal$ term arises from the area element $\dmun$.  The
Haar approximation and wavelet spaces that we have constructed
therefore satisfy the requirements of a orthonormal multiresolution
analysis on the sphere.  Although the orthogonal nature of these
spaces is important, a different normalisation could be chosen.  It is
now possible to define the analysis and synthesis of a function on the
sphere in this Haar wavelet multiresolution framework.

\begin{figure*}
\centering
\includegraphics[height=120mm]{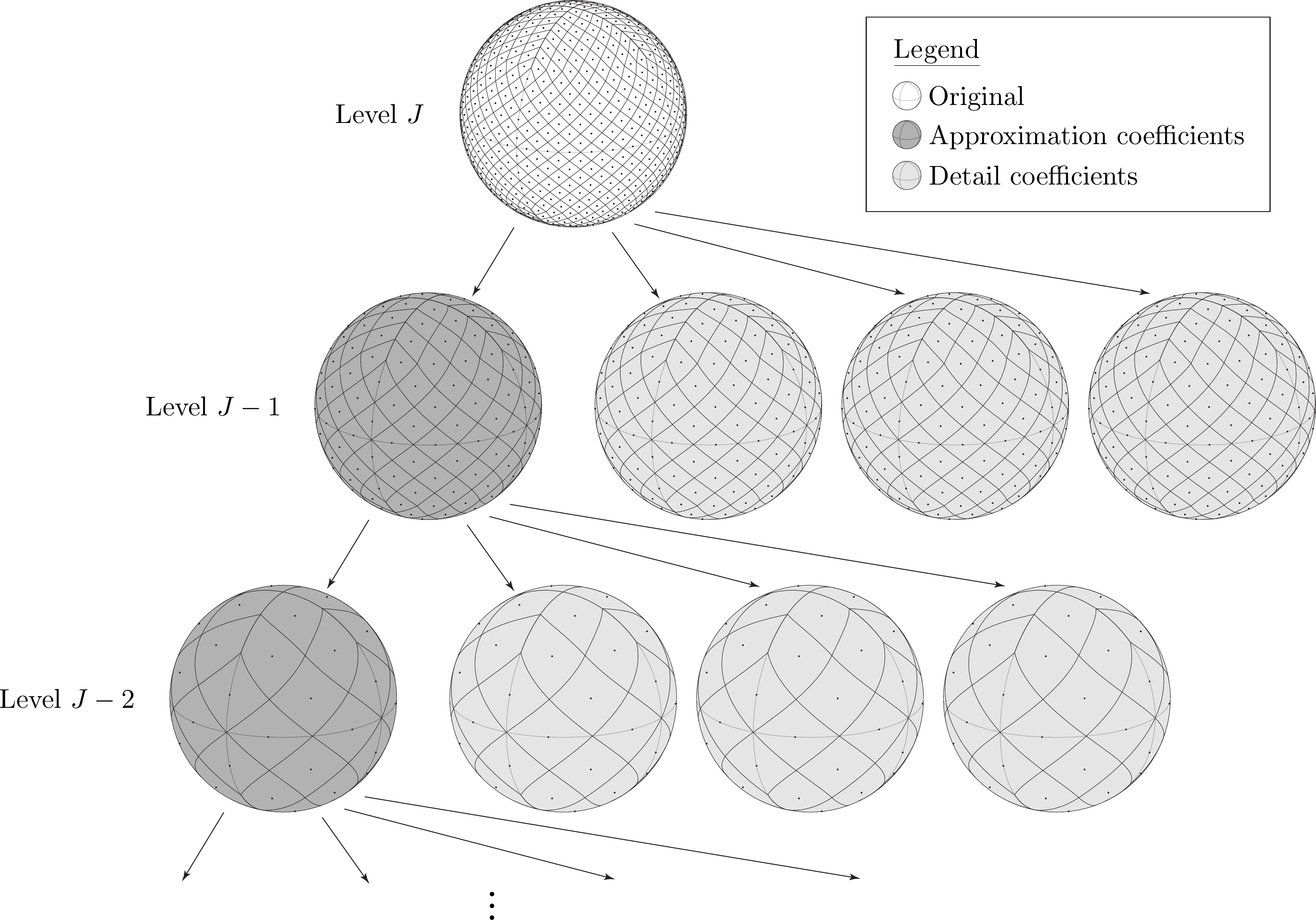}
\caption{Haar multiresolution decomposition.  Starting at the finest
  level $\scalmax$ (the original data-sphere), the approximation and
  detail coefficients at level $\scalmax-1$ are computed.  This
  procedure is repeated to decompose the approximation coefficients at
  level $\scalmax-1$ (\ie\ the approximation function
  $f_{\scalmax-1}$), into approximation and detail coefficients at the
  coarser level $\scalmax-2$.  Repeating this procedure continually,
  one recovers the multiresolution representation of $f_\scalmax$ in
  terms of the coarsest level approximation $f_{\scalmin}$ and all of
  the detail coefficients.}
\label{fig:multiresolution}
\end{figure*}

The decomposition of a function defined on a \healpix\ data-sphere at
resolution $\scalmax$, \ie\ $f_\scalmax\in\approxspace_\scalmax$, into
its wavelet and scaling coefficients proceeds as follows.  Consider an
intermediate level $\scal+1<\scalmax$ and let $f_{\scal+1}$ be the
approximation of $f_\scalmax$ in $\approxspace_{\scal+1}$.  The
scaling coefficients at the coarser level $\scal$ are given by the
projection of $f_{\scal+1}$ onto the scaling functions
$\scalefun_{\scal,\locat}$:
\begin{align}
\acoeff_{\scal,\locat} &\equiv \int_\sphere
f_{\scal+1}(\sa) 
\scalefun_{\scal,\locat}(\sa) \dmun \nonumber \\
& = \bigl (
\acoeff_{\scal+1,\locat_0} +
\acoeff_{\scal+1,\locat_1} +
\acoeff_{\scal+1,\locat_2} +
\acoeff_{\scal+1,\locat_3}
\bigr ) \sqrt{\area_\scal}/4
\spcend , \nonumber
\end{align}
where we call $\acoeff_{\scal,\locat}$ the approximation coefficients
since they define the approximation function
$f_\scal\in\approxspace_\scal$.  At the finest level $\scalmax$, we
naturally associate the function values of $f_\scalmax$ with the
approximation coefficients of this level.  The wavelet coefficients at
level $\scal$ are given by the projection of $f_{\scal+1}$ onto the wavelets
$\wav_{\scal,\locat}^\wavtype$:
\begin{equation*}
\dcoeff_{\scal,\locat}^\wavtype \equiv
\int_\sphere
f_{\scal+1}(\sa) 
\wav_{\scal,\locat}^\wavtype(\sa) \dmun 
\spcend ,
\end{equation*}
giving
\begin{equation*}
\dcoeff_{\scal,\locat}^0 = \bigl (
\acoeff_{\scal+1,\locat_0} -
\acoeff_{\scal+1,\locat_1} +
\acoeff_{\scal+1,\locat_2} -
\acoeff_{\scal+1,\locat_3}
\bigr ) \sqrt{\area_\scal}/4
\spcend ,
\end{equation*}
\begin{equation*}
\dcoeff_{\scal,\locat}^1 = \bigl (
\acoeff_{\scal+1,\locat_0} +
\acoeff_{\scal+1,\locat_1} -
\acoeff_{\scal+1,\locat_2} -
\acoeff_{\scal+1,\locat_3}
\bigr ) \sqrt{\area_\scal}/4
\spcend 
\end{equation*}
and
\begin{equation*}
\dcoeff_{\scal,\locat}^2 = \bigl (
\acoeff_{\scal+1,\locat_0} -
\acoeff_{\scal+1,\locat_1} -
\acoeff_{\scal+1,\locat_2} +
\acoeff_{\scal+1,\locat_3}
\bigr ) \sqrt{\area_\scal}/4
\spcend ,
\end{equation*}
where we call $\dcoeff_{\scal,\locat}^\wavtype$ the detail coefficients
of type $\wavtype$.
Starting from the finest level $\scalmax$, we compute the
approximation and detail coefficients at level $\scalmax-1$ as
outlined above.  We then repeat this procedure to decompose the
approximation coefficients at level $\scalmax-1$ (\ie\ the
approximation function $f_{\scalmax-1}$), into approximation and
detail coefficients at the coarser level $\scalmax-2$.  Repeating this
procedure continually, we recover the multiresolution representation
of $f_\scalmax$ in terms of the coarsest level approximation $f_1$ and
all of the detail coefficients, as specified by
\eqn{\ref{eqn:multires}} and illustrated in \fig{\ref{fig:multiresolution}}.  In
general it is not necessary to continue the multiresolution
decomposition down to the coarsest level $\scal=1$; one may choose to
stop at the intermediate level $\scalmin$, where $1\leq \scalmin <
\scalmax$.

The function $f_\scalmax\in\approxspace_\scalmax$ may then be
synthesised from its approximation and detail coefficients.  Due to
the orthogonal nature of the Haar basis, the approximation
coefficients at level $\scal+1$ may be reconstructed from the weighted
expansion of the scaling function and wavelets at the coarser level $\scal$,
where the weights are given by the approximation and detail
coefficients respectively.  Writing this expansion explicitly, the
approximation coefficients at level $\scal+1$ are given in terms of the
approximation and detail coefficients of the coarser level $\scal$:
\begin{equation*}
\acoeff_{\scal+1,\locat_0} = \bigl (
\acoeff_{\scal,\locat} +
\dcoeff_{\scal,\locat}^0 +
\dcoeff_{\scal,\locat}^1 +
\dcoeff_{\scal,\locat}^2
\bigr ) / \sqrt{\area_\scal}
\spcend ;
\end{equation*}
\begin{equation*}
\acoeff_{\scal+1,\locat_1} = \bigl (
\acoeff_{\scal,\locat} -
\dcoeff_{\scal,\locat}^0 +
\dcoeff_{\scal,\locat}^1 -
\dcoeff_{\scal,\locat}^2
\bigr ) / \sqrt{\area_\scal}
\spcend ;
\end{equation*}
\begin{equation*}
\acoeff_{\scal+1,\locat_2} = \bigl (
\acoeff_{\scal,\locat} +
\dcoeff_{\scal,\locat}^0 -
\dcoeff_{\scal,\locat}^1 -
\dcoeff_{\scal,\locat}^2
\bigr ) / \sqrt{\area_\scal}
\spcend ;
\end{equation*}
\begin{equation*}
\acoeff_{\scal+1,\locat_3} = \bigl (
\acoeff_{\scal,\locat} -
\dcoeff_{\scal,\locat}^0 -
\dcoeff_{\scal,\locat}^1 +
\dcoeff_{\scal,\locat}^2
\bigr ) / \sqrt{\area_\scal}
\spcend .
\end{equation*}
Repeating this procedure from level $\scal=\scalmin$ up to
$\scal=\scalmax$, one finds that the signal
$f_\scalmax\in\approxspace_\scalmax$ may be written
\begin{equation*}
f_\scalmax(\sa) =
\sum_{\locat=0}^{\npix_{\scalmin} - 1} 
\acoeff_{\scalmin,\locat} \scalefun_{\scalmin,\locat}(\sa)
+
\sum_{\scal=\scalmin}^{\scalmax-1}
\sum_{\locat=0}^{\npix_\scal - 1}
\sum_{\wavtype=0}^{2}
\dcoeff_{\scal,\locat}^\wavtype
\wav_{\scal,\locat}^\wavtype(\sa)
\spcend .
\end{equation*}

\subsection{Lossless compression}
\label{sec:algorithm_lossless}

The Haar wavelet transform on the sphere defined in the previous
section is used as the first stage of the lossless compression
algorithm.  The purpose of this stage is to compress the energy
content of the original data.  In order to recover the original data
from its compressed representation, the energy compression stage must
be reversible.  This requirement limits candidate wavelet transforms
on the sphere to those that allow the exact reconstruction of a signal
from its wavelet coefficients.  We choose the Haar wavelet transform
on the sphere since it satisfies this requirement and also because of
its simplicity and computational efficiency.

We demonstrate the energy compression achieved by the Haar wavelet
transform with an example.  In \fig{\ref{fig:hist}(a)} we show a
histogram of the value of each datum contained in a data-sphere that
we wish to compress (although the particular data-sphere examined here
is not of considerable importance, for the purpose of this
demonstration we use the \cmb\ data described in
\sectn{\ref{sec:applications_lossless}}).  In \fig{\ref{fig:hist}(b)}
we show a histogram of the value of the Haar approximation and detail
coefficients for the same data-sphere.  Notice how the wavelet
transform has compressed the energy of the signal so that it is
contained predominantly within a smaller range of values.  Entropy
provides a measure of the information content of a signal and is
defined by $H=-\sum_i P_i \log_2 P_i$, where $P_i$ is the probability
of symbol $i$ occurring in the data.  By compressing the energy of the
data so that certain symbols will have higher probability, we reduce
its entropy.  The aforementioned entropy value also provides a
theoretical limit on the best compression of data attainable with
entropy encoding, \update{hence by reducing the entropy of data the
performance of any subsequent compression is improved.}

Following the wavelet transform stage of the compression algorithm, we
perform an entropy encoding stage to compress the data.  Entropy
encoding is a type of variable length encoding, where symbols that
occur frequently are given short codewords.  The entropy $H$ of the
data gives the mean number of bits per datum required to encode the
data using an ideal variable length entropy code.  We adopt Huffman
encoding, which produces a code that closely approximates the
performance of the ideal entropy code.

A compression algorithm consisting of the wavelet transform and
entropy encoding stages described above would work, however its
performance would be limited.  Although the wavelet transform
compresses the energy of the data, coefficient values that are
extremely close to one another may take distinct machine
representations.  In order to achieve good compression ratios,
one requires a compressed energy representation of the data with a
relatively small number of unique symbols.  To satisfy this
requirement we introduce a quantisation stage in our compression
algorithm before the Huffman encoding.  By quantising we map similar
coefficient values to the same machine representation, thus reducing
the number of unique symbols contained in the data.  The quantisation
stage does introduce some distortion and so the resulting compression
algorithm is no longer perfectly lossless, but is lossless only to a
user specified numerical precision.  As one increases the precision
parameter, lossless compression is achieved in the limit.  The user
specified precision parameter $\precision$ defines the number of
significant figures to retain in the wavelet detail coefficients
(approximation coefficients are kept to the full number of significant
figures provided by the machine representation).  The
precision parameter trades-off decompression fidelity with compression
ratio.  The effect of quantisation on compression performance is
evaluated in \sectn{\ref{sec:applications}}.  

\update{For this lossless compression algorithm, data are decompressed
  simply by decoding the Huffman encoding of the wavelet coefficients,
  followed by application of the inverse Haar wavelet transform on the
  sphere.}

\newlength{\histplotheight}
\setlength{\histplotheight}{45mm}

\begin{figure}
\centering
\subfigure[\subfigcapsize Original
data]{\includegraphics[height=\histplotheight]{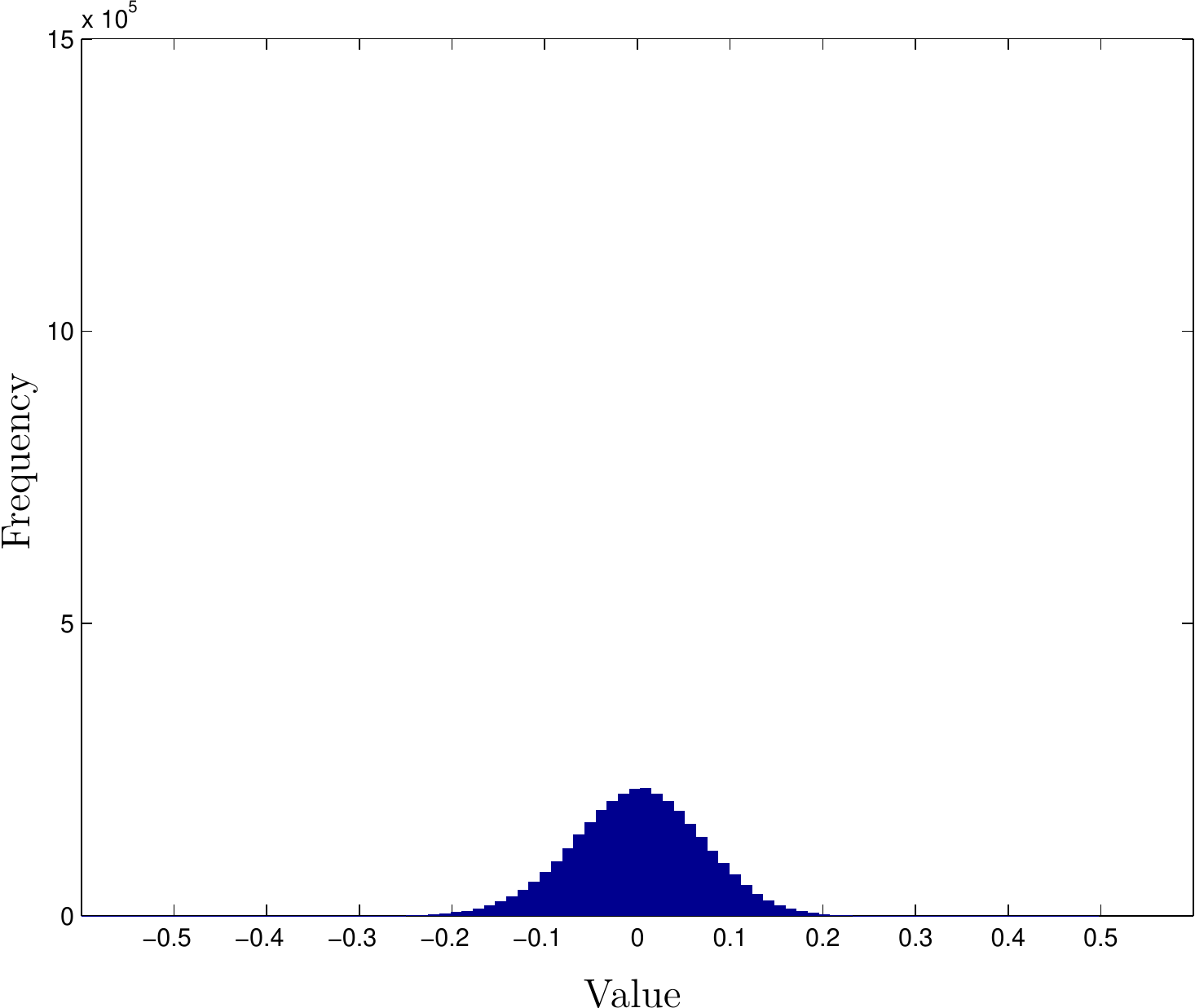}} 
\ifmnras 
  \\
\fi\ifspie
  \quad
\fi
\subfigure[\subfigcapsize Wavelet coefficients]{\includegraphics[height=\histplotheight]{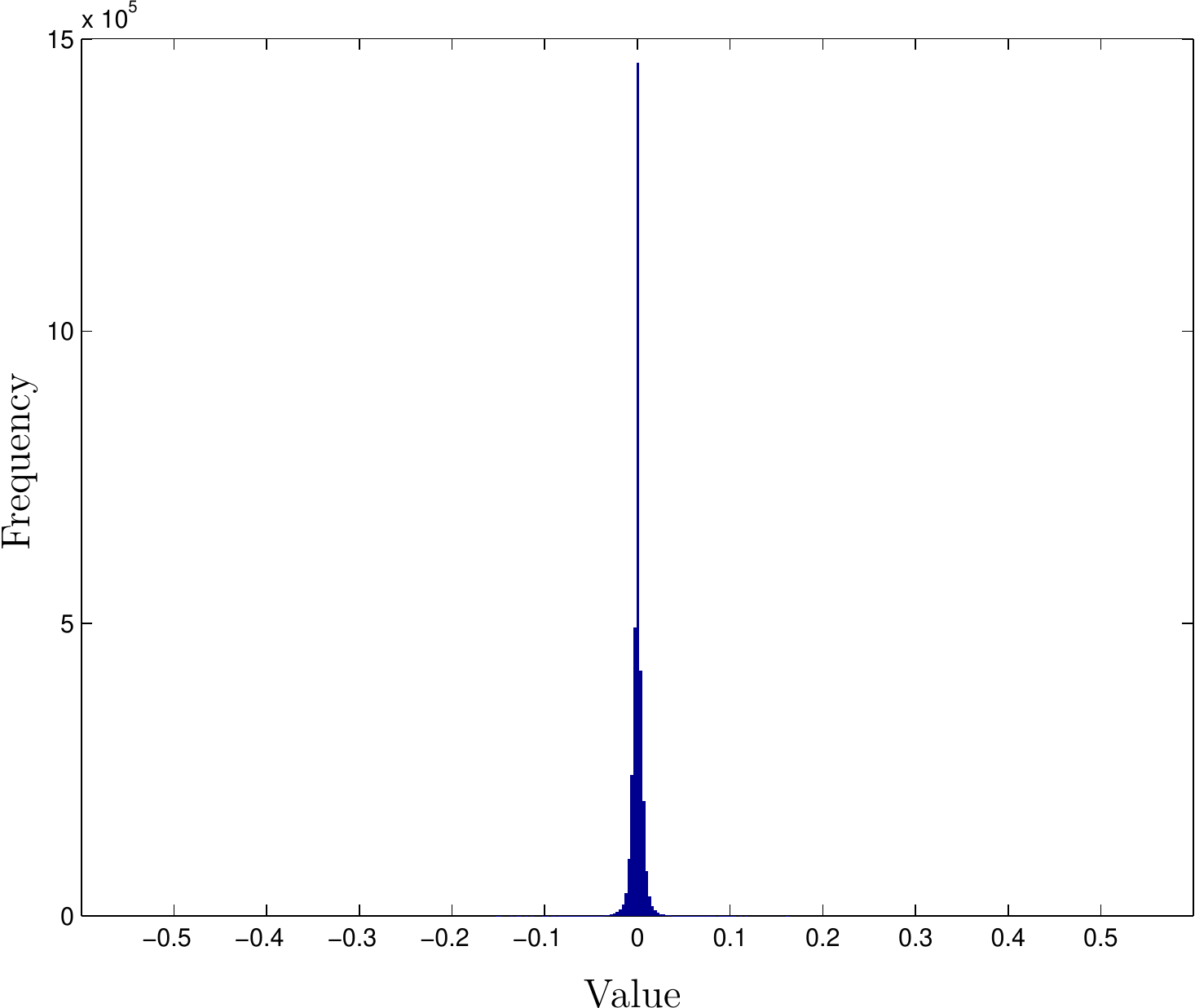}}
\caption{Histograms of original data and wavelet (approximation and
  detail) coefficient values.  Although the particular data-sphere
  considered here is not important for the purpose of this
  demonstration, these histograms correspond to the \cmb\ data
  described in \sectn{\ref{sec:applications_lossless}}.  Notice how
  the Haar wavelet transform has compressed the energy content of the
  signal, thereby reducing its entropy and allowing for greater
  compression performance.  }
\label{fig:hist}
\end{figure}

\subsection{Lossy compression}
\label{sec:algorithm_lossy}

If we allow degradation to the quality of the
decompressed data it is possible to achieve higher compression
ratios.  In this section we describe a lossy compression
algorithm that trades-off losses in decompression fidelity against
compression ratio in a natural manner.  

The Haar wavelet representation of a data-sphere decomposes the data
into an approximation sphere and detail coefficients that encode the
differences between the approximation sphere and the original
data-sphere.  Many of these detail coefficients are often close to
zero (as illustrated by the histogram shown in
\fig{\ref{fig:hist}(b)}).  If we discard those detail coefficients
that are near zero, by essentially setting their value to zero, then
we lose only a small amount of accuracy in the representation of the
original data but reduce the entropy considerably.  By
increasing the proportion of detail coefficients neglected, we
improve the compression ratio of the compressed data while
reducing its fidelity in a natural manner.

Our lossy compression algorithm is identical to the lossless algorithm
described in \sectn{\ref{sec:algorithm_lossless}} but with two
additional stages included.  Firstly, we introduce a thresholding
stage after the quantisation and before the Huffman encoding stage.
The threshold level is determined by choosing the proportion of detail
coefficients to retain.  We treat the detail coefficients on all
levels identically.  More sophisticated thresholding algorithms could
treat the detail coefficients on each level $\scal$ differently,
perhaps using an annealing scheme to specify the proportion of detail
coefficients to retain at each level.  However, we demonstrate in
\sectn{\ref{sec:applications_lossy}} that the na\"{\i}ve thresholding
scheme outlined here performs very well in practice and so we do not
investigate more sophisticated strategies.  Once the threshold level
is determined we perform hard thresholding so that all detail
coefficients below this value are set to zero, while coefficients
above the threshold remain unaltered.  The thresholding stage reduces
the number of unique symbols in the data by replacing many unique
values with zero, hence reducing the entropy of the data and enabling
greater compression.  Furthermore, since many of the data are now
zero, it is worthwhile to incorporate a run length encoding (RLE)
stage so that long runs of zeros are encoded efficiently.  The RLE
stage is included after the thresholding and before the Huffman
encoding stage.  RLE introduces some additional encoding overhead,
thus it only improves the compression ratio for cases where there are
sufficiently long runs of zeros.  In
\sectn{\ref{sec:applications_lossy}} we evaluate the performance of
the lossy compression algorithm described here and examine the
trade-off between compression ratio and fidelity.  Moreover, we also
examine cases where the additional overhead due to RLE acts to increase
the compression ratio.

\update{For this lossy compression algorithm, data are decompressed
  simply by decoding the RLE and Huffman encoding of the wavelet
  coefficients, followed by application of the inverse Haar wavelet
  transform on the sphere.}

\section{Applications}
\label{sec:applications}

In this section we evaluate the performance of the lossless and lossy
compression algorithms on data-spheres that arise in a range of 
applications.  The trade-off between compression ratio and the
fidelity of the decompressed data is examined in detail.  We begin by
considering applications where lossless compression is required,
before then considering applications when lossy compression is appropriate.

\subsection{Lossless compression}
\label{sec:applications_lossless}

The algorithms developed here to compress data defined on the sphere
were driven primarily by the need to compress large \cmb\
data-spheres.  \cmb\ data are used to study cosmological models of the
Universe.  Any errors introduced in the data may alter cosmological
inferences drawn from it, hence the introduction of large errors in
the compression of \cmb\ data will not be tolerated.  The lossless
compression algorithm is therefore required for this application.  Our
lossless compression algorithm is lossless only to a user specified
numerical precision (as described in detail in
\sectn{\ref{sec:algorithm_lossless}}).  It is therefore important to
ascertain whether the small quantisation errors that are introduced by
this limited precision could affect cosmological
inferences drawn from the data.  We first evaluate the performance of
the lossless compression of \cmb\ data, before investigating the
impact of errors on the cosmological information content of the data.

To evaluate our lossless compression algorithm we use simulated \cmb\
data.  In the simplest inflationary models of the Universe, the \cmb\
is fully described by its angular power spectrum.  Using the
theoretical angular power spectrum that best fits the three-year
\wmap\ observations (\ie\ the power spectrum defined by the
cosmological parameters specified in \tbl{2} of \citet{spergel:2006}),
we simulate a Gaussian realisation of the \cmb\ temperature
anisotropies.  \update{Foreground emissions contaminate real observations of
the \cmb, hence we also consider simulated maps where a mask is
applied to remove regions of known Galactic and point source
contamination.  We apply the conservative Kp0 mask associated with the
three-year \wmap\ observations \citep{hinshaw:2006}.  These simulated \cmb\
data, with and without application of the Kp0 mask, are illustrated at
resolutions $\nside=512$ ($\scalmax=9$; $\npix\simeq3\times10^6$) and
$\nside=1024$ ($\scalmax=10$; $\npix\simeq13\times10^6$) in the first
column of panels of \fig{\ref{fig:cmb}}.}

The simulated \cmb\ data are compressed using the lossless compression
algorithm for a range of precision values $\precision$.  \update{RLE
  is applied when compressing the masked data to efficiently compress
  the runs of zeros associated with the masked regions but not for the
  unmasked data (since the encoding overhead does not make it
  worthwhile).}  For each precision value, we compute the compression
ratio achieved and the relative error between the decompressed data
and the original data.  The compression ratio is defined by the ratio
of the size of the compressed data (including the Huffman encoding
table) relative to the size of the original data, expressed as a
percentage.  The error used to evaluate the fidelity of the compressed
data is defined by the ratio of the mean-square-error (MSE) between
the original and decompressed data-spheres relative to the
root-mean-squared (RMS) value of the original data-sphere, expressed
as a percentage.  These values are plotted for a range of precision
values in the third column of panels of \fig{\ref{fig:cmb}}.
\update{In the second column of panels of \fig{\ref{fig:cmb}} residual
  errors between the original data and decompressed data reconstructed
  for a precision parameter $\precision=3$ are shown, where a
  compression ratio of 18\% is achieved for resolution $\nside=1024$.
  Although the precision level $\precision=3$ introduces some error in
  the reconstructed data (2.7\% for $\nside=1024$), the error on each
  pixel is reassuringly small at typically a factor of $\sim 100$
  smaller than the corresponding data value.}  For the precision
parameter $\precision=5$, a compression ratio of 40\% and an error of
0.03\% is obtained for resolution $\nside=1024$.  The error introduced
by the compression for this case is sufficiently small that one might
hope that no significant cosmological information content is lost in
the compressed data.  We investigate the cosmological information
content of the compressed data in detail next.

\newlength{\cmbplotheight}
\setlength{\cmbplotheight}{33mm}

\newlength{\performanceplotheight}
\setlength{\performanceplotheight}{35mm}

\begin{figure*}
\centering
\mbox{
\subfigure[\subfigcapsize \cmb\ at $\nside=512$ (13MB)]{\includegraphics[height=\cmbplotheight]{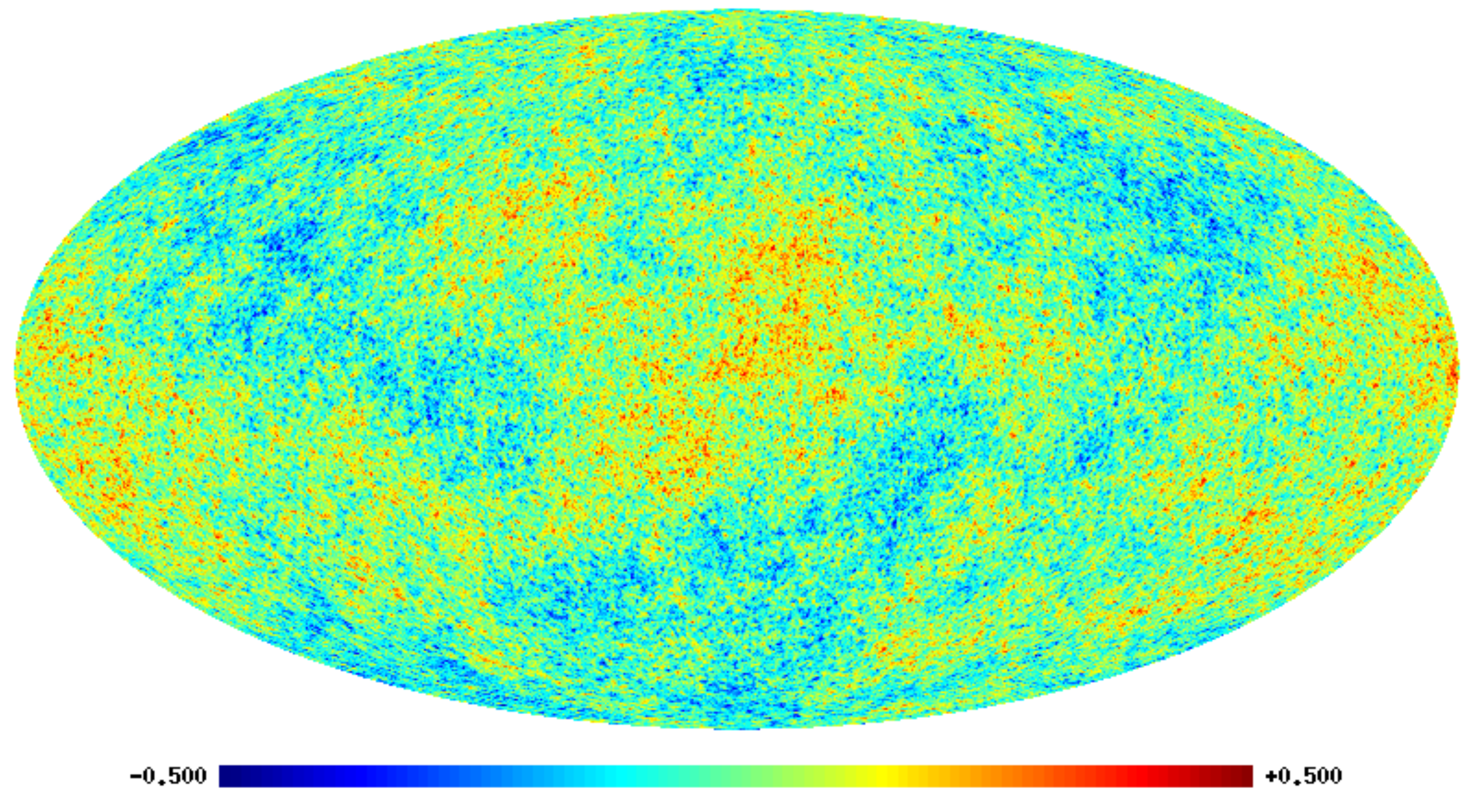}} \quad
\subfigure[\subfigcapsize Residual for $\precision=3$ at $\nside=512$ (2.5MB)]{\includegraphics[height=\cmbplotheight]{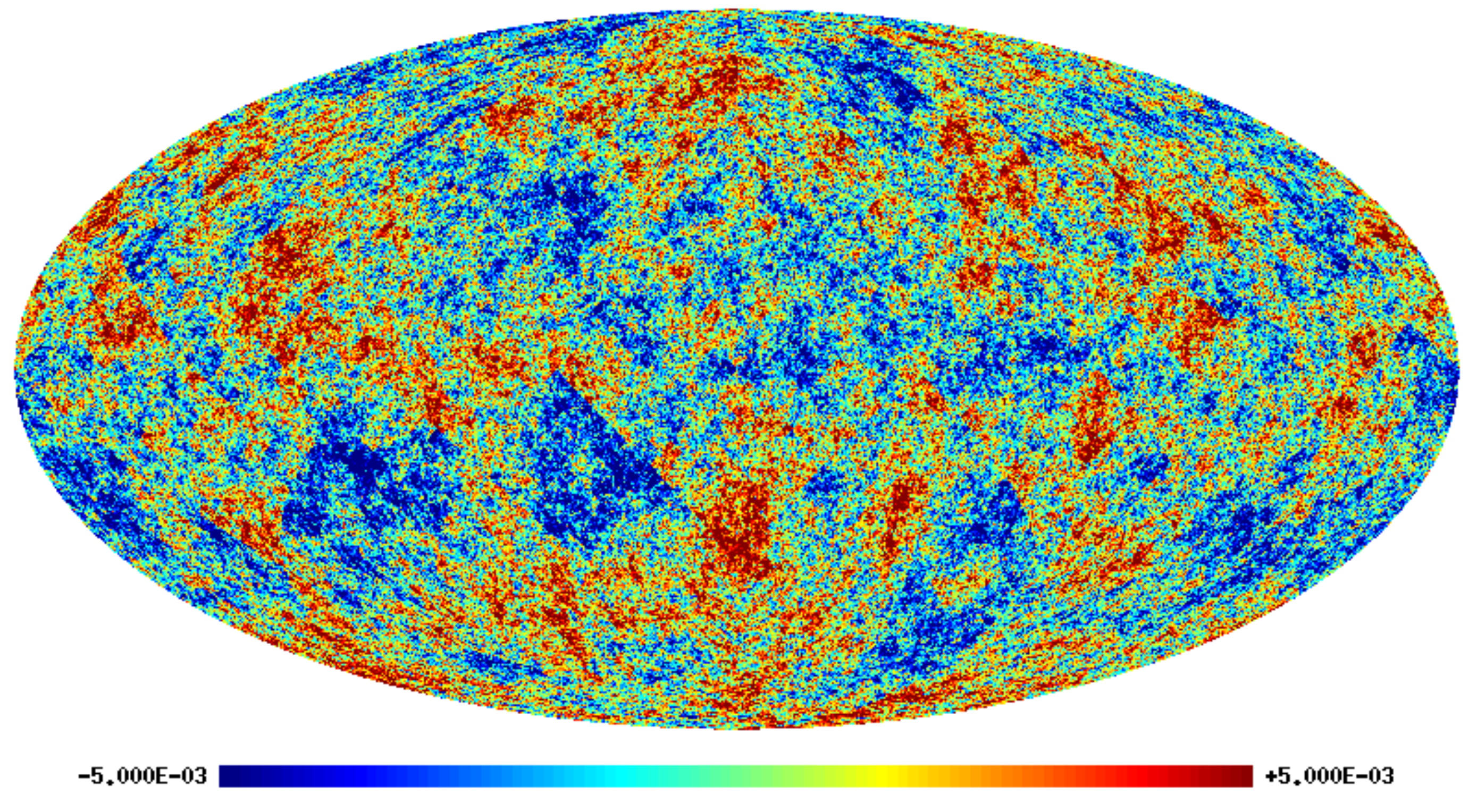}} 
\subfigure[\subfigcapsize Performance at \mbox{$\nside=512$}]{\quad\includegraphics[height=\performanceplotheight]{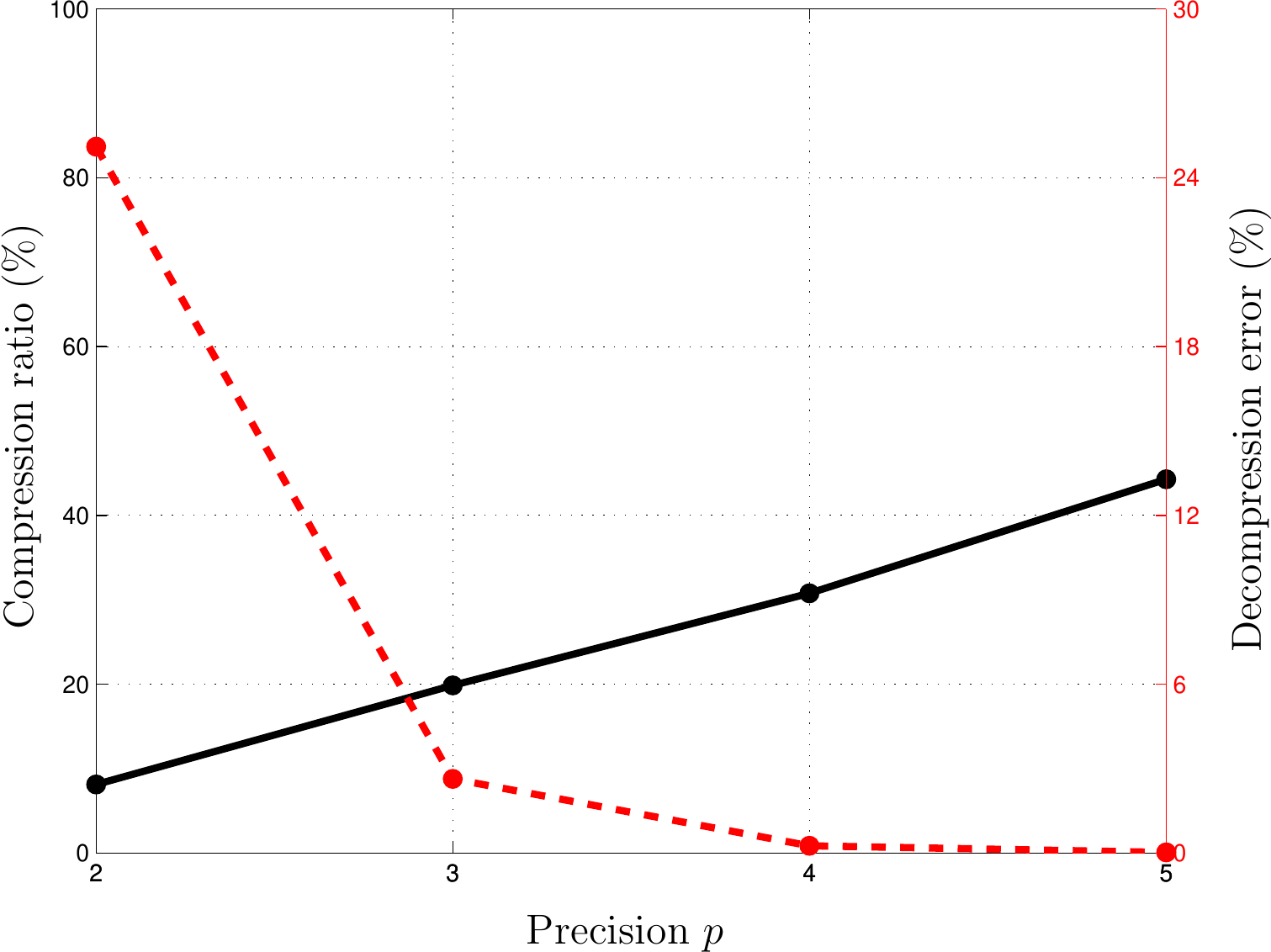}}
}\\
\mbox{
\subfigure[\subfigcapsize Masked \cmb\ at $ \nside=512$ (13MB)]{\includegraphics[height=\cmbplotheight]{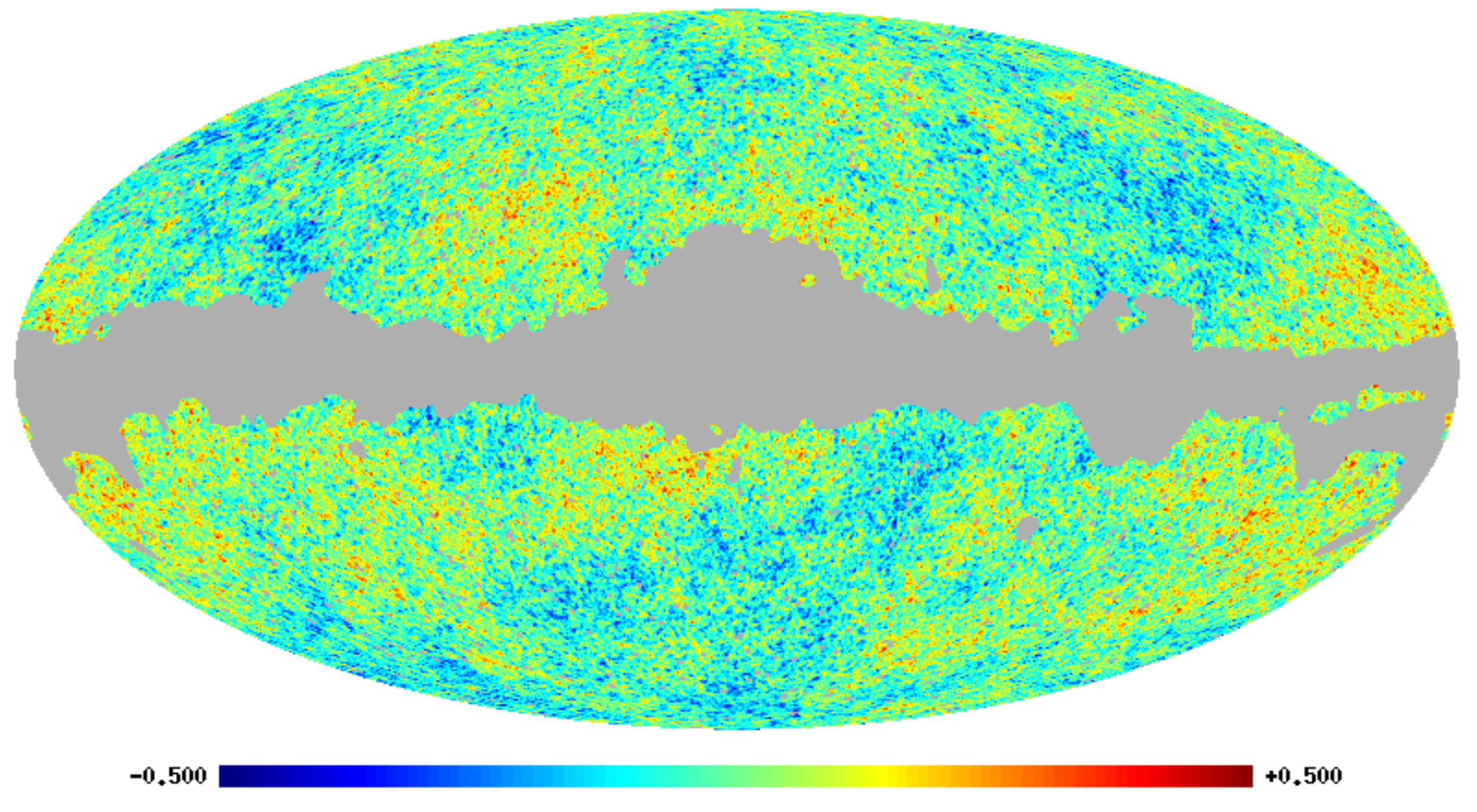}} \quad
\subfigure[\subfigcapsize Masked residual for $\precision=3$ at $\nside=512$ (2.1MB)]{\includegraphics[height=\cmbplotheight]{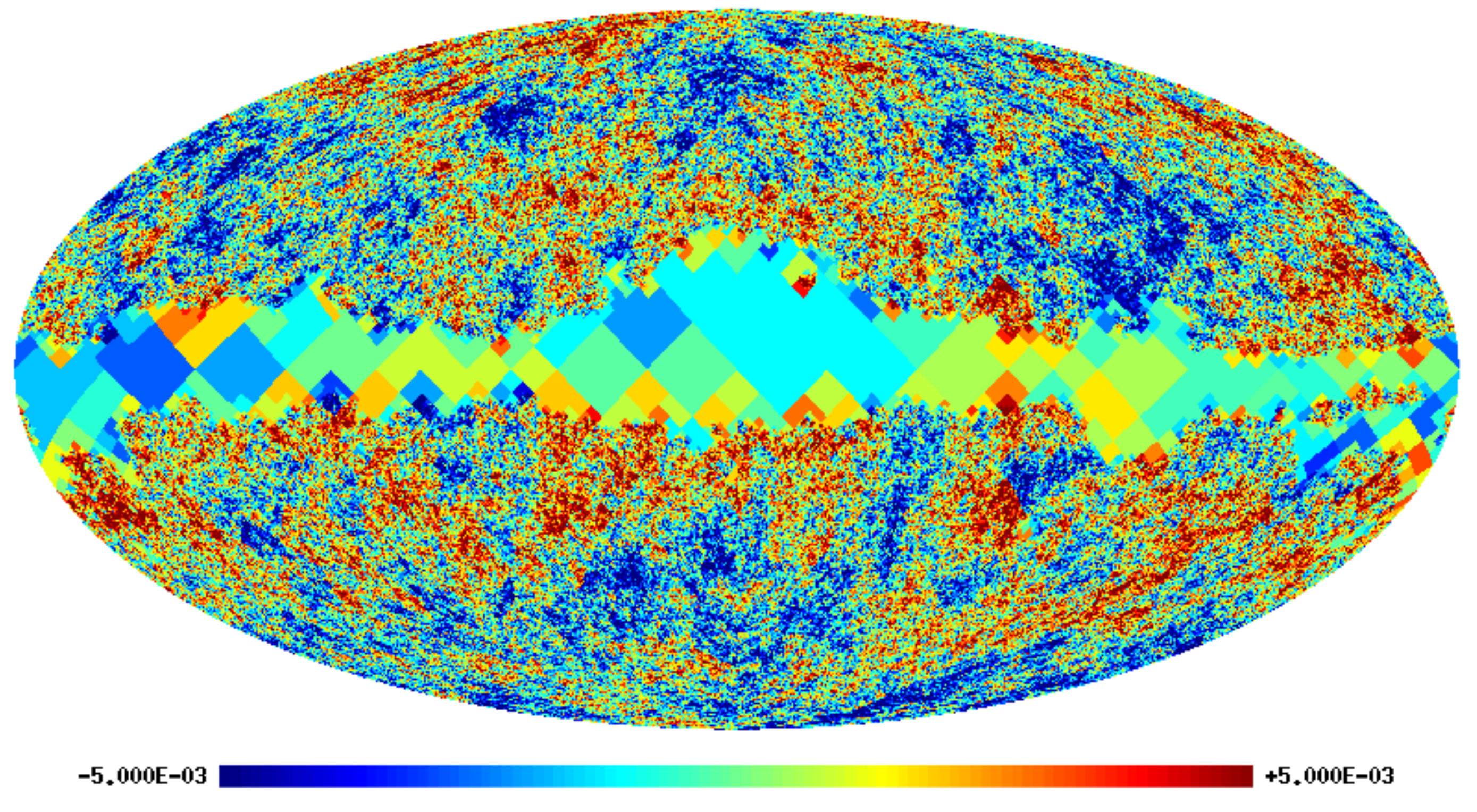}} 
\subfigure[\subfigcapsize Masked performance at \mbox{$\nside=512$}]{\quad\includegraphics[height=\performanceplotheight]{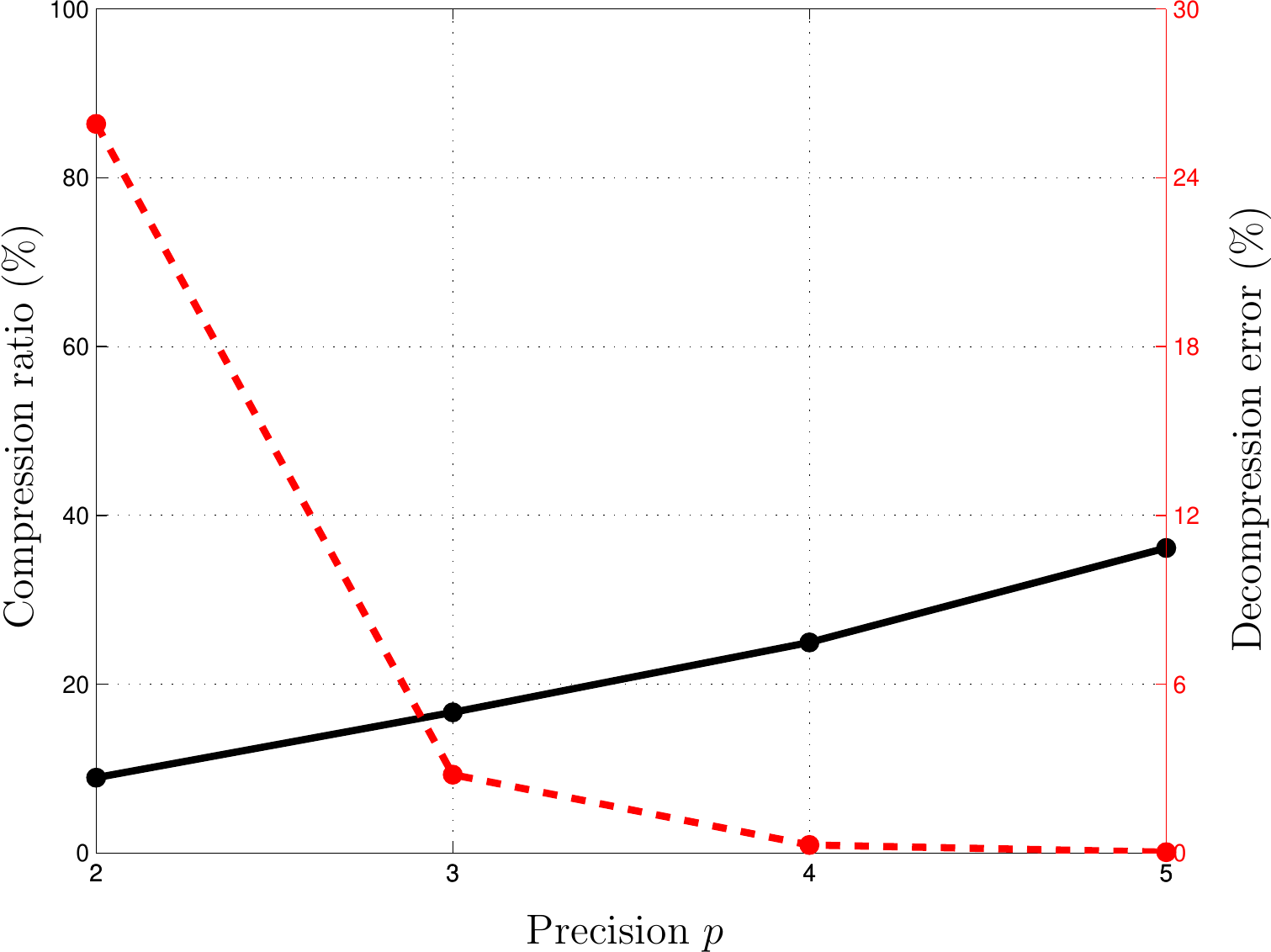}}
}\\
\mbox{
\subfigure[\subfigcapsize \cmb\ at $\nside=1024$ (50MB)]{\includegraphics[height=\cmbplotheight]{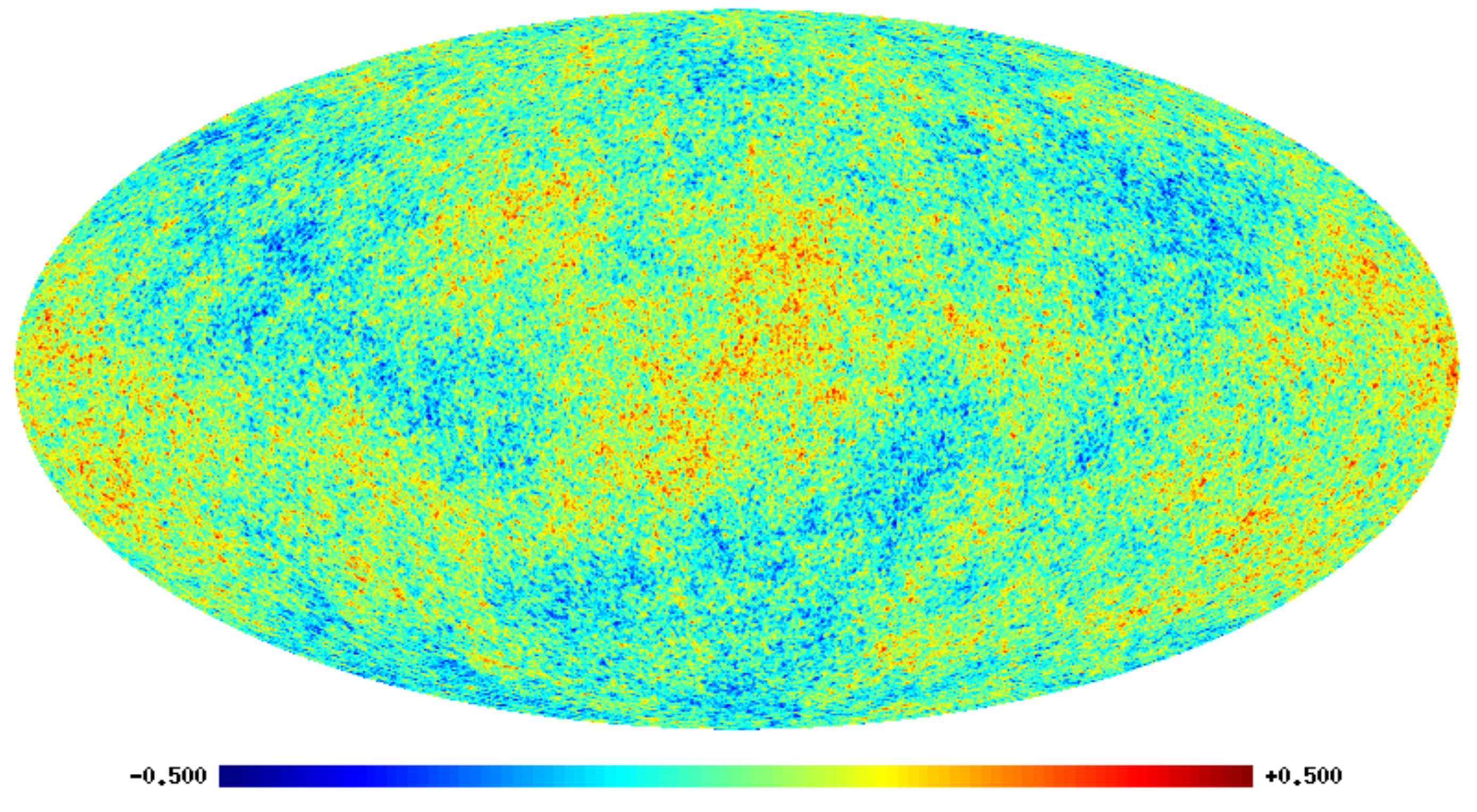}} \quad
\subfigure[\subfigcapsize Residual for $\precision=3$ at $\nside=1024$ (9.1MB)]{\includegraphics[height=\cmbplotheight]{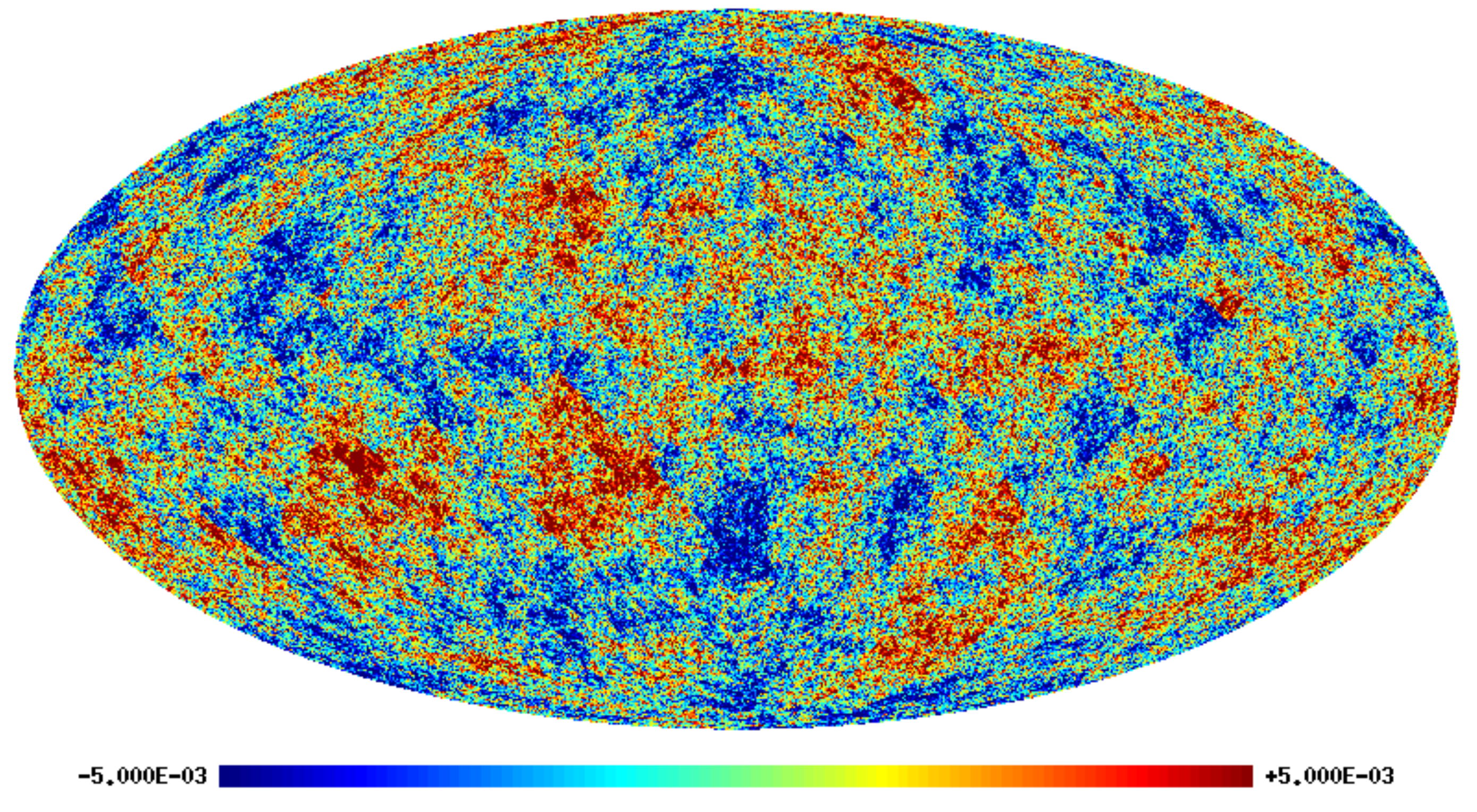}} 
\subfigure[\subfigcapsize Performance at \mbox{$\nside=1024$}]{\quad\includegraphics[height=\performanceplotheight]{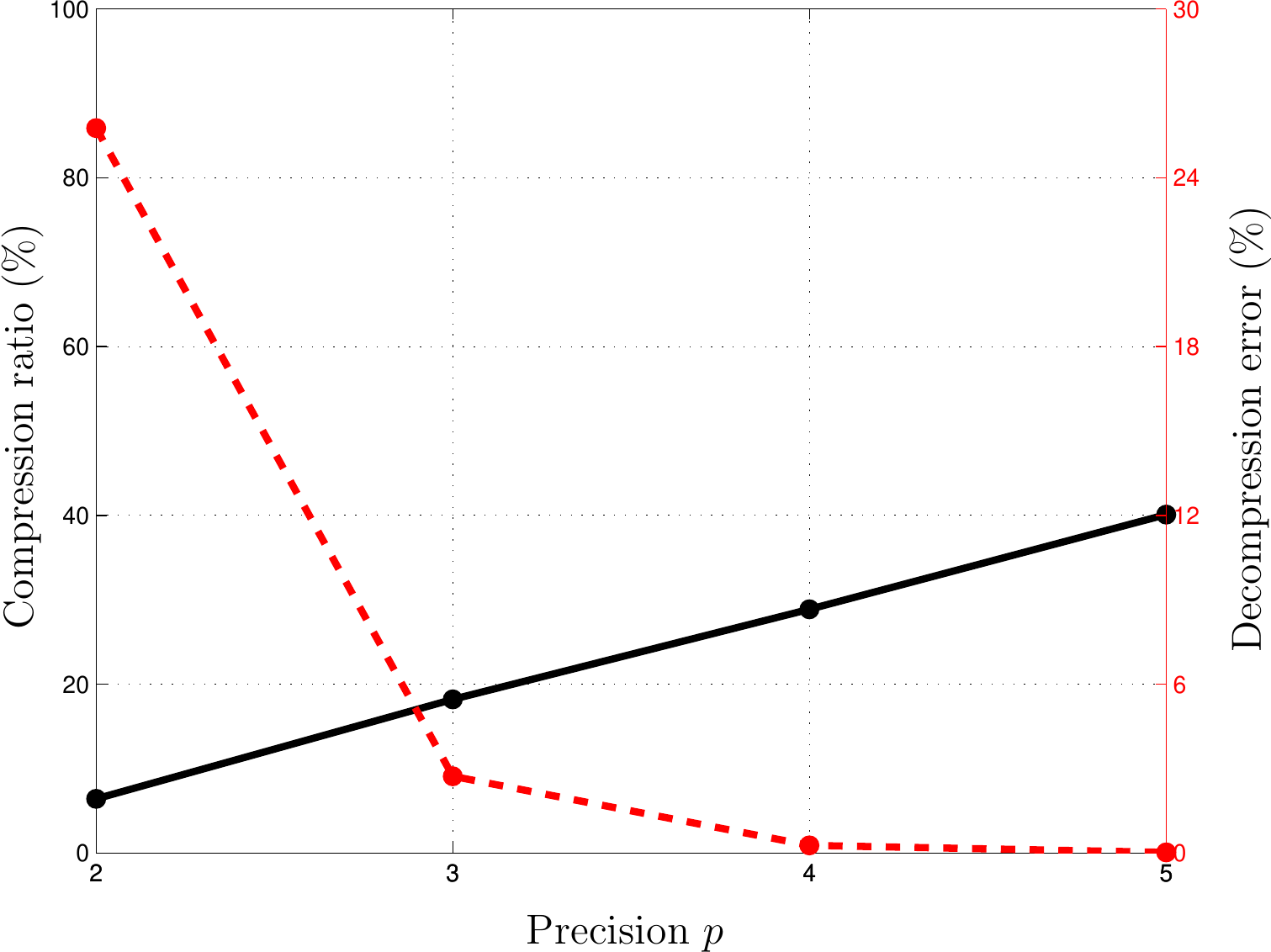}}
}
\mbox{
\subfigure[\subfigcapsize Masked \cmb\ at $\nside=1024$ (50MB)]{\includegraphics[height=\cmbplotheight]{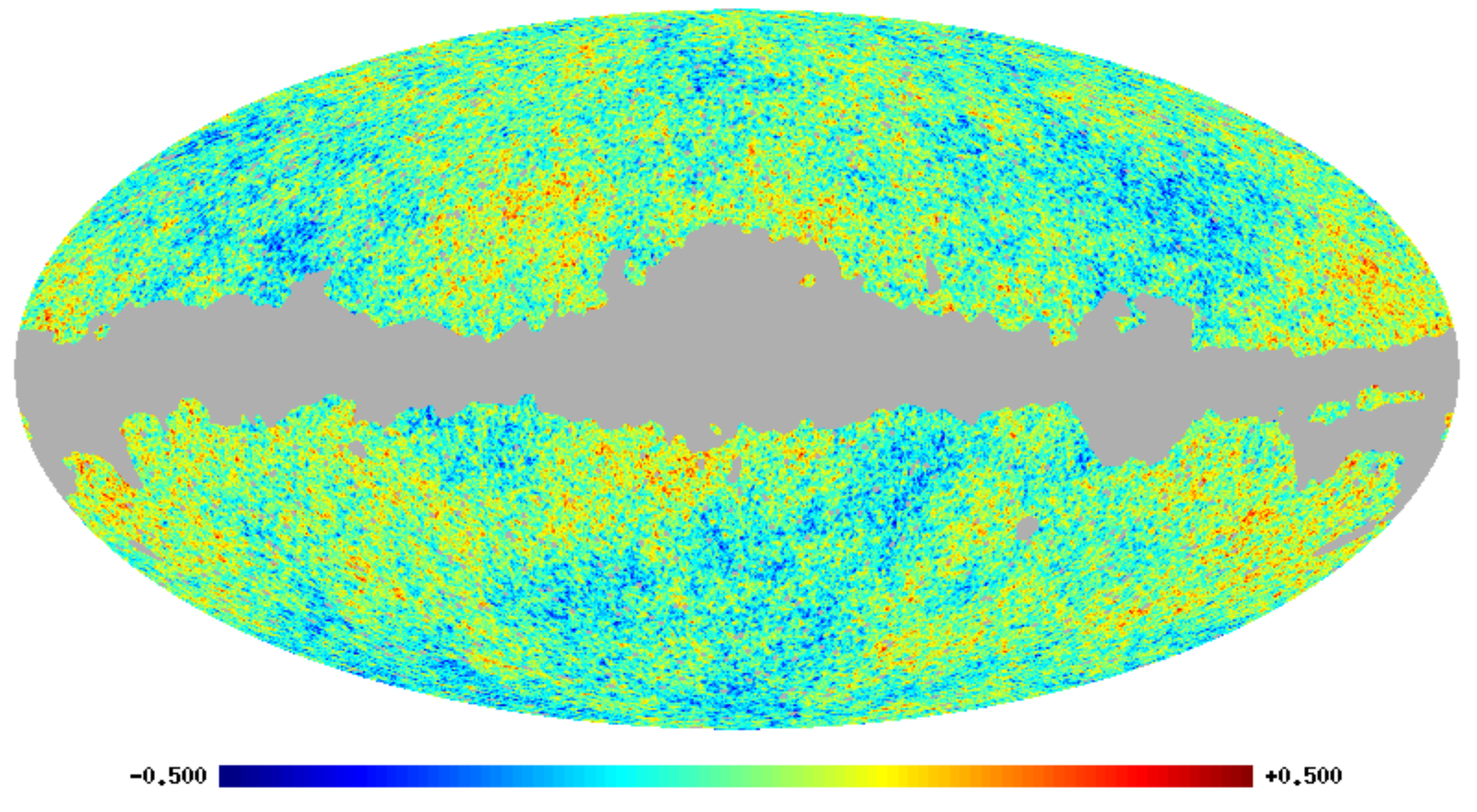}} \quad
\subfigure[\subfigcapsize Masked residual for $\precision=3$ at $\nside=1024$ (7.6MB)]{\includegraphics[height=\cmbplotheight]{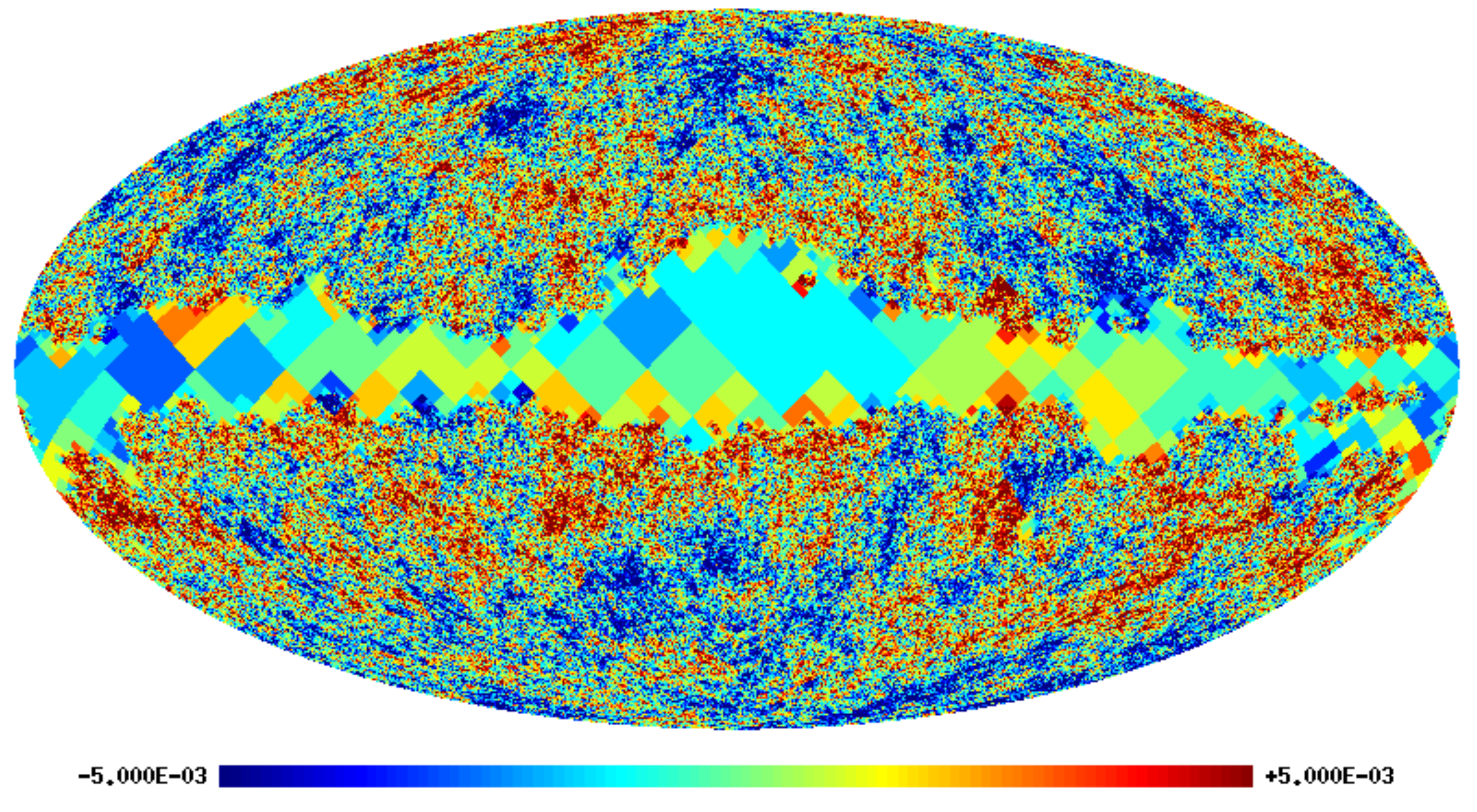}} 
\subfigure[\subfigcapsize Masked performance at \mbox{$\nside=1024$}]{\quad\includegraphics[height=\performanceplotheight]{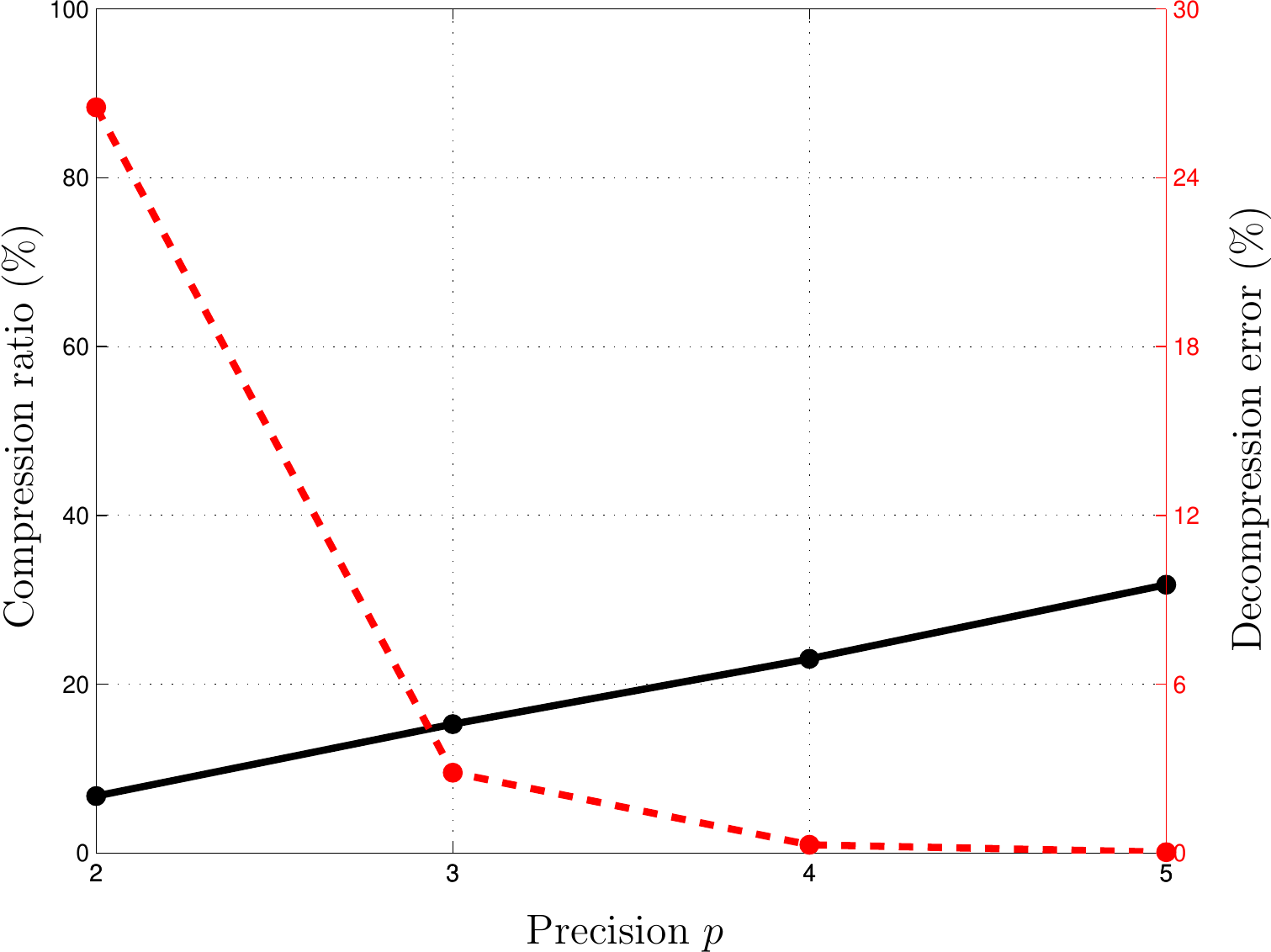}}
}
\caption{Lossless compression of simulated Gaussian \cmb\ data, \update{with
  and without application of the Kp0 mask (data-spheres are
  displayed using the Mollweide projection).}
  The first column of panels shows the original simulated \cmb\ data,
  with corresponding file sizes also specified.  \update{The second column of
  panels shows the residual of the original and decompressed \cmb\
  data reconstructed for a precision parameter of $\precision=3$, with
  corresponding file sizes of the compressed data also specified.  Note that the colour scale
  between the first and second column of panels is scaled by a factor
  of 100.  RLE is applied when compressing the masked data to
  efficiently compress the runs of zeros associated with the masked
  regions.}  The third column of panels shows the trade-off between
  compression ratio and decompression fidelity with precision
  parameter $\precision$.  Compression ratio (solid black line; left
  axis) is defined by the ratio of the compressed file size relative
  to the original file size, expressed as a percentage.  The
  decompression error (dashed red line; right axis) is defined by the
  ratio of the MSE between the original and decompressed data relative
  to the RMS value of the original data, expressed as a percentage.  }
\label{fig:cmb}
\end{figure*}

In the simplest inflationary scenarios the cosmological information
content of the \cmb\ is contained fully in its angular power spectrum.
Although the angular power spectrum does not contain all cosmological
information in non-standard inflationary settings, anisotropic models
of the Universe or various cosmic defect scenarios, we nevertheless
use it as a figure of merit to determine any errors in the
cosmological information content of \cmb\ data.  To evaluate any loss
to the cosmological information content of compressed \cmb\ data, we
examine the errors that are induced in the angular power spectrum of
the compressed data.  \update{We consider here unmasked \cmb\ data only, which
simplifies the estimation of the angular power spectrum.}
Before proceeding, we briefly define the angular power spectrum of the
\cmb\ and the estimator that we use to compute the power spectrum from
\cmb\ data.  The angular power spectrum $C_\el$ is given by the
variance of the spherical harmonic coefficients of
the \cmb, \ie\ $\langle a_{\el m}^{} a_{\el m}^\cconj \rangle = C_\el
\kron{\el}{\el^\prime} \kron{m}{m^\prime}$, where $\kron{i}{j}$ is the
Kronecker delta symbol and the spherical harmonic coefficients $a_{\el
  m}$ are given by the projection of the \cmb\ anisotropies $\Delta
T(\sas)$ onto the spherical harmonic functions $Y_{\el m}(\sas)$
through $a_{\el m} = \langle \Delta T | Y_{\el m} \rangle$.
If the \cmb\ is assumed to be isotropic, then for a given $\el$ the
$m$-modes of the spherical harmonic coefficients are independent and
identically distributed.  The underlying $C_\el$ spectrum may therefore be
estimated by the quantity
$\hat{C}_\el \equiv \sum_{m=-\el}^\el | a_{\el m} | / (2\el+1)$.
Since more $m$-modes are available at higher $\el$, the error on this
estimator reduces as $\el$ increases.  This phenomenon is termed
cosmic variance and arises since we may observe one realisation
of the \cmb\ only.  Cosmic variance is given by 
$(\Delta \hat{C}_\el)^2 = 2 C_\el^2 / (2\el+1)$
and provides a natural uncertainty level for power spectrum estimates
made from \cmb\ data.  Any errors introduced in the angular power
spectrum of compressed \cmb\ data may therefore be related to cosmic
variance to determine the cosmological implication of these errors.
In \fig{\ref{fig:cl}} we show the angular power spectrum computed for
the original and compressed \cmb\ data for $\nside=1024$, and errors
between these spectra, for a range of precision parameter values.  In
the first two columns of panels we also highlight the three standard
deviation confidence interval due to cosmic variance.  For the
precision parameter $\precision=5$, we find that essentially no
cosmological information content is lost in the compressed data.  Even
for large values of $\el$, for which cosmic variance is very small,
the error in the recovered power spectrum relative to cosmic variance
is of the order of a few percent only.  For the case $\precision=4$,
still only minimal cosmological information content is lost, while for
the case $\precision=3$ we begin to see a moderate loss of
cosmological information.  Obviously the degree of cosmological
information content loss that may be tolerated depends on the
application at hand.  However, we have demonstrated that it is
possible to compress \cmb\ data to 40\% of its original size while
ensuring that essentially no cosmological information content is lost
(corresponding to $\precision=5$).  If one tolerates a moderate loss of
cosmological information then the data may be compressed to 18\% of
its original size (corresponding to $\precision=3$).

\newlength{\clplotwidth}
\setlength{\clplotwidth}{55mm}

\begin{figure*}
\centering
\mbox{
\subfigure[\subfigcapsize Angular power spectra for $p=5$]
  {\includegraphics[clip=,width=\clplotwidth]{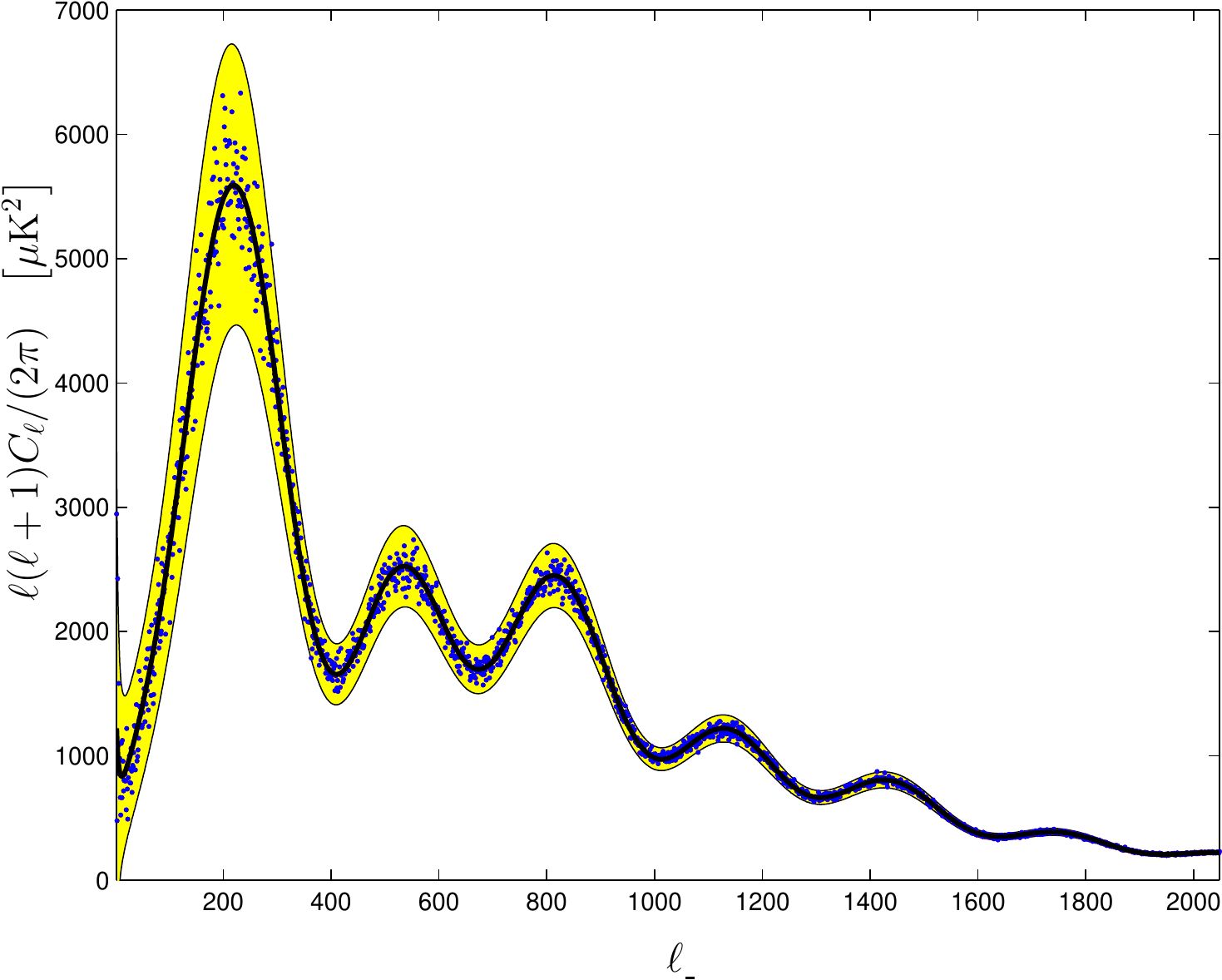}} \quad
\subfigure[\subfigcapsize Absolute error for $p=5$]
  {\includegraphics[clip=,width=\clplotwidth]{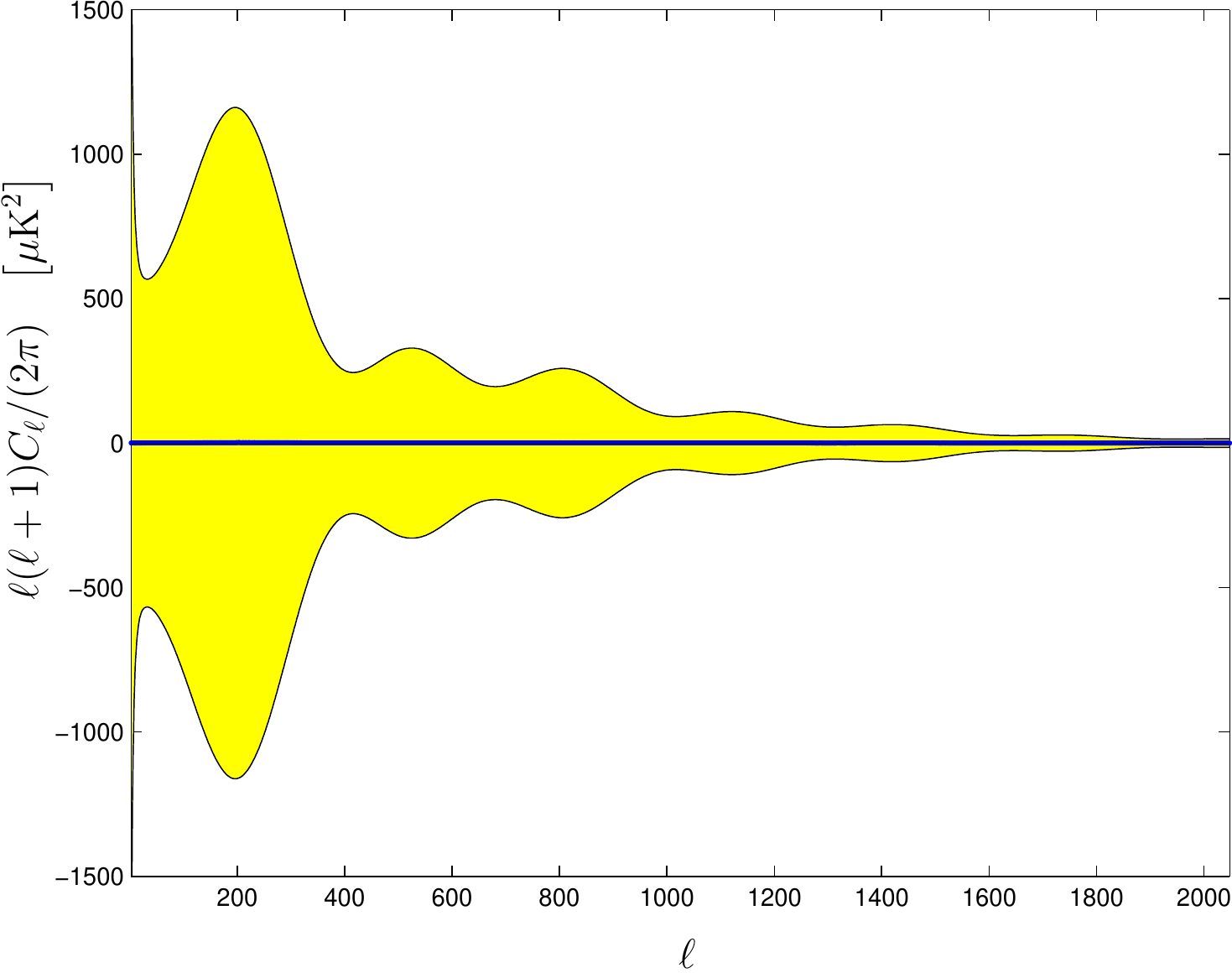}} 
\subfigure[\subfigcapsize Error relative to cosmic variance for $p=5$]
  {\quad \includegraphics[clip=,width=\clplotwidth]{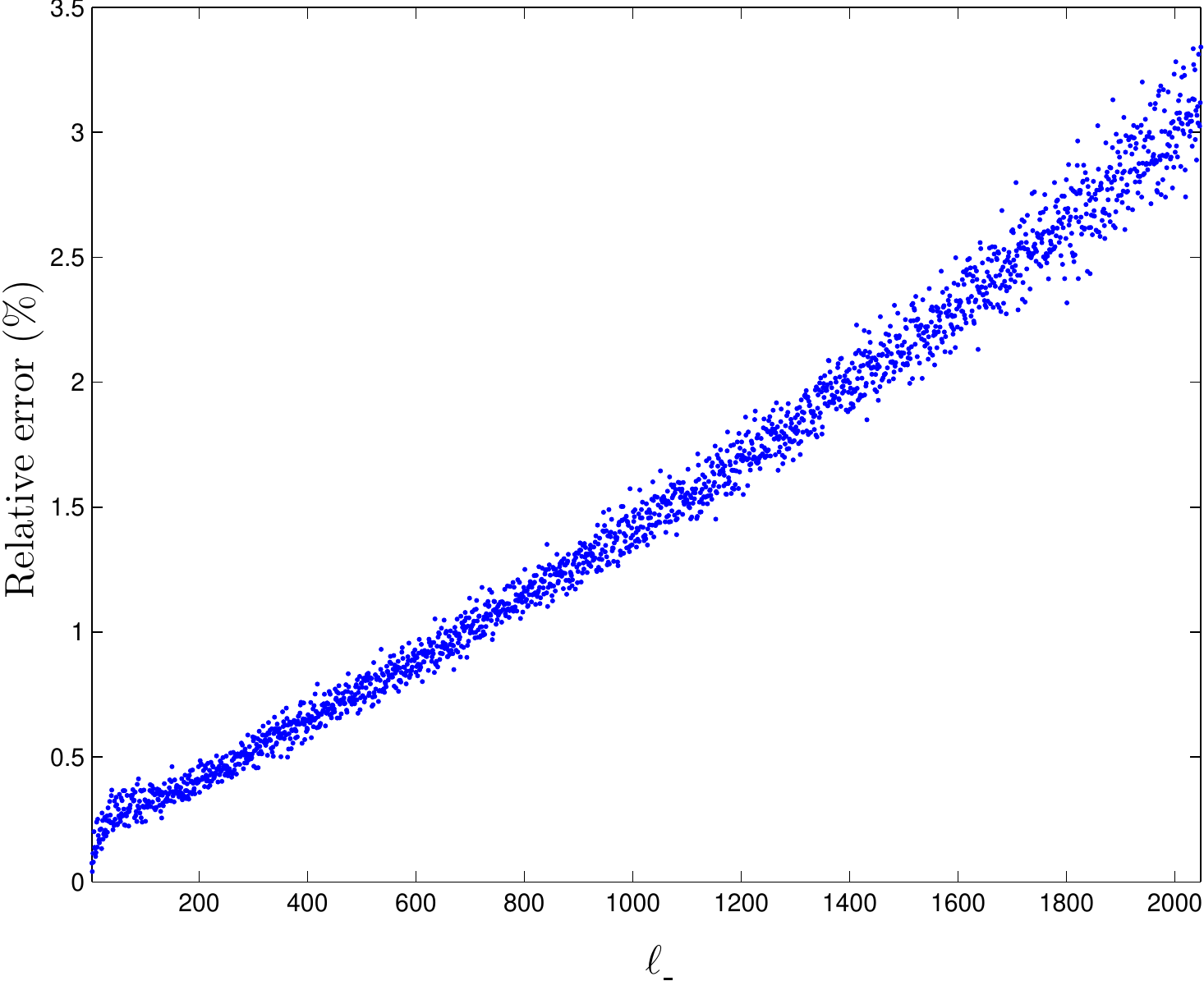}}
}
\mbox{
\subfigure[\subfigcapsize Angular power spectra for $p=4$]
  {\includegraphics[clip=,width=\clplotwidth]{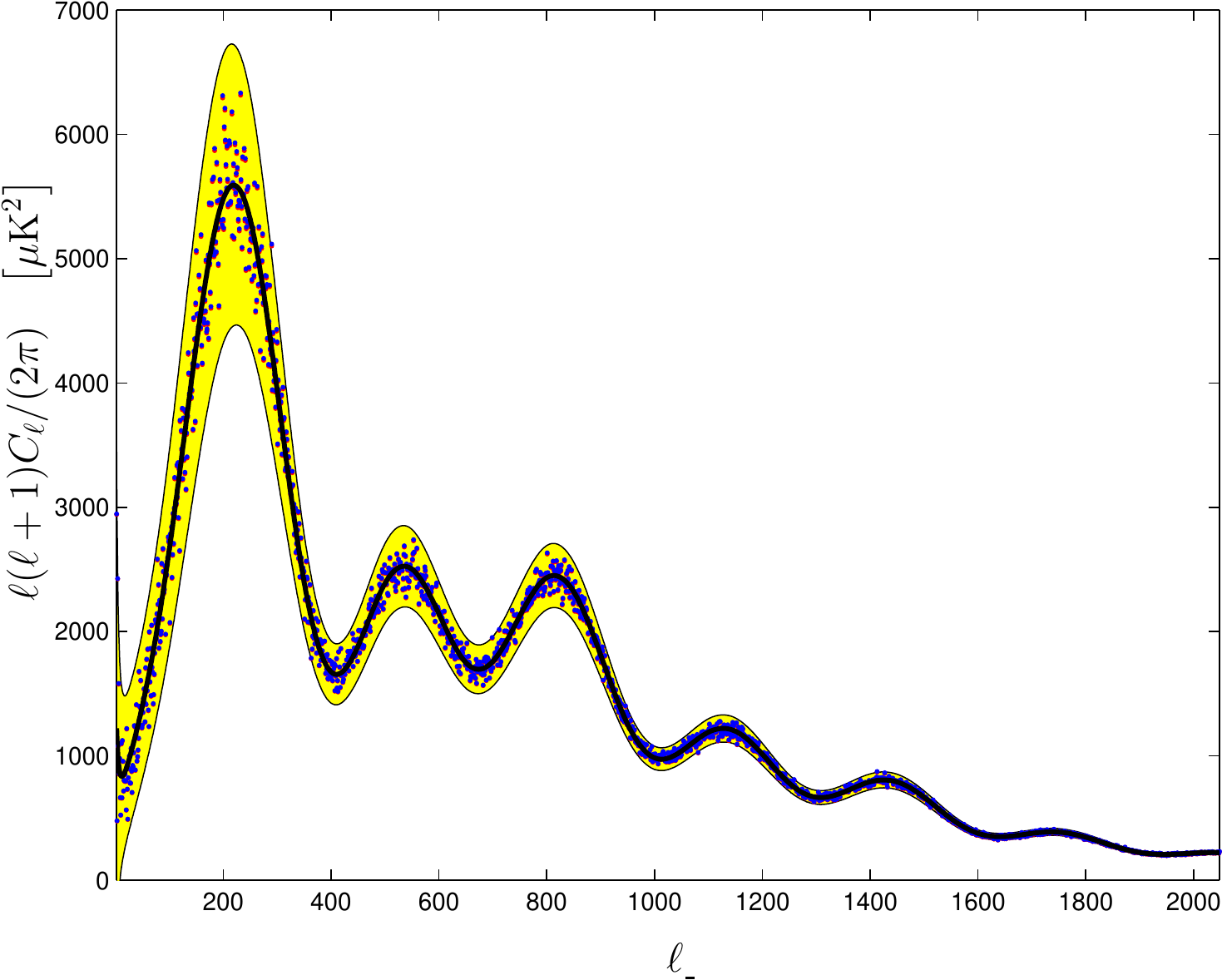}} \quad
\subfigure[\subfigcapsize Absolute error for $p=4$]
  {\includegraphics[clip=,width=\clplotwidth]{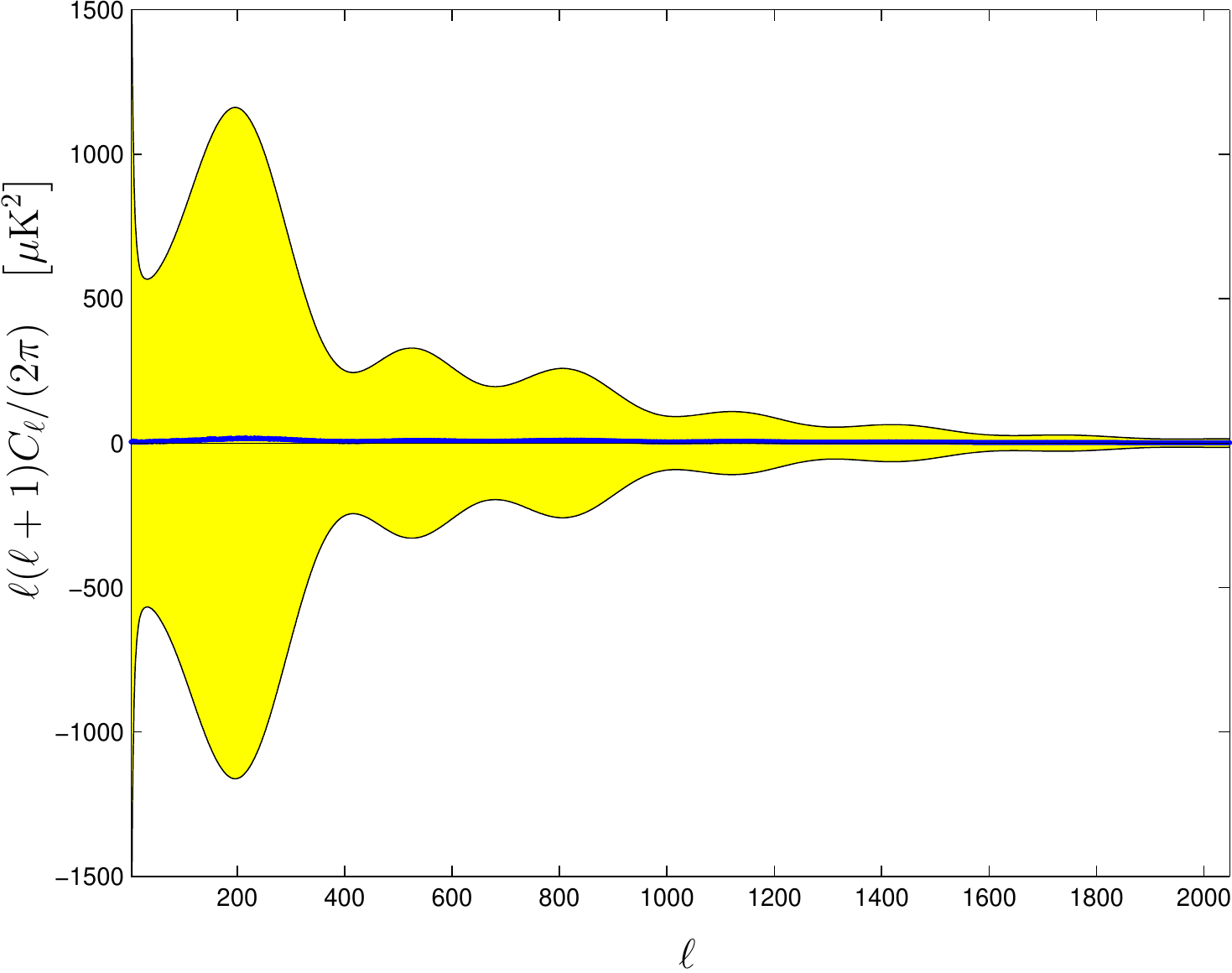}}
\subfigure[\subfigcapsize Error relative to cosmic variance for $p=4$]
  {\quad\includegraphics[clip=,width=\clplotwidth]{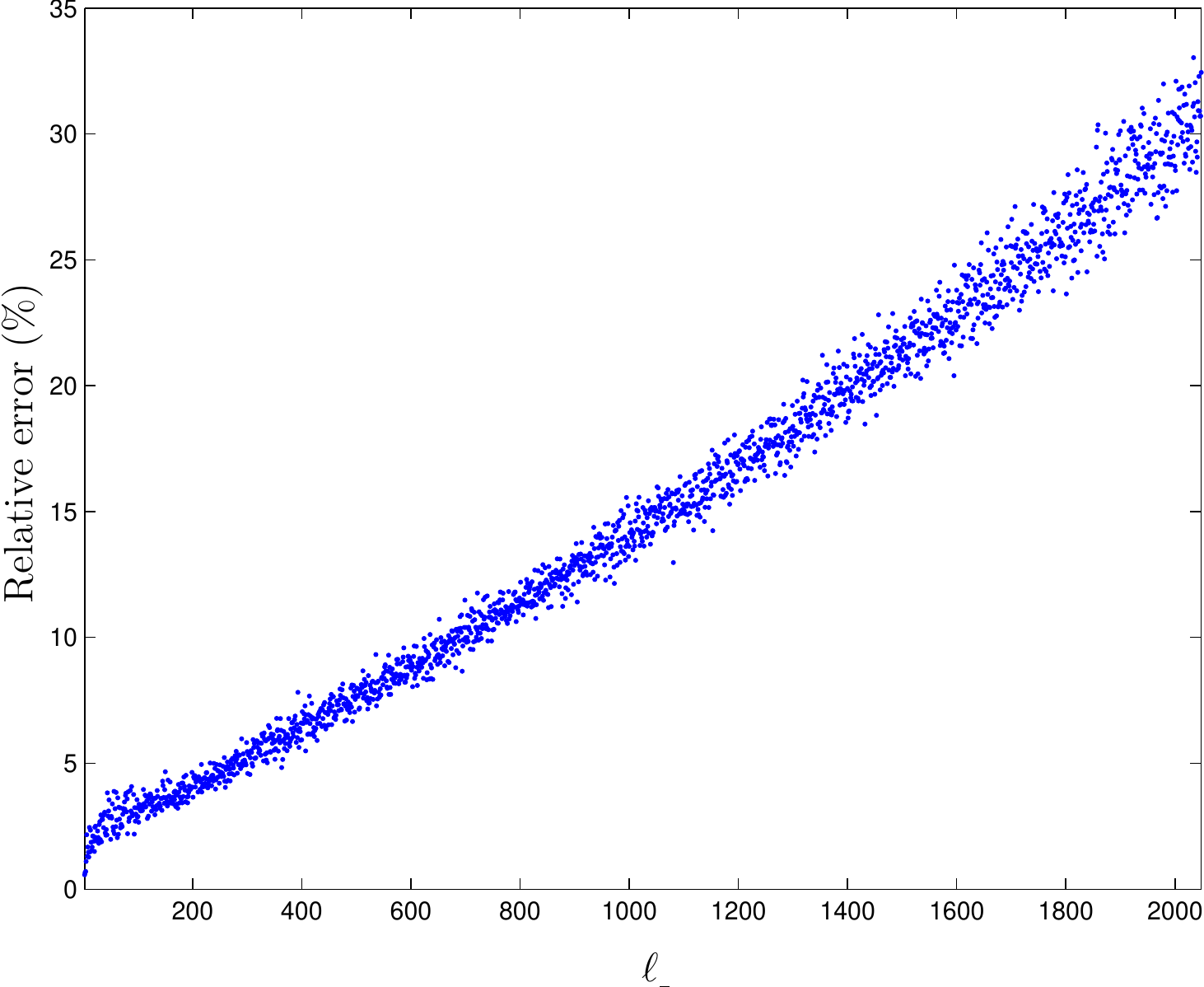}}
}
\mbox{
\subfigure[\subfigcapsize Angular power spectra for $p=3$]
  {\includegraphics[clip=,width=\clplotwidth]{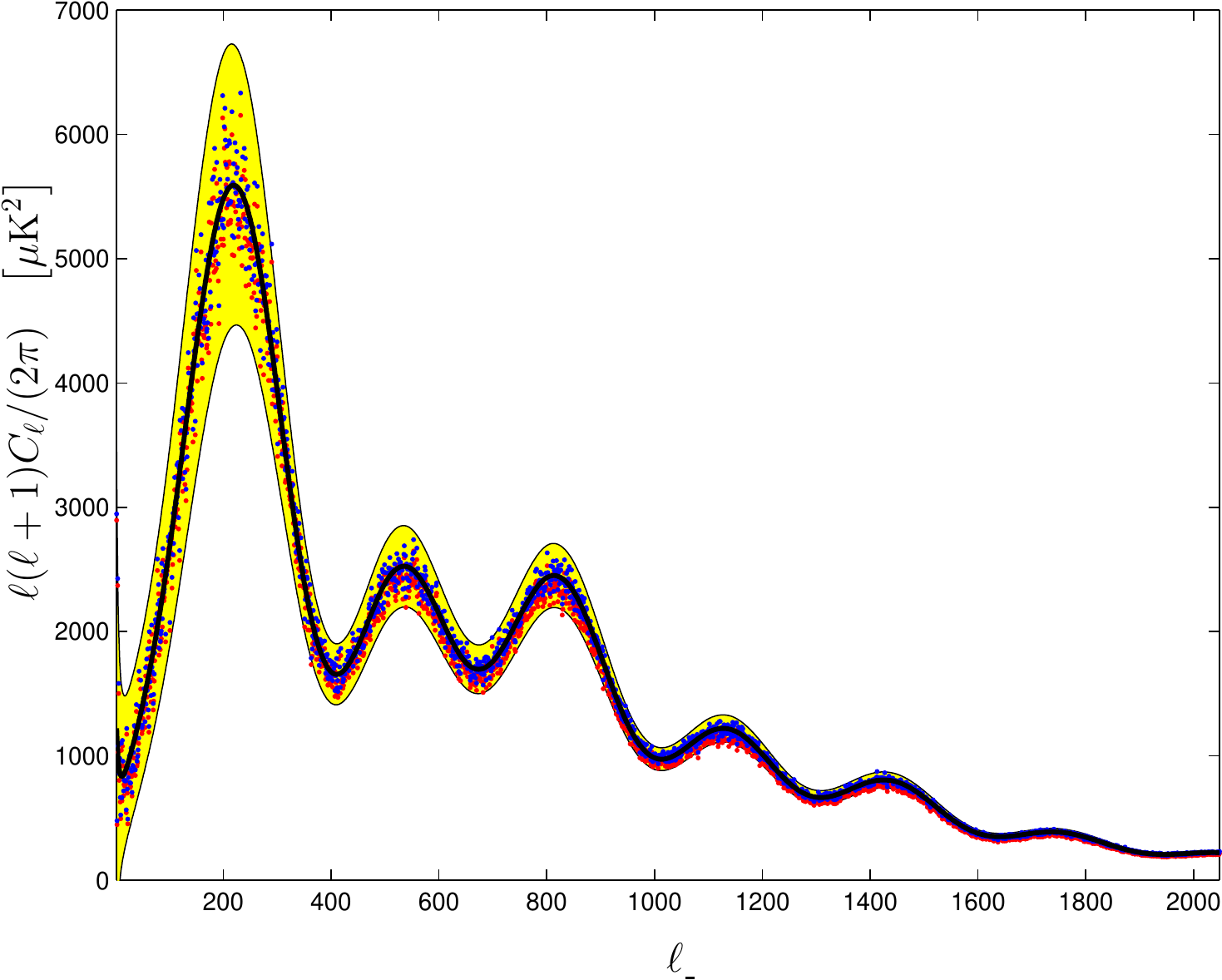}} \quad
\subfigure[\subfigcapsize Absolute error for $p=3$]
  {\includegraphics[clip=,width=\clplotwidth]{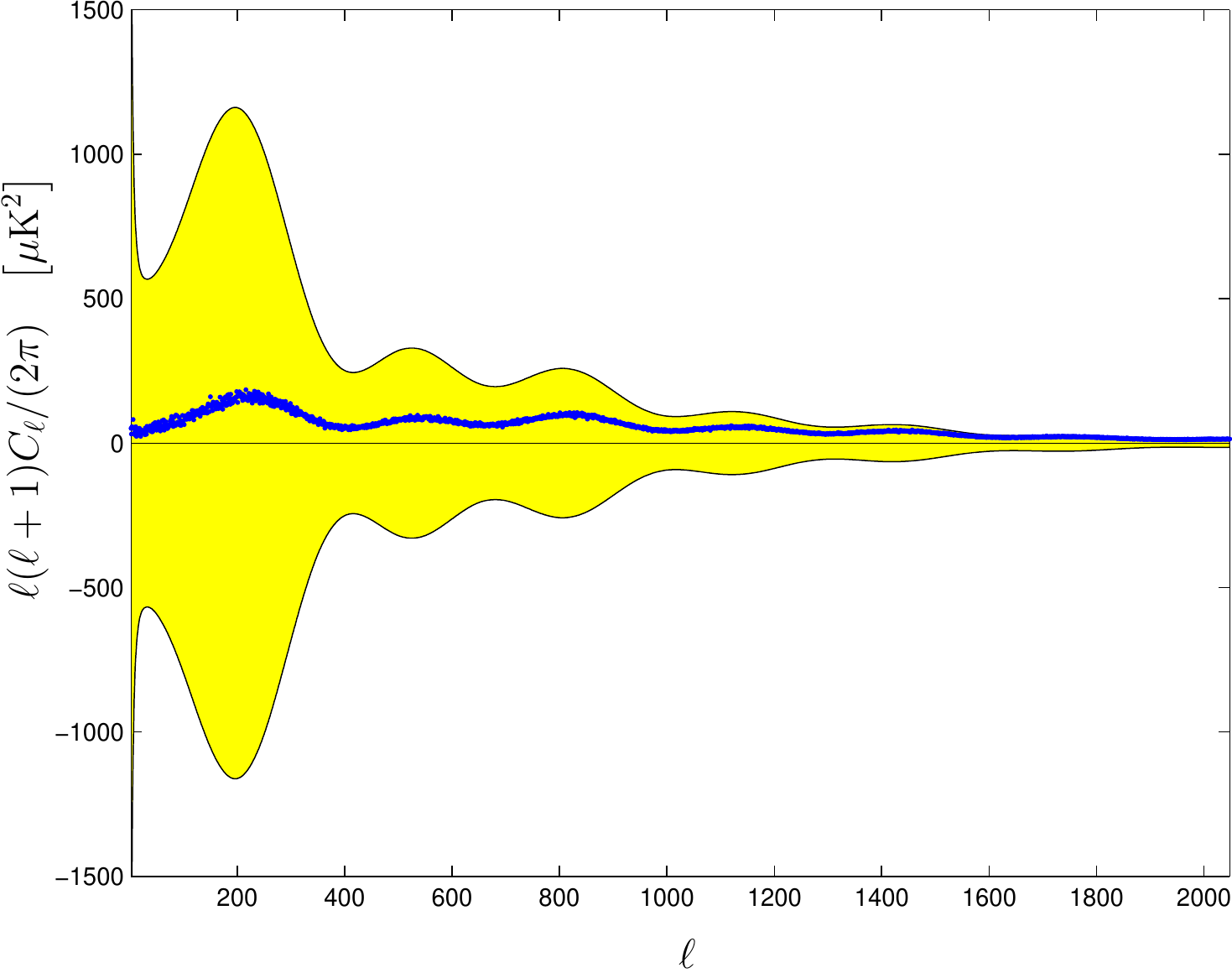}}
\subfigure[\subfigcapsize Error relative to cosmic variance for $p=3$]
  {\quad \includegraphics[clip=,width=\clplotwidth]{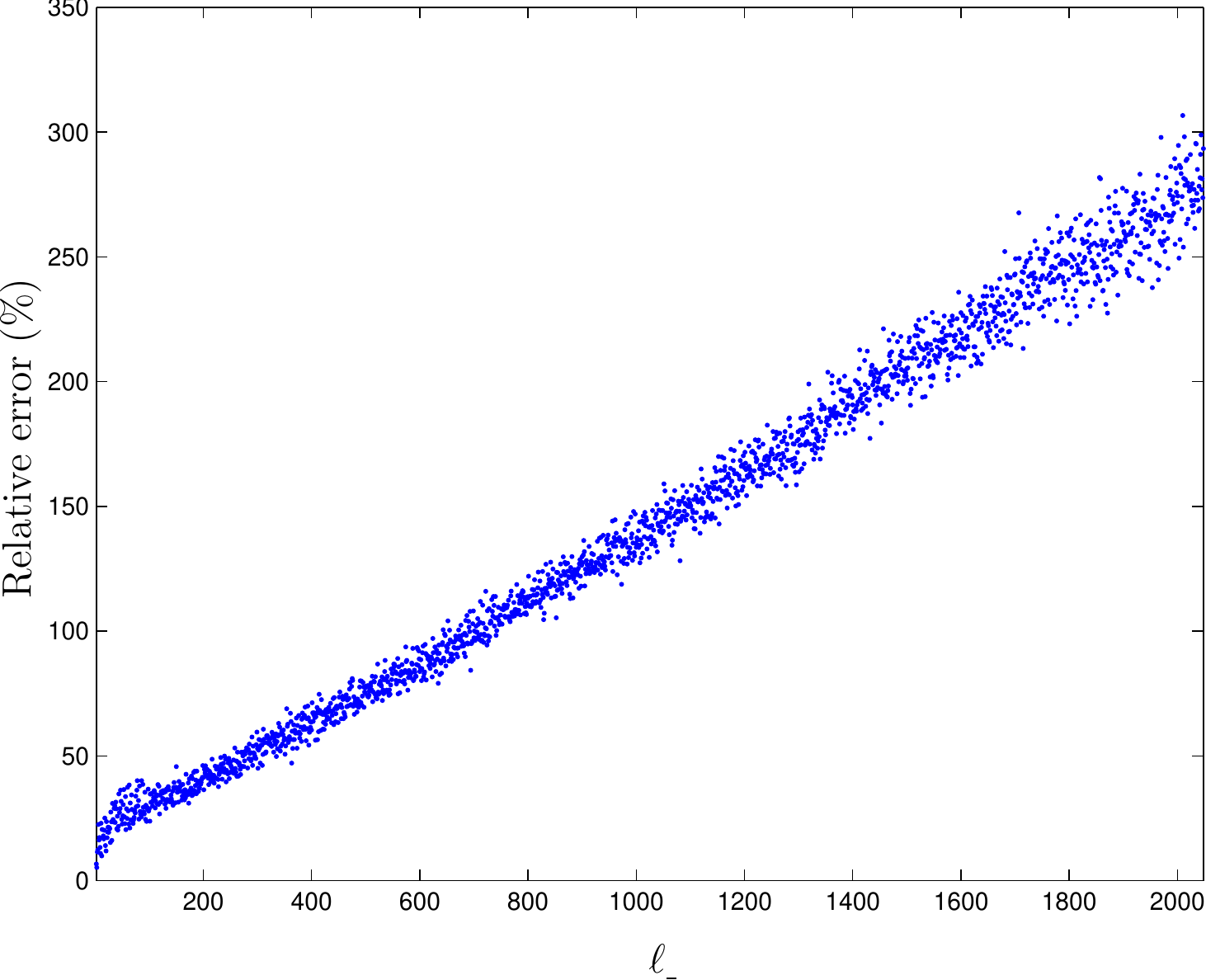}}
}
\caption{Reconstructed angular power spectrum of compressed \cmb\
  data at resolution $\nside=1024$.  Each row of panels shows the reconstructed power spectrum and
  errors for a particular compression precision parameter $p$.  In the
  first column of panels, the power spectrum reconstructed from the
  original \cmb\ data are given by the red dots, the power spectrum
  reconstructed from the compressed \cmb\ data are given by the blue
  dots and the underlying power spectrum of the simulated model is
  shown by the solid black line, with three standard deviation cosmic variance
  regions shaded in yellow.  Note that in some instances the red and
  blue dots align closely and may not both be visible.  In the second
  column of panels, the absolute error between the power spectra
  reconstructed from the original and compressed \cmb\ data is given by
  the blue dots, with three standard deviation cosmic variance regions shaded in
  yellow.  In the third column of panels, the absolute error between
  the power spectra reconstructed from the original and compressed
  \cmb\ data is expressed as a percentage of cosmic variance.  Note
  that the scale of the vertical axis changes by an order of magnitude
  between each row of the third column of panels.
}
\label{fig:cl}
\end{figure*}

\begin{figure}
\centering
\includegraphics[height=50mm]{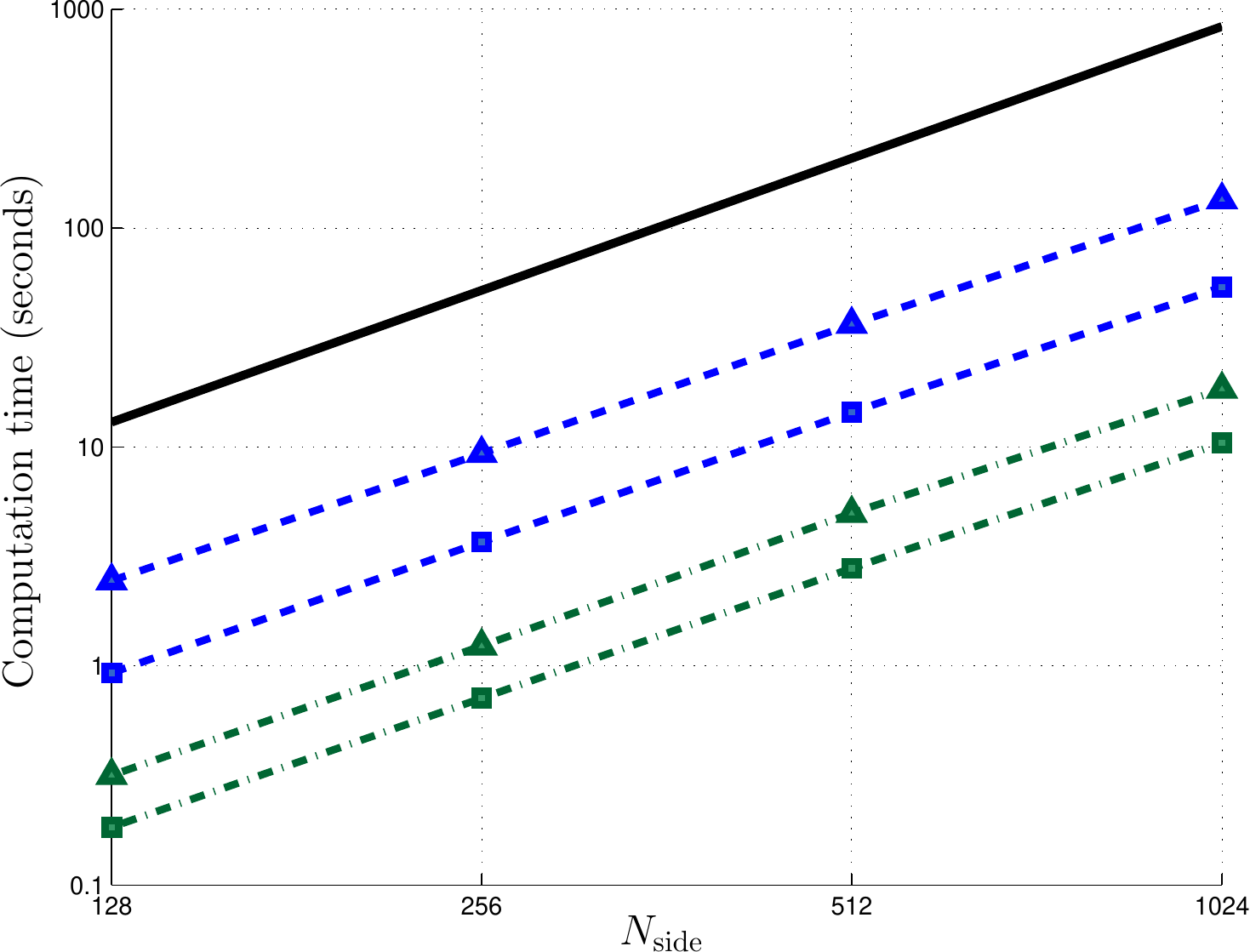}
\caption{\update{Computation time required to compress (blue/dashed line) and
  decompress (green/dot-dashed line) simulated Gaussian CMB data of
  various resolution \nside.  Computation times are averaged over five
  simulated Gaussian \cmb\ maps and are shown for precisions parameters
  $\precision=3$ (squares) and $\precision=5$ (triangles).
  $\order(\nside^2)$ scaling is shown by the heavy black/solid line.}}
\label{fig:timing}
\end{figure}

\update{

  Finally, we measure the CPU time required to compress and decompress
  \cmb\ maps.  All timing tests are performed on a laptop with a
  2.66GHz Intel Core 2 Duo processor and 4GiB of RAM.  We restrict our
  attention to unmasked data and to precision parameters
  $\precision=3$ and $\precision=5$ only.  Computation times are
  plotted in \fig{\ref{fig:timing}} for a range of resolutions, where
  all measurements are averaged over five random Gaussian \cmb\
  simulations.  Note that computation time increases with precision
  parameter $\precision$ since the number of unique wavelet
  coefficients requiring encoding also increases with \precision.  All
  stages of our compression and decompression algorithms are linear in
  the number of data samples, hence the computation time of our
  algorithms scales linearly with the number of samples on the sphere,
  \ie\ as $\order(\nside^2)$, as also apparent from
  \fig{\ref{fig:timing}}.

}

\subsection{Lossy compression}
\label{sec:applications_lossy}

In certain applications the loss of a small amount of information from
a data-sphere is not catastrophic.  For example, in computer graphics
environmental illumination maps and reflectance functions that are
defined on the sphere are used in rendering synthetic images (\eg\
\citealt{ramamoorthi:2004}).  In this application accuracy is
determined by human perception, hence errors may be tolerated if they
are not suitably noticeable.  Moreover, data-spheres that are input to
reflectance algorithms are not viewed directly, thus moderate errors
in these data may not necessarily produce noticeable errors in
rendered images.  Lossy representations of data-spheres in computer
graphics are therefore not only tolerated, but are often desired as
they may improve the computational efficiency of rendering algorithms
(\eg\ \citealt{ng:2004}).  For data compression purposes, our lossy
compression algorithm is certainly appropriate and may be applied in
order to achieve higher compression ratios.  In addition to the
environmental illumination maps discussed previously, we also compress
Earth topography data and evaluate the performance of both our
lossless and lossy compression algorithms on both of these types of
data.

The Earth topography and environmental illumination data-spheres considered
here are all obtained from real-world observations.  The original data
are illustrated in the first column of panels in
\fig{\ref{fig:lossy_maps}}.  The topographical data are represented at
a \healpix\ resolution of $\nside=512$.  The environmental illumination
data-spheres were constructed by \citet{debevec:1998} and are
available publicly\footnote{\url{http://www.debevec.org/Probes/}}.
These data defined on the sphere were constructed by taking two
photographic images of a mirrored ball from different locations, and
mapping the observed intensities of the images onto the surface of a
sphere.  The illumination maps are available in a cross-cube format
and have been converted to a \healpix\ data-sphere at resolution
$\nside=256$ ($\scalmax=8$; $\npix\simeq0.8\times10^6$).  We consider environmental illumination data that was captured
in this manner from within Galileo's Tomb in Santa Croce, Florence,
St.\ Peter's Basilica in Rome and the Uffizi Gallery in Florence.

Lossless and lossy compressed versions of the data are illustrated in
\fig{\ref{fig:lossy_maps}}.  For the lossless compression we use a
precision parameter of $\precision=3$ in the quantisation stage of the
compression since ultimately we are concerned with achieving a high
compression ratio and will allow some quantisation error.  For the
lossy compression we again use a precision parameter of $\precision=3$
and retain only 5\% of the detail coefficients in the thresholding
stage of the compression algorithm.  Compression ratios of
approximately 40:1 are achieved for the lossy compression of both the
topographic and environmental illumination data.  Although it is
possible to discern errors in the lossy compressed data, the overall
structure and many of the details of the original data are well
approximated in this highly compressed representation.

\newlength{\mapplotwidth}
\setlength{\mapplotwidth}{55mm}

\begin{figure*}
\centering
\mbox{
\subfigure[\subfigcapsize Earth: original (13MB)]
  {\includegraphics[clip=,width=\mapplotwidth]{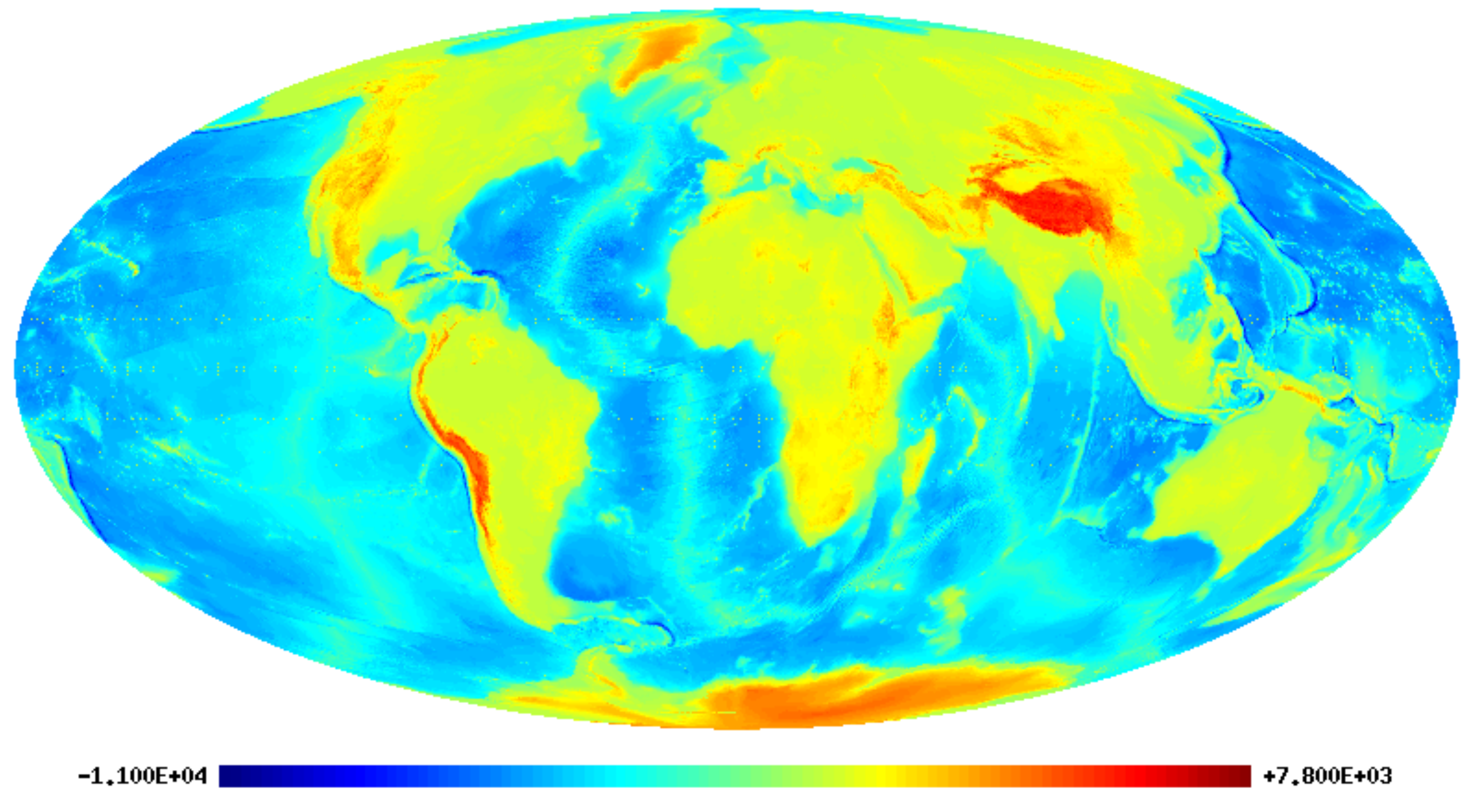}} \quad
\subfigure[\subfigcapsize Earth: lossless compressed (1.4MB)]
  {\includegraphics[clip=,width=\mapplotwidth]{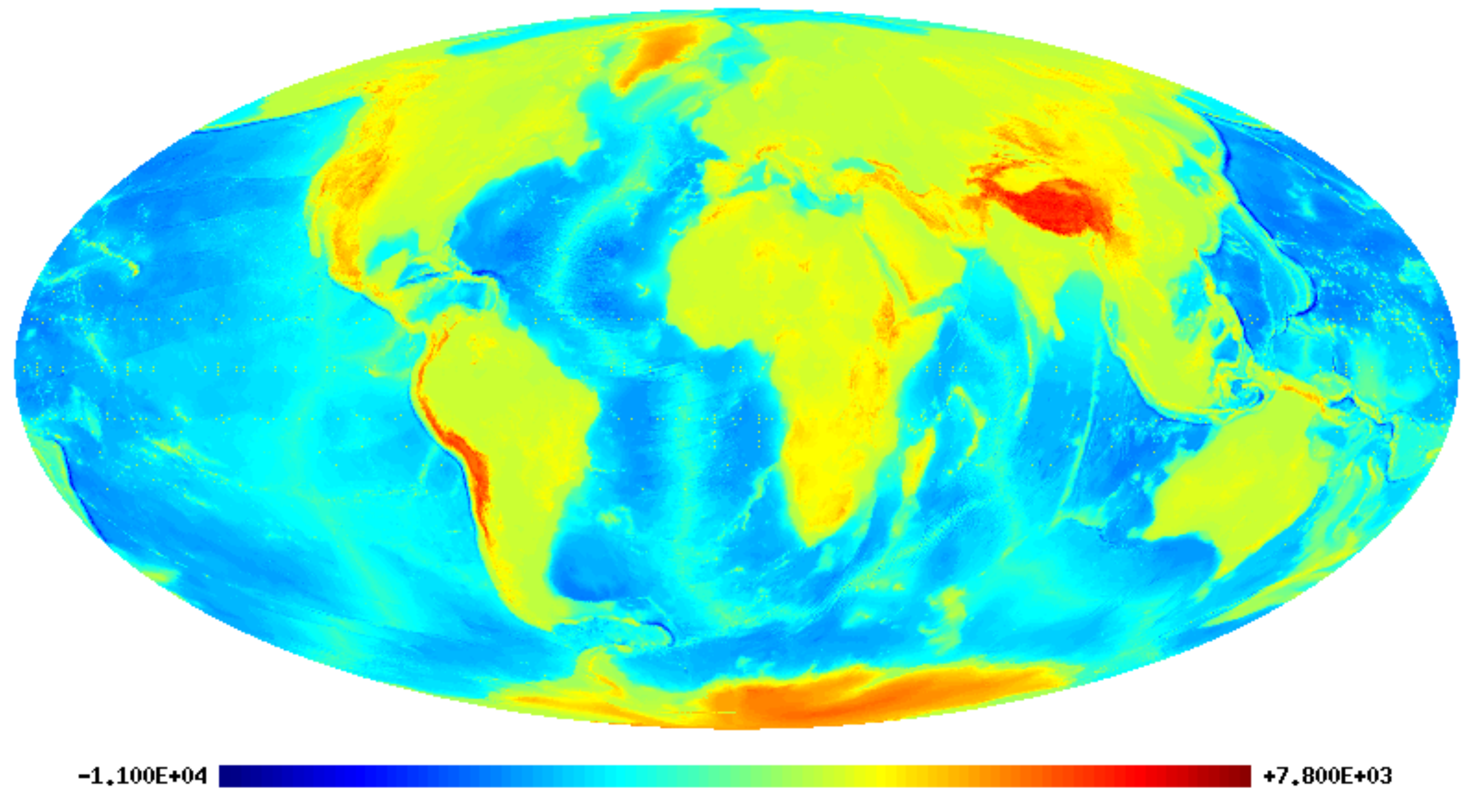}} \quad
\subfigure[\subfigcapsize Earth: lossy compressed (0.33MB)]
  {\includegraphics[clip=,width=\mapplotwidth]{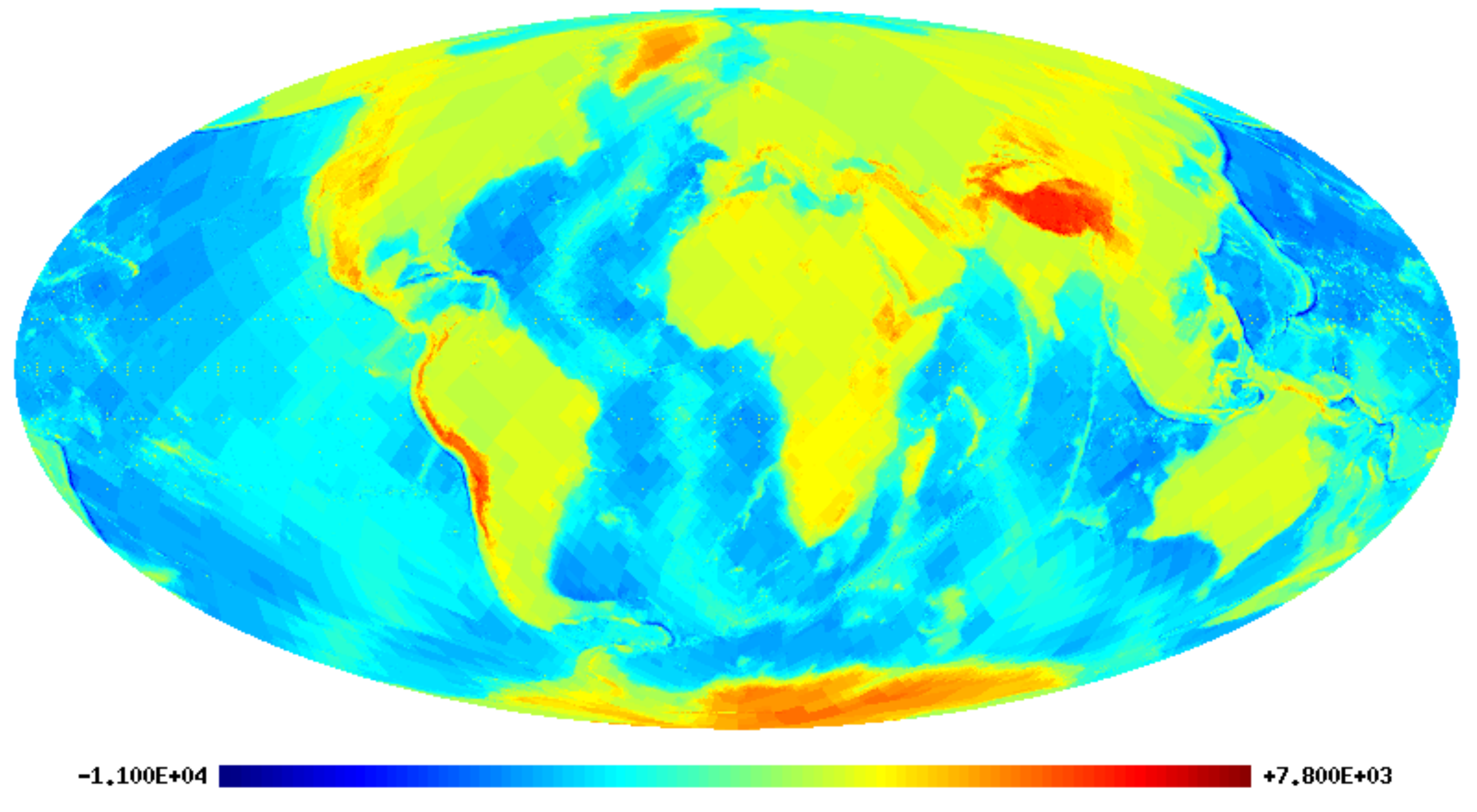}}
}
\mbox{
\subfigure[\subfigcapsize Galileo: original (3.2MB)]
  {\includegraphics[clip=,width=\mapplotwidth]{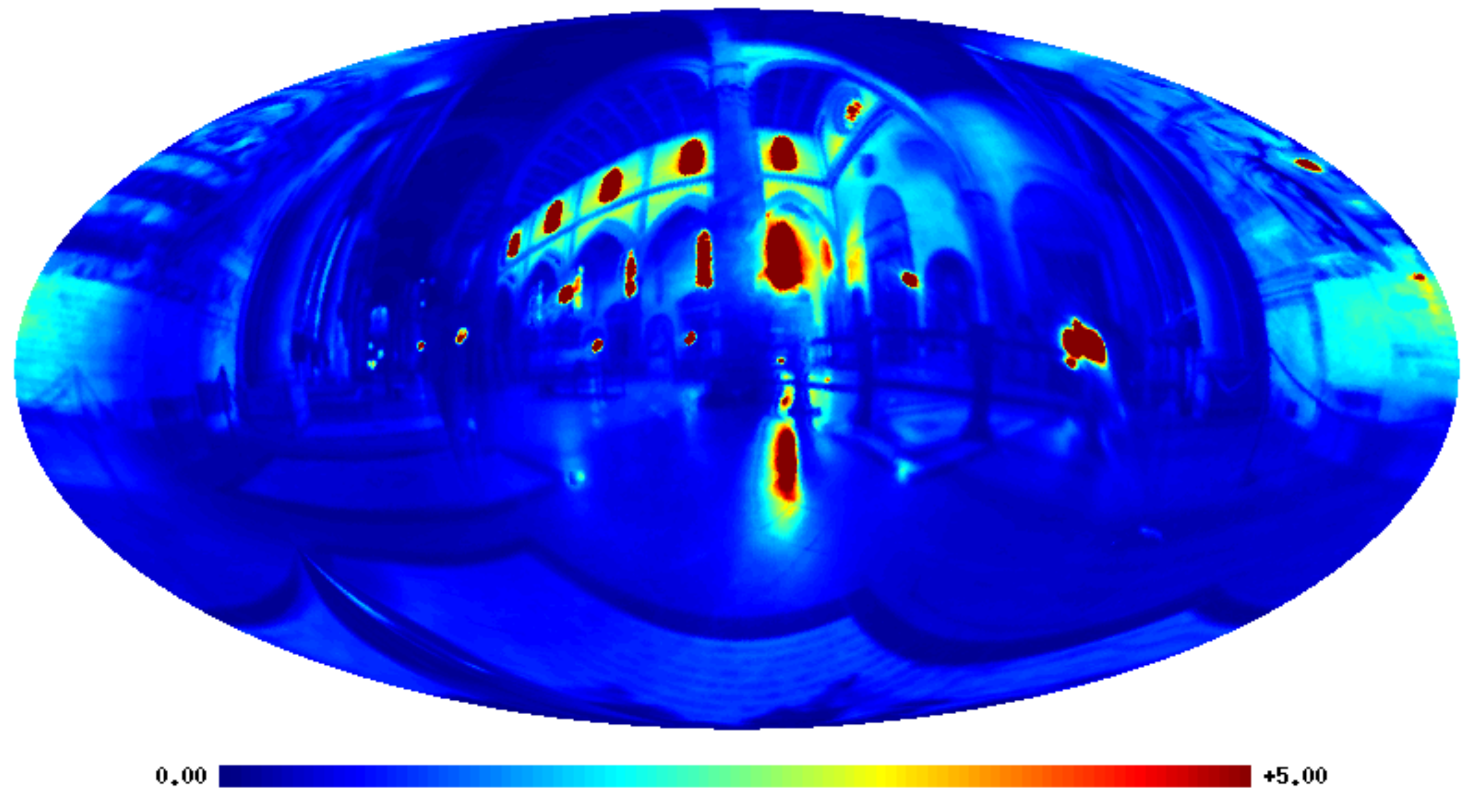}} \quad
\subfigure[\subfigcapsize Galileo: lossless compressed (0.21MB)]
  {\includegraphics[clip=,width=\mapplotwidth]{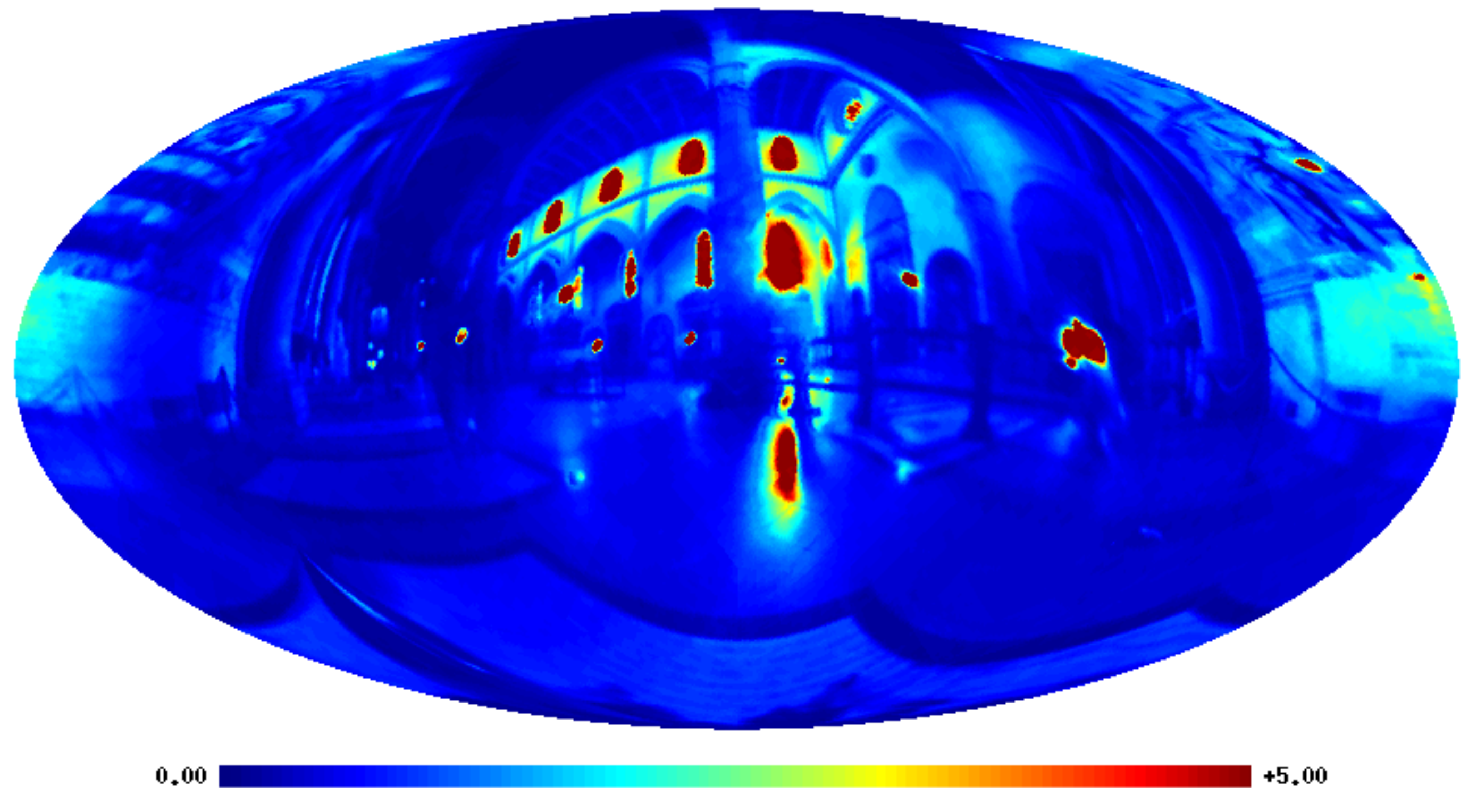}} \quad
\subfigure[\subfigcapsize Galileo: lossy compressed (0.07MB)]
  {\includegraphics[clip=,width=\mapplotwidth]{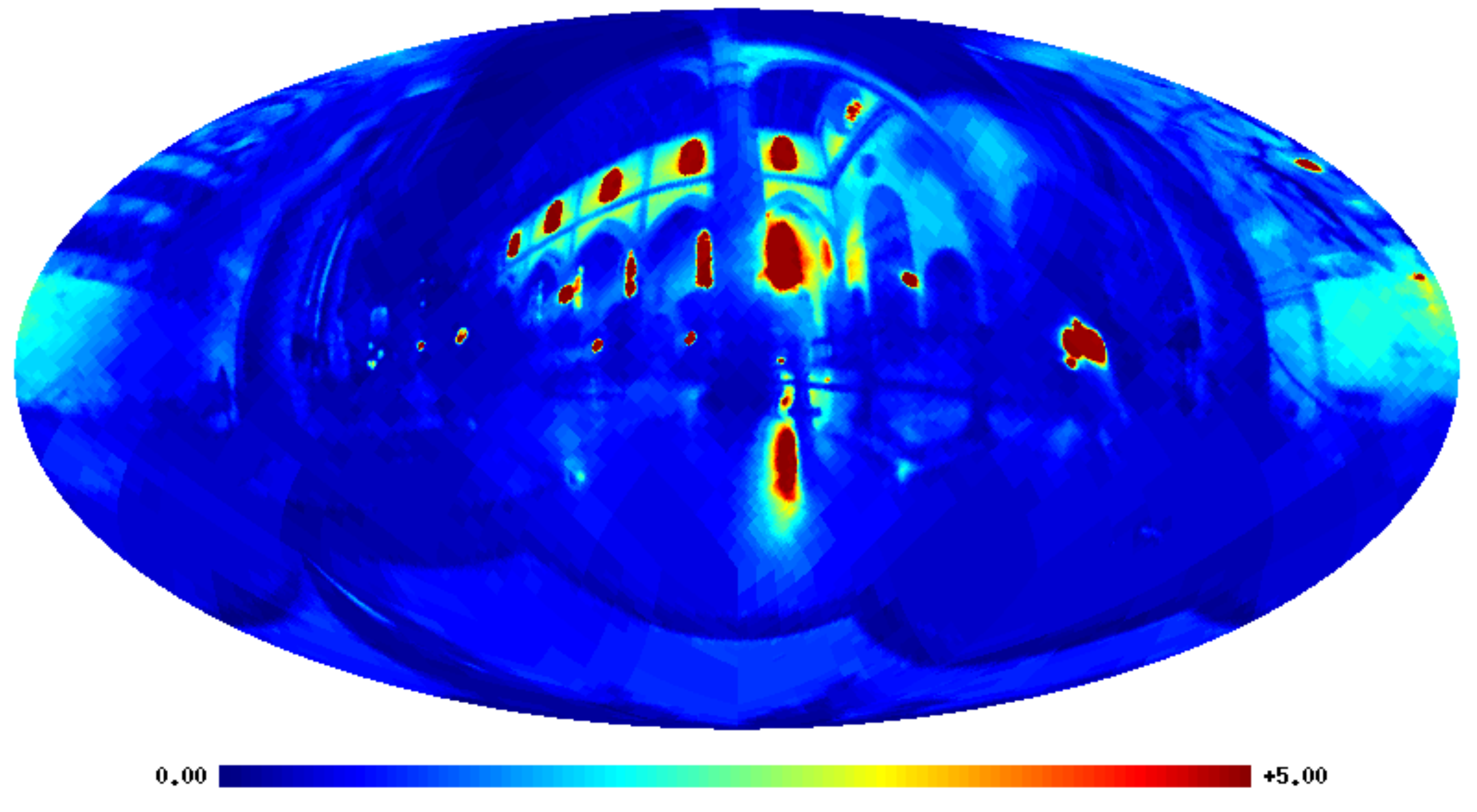}}
}
\mbox{
\subfigure[\subfigcapsize St Peter's: original (3.2MB)]
  {\includegraphics[clip=,width=\mapplotwidth]{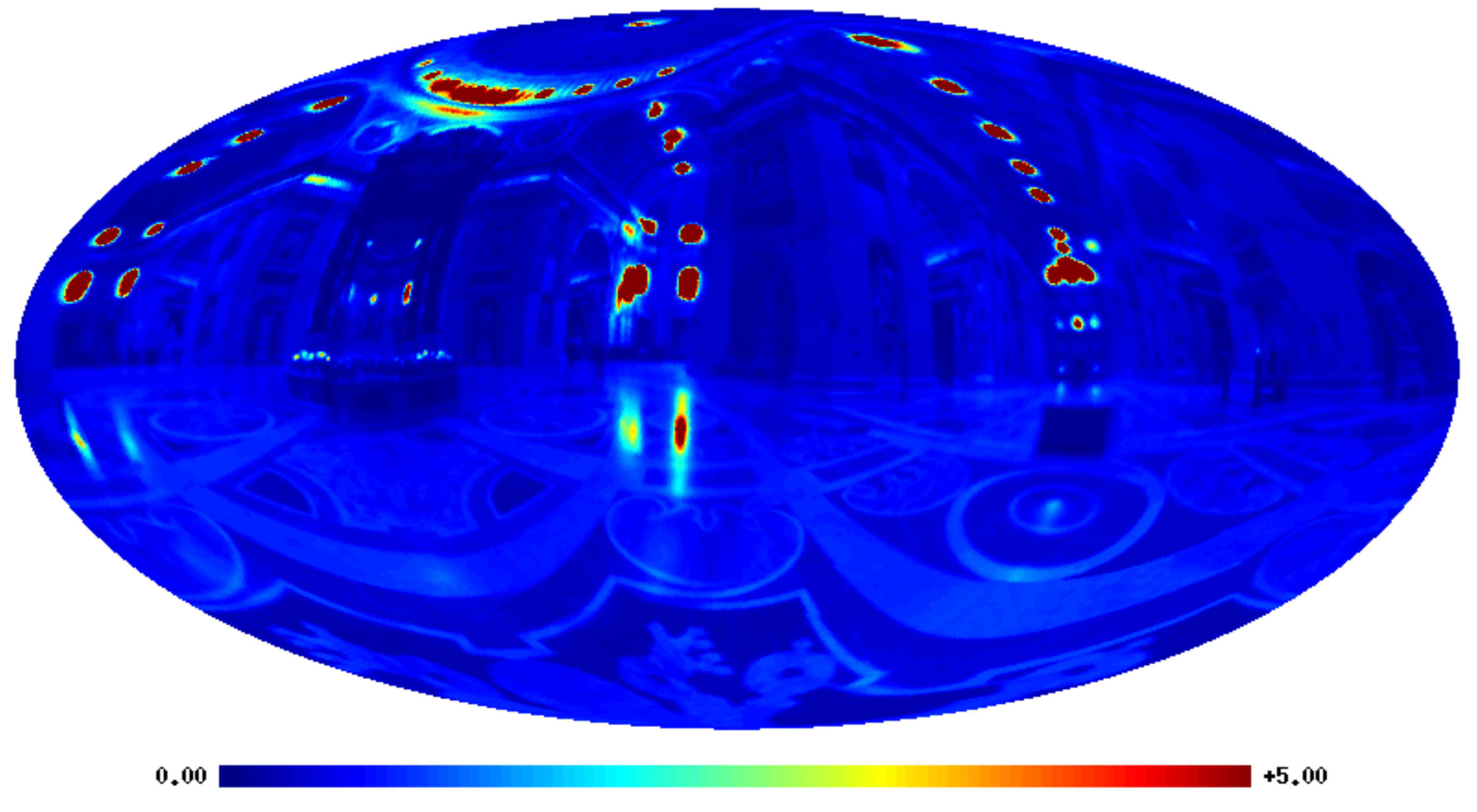}} \quad
\subfigure[\subfigcapsize St Peter's: lossless compressed (0.20MB)]
  {\includegraphics[clip=,width=\mapplotwidth]{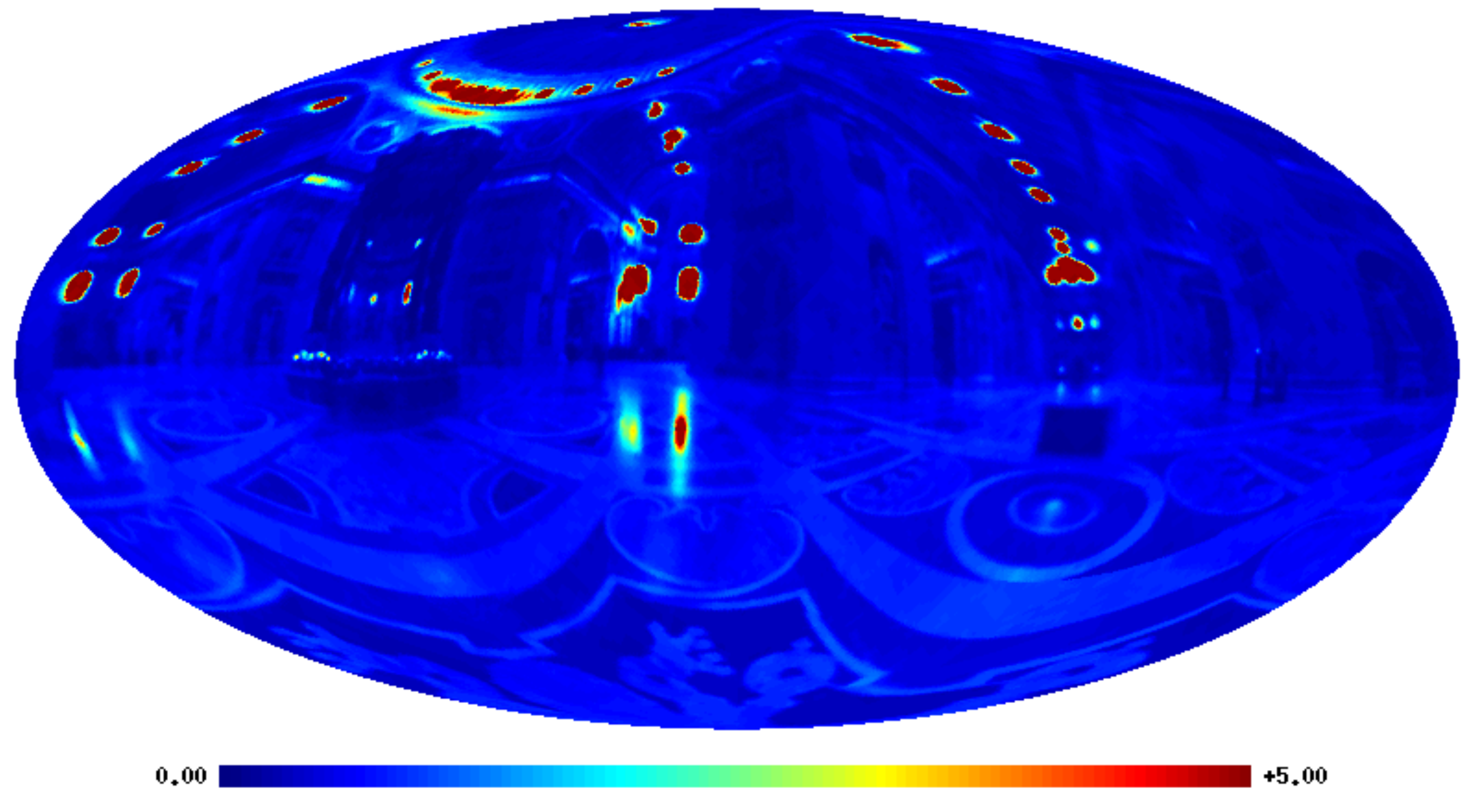}} \quad
\subfigure[\subfigcapsize St Peter's: lossy compressed (0.08MB)]
  {\includegraphics[clip=,width=\mapplotwidth]{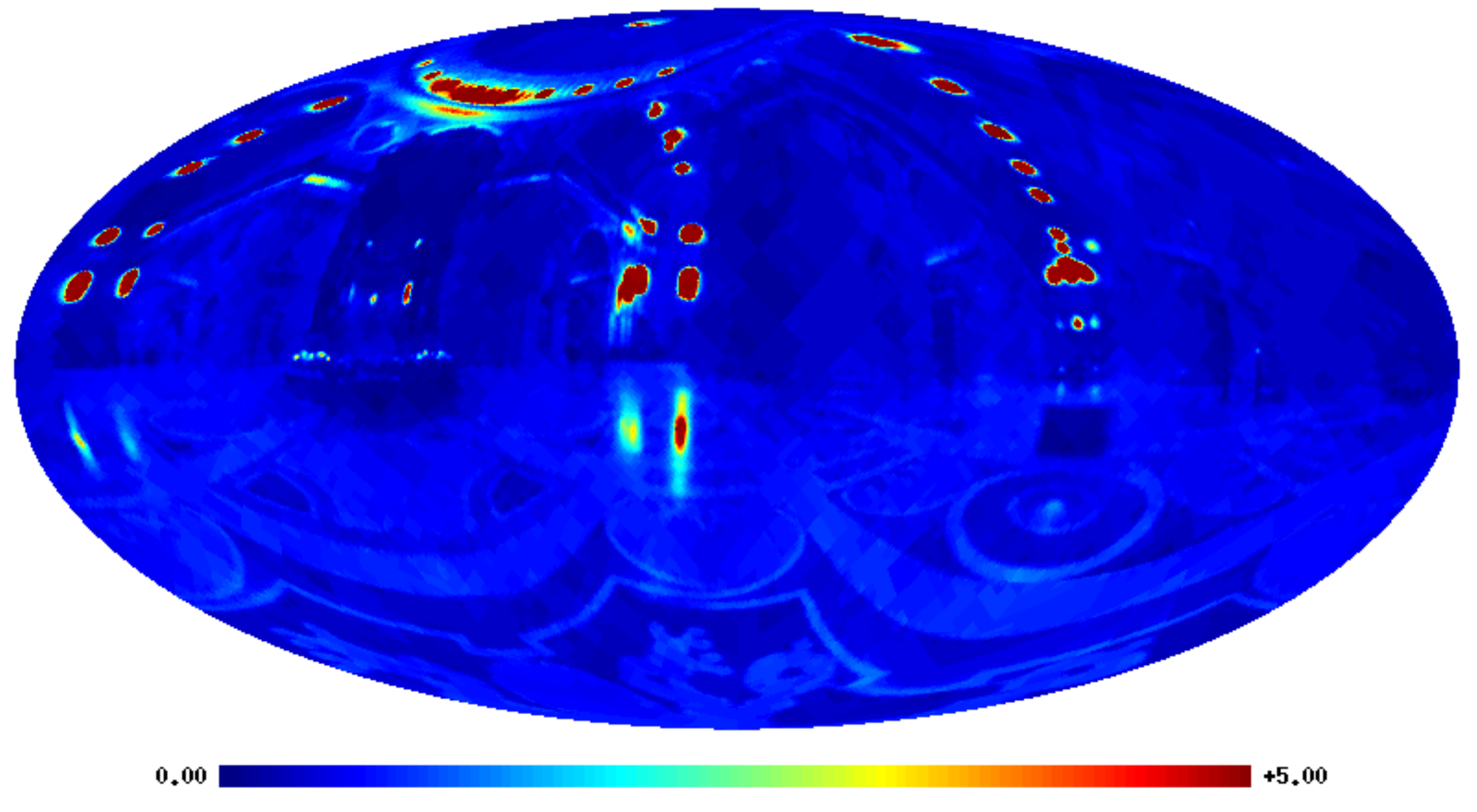}}
}
\mbox{
\subfigure[\subfigcapsize Uffizi: original (3.2MB)]
  {\includegraphics[clip=,width=\mapplotwidth]{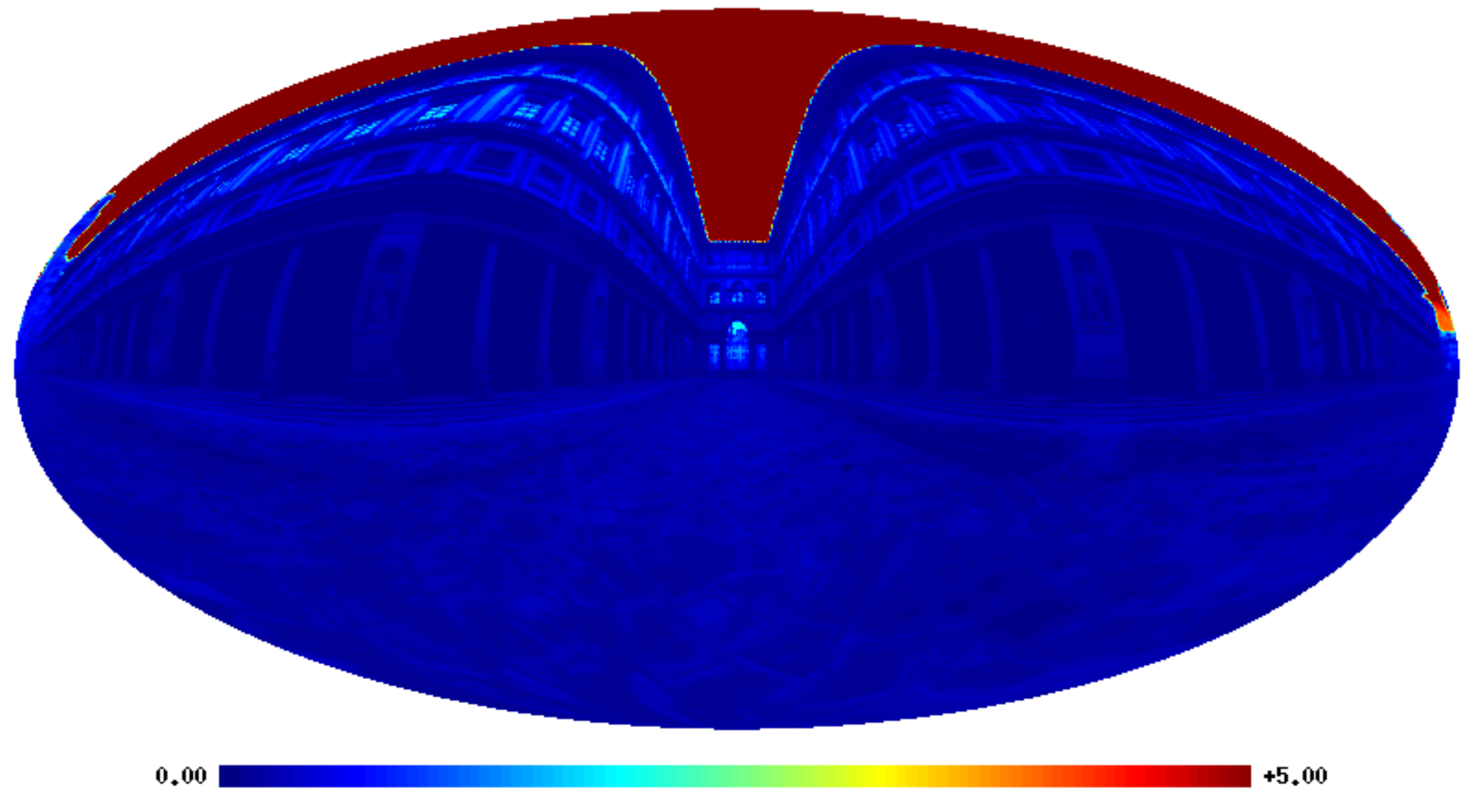}} \quad
\subfigure[\subfigcapsize Uffizi: lossless compressed (0.19MB)]
  {\includegraphics[clip=,width=\mapplotwidth]{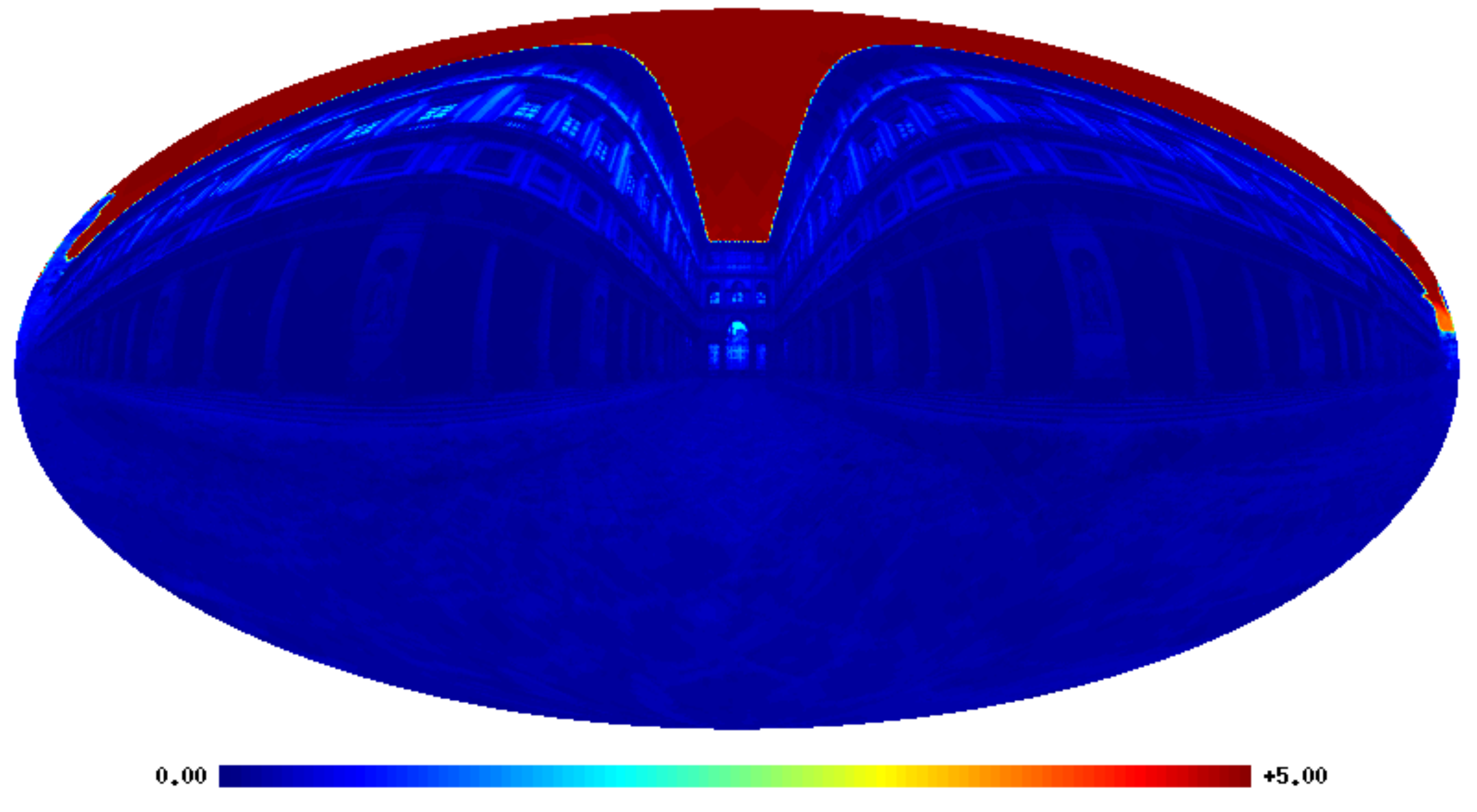}} \quad
\subfigure[\subfigcapsize Uffizi: lossy compressed (0.10MB)]
  {\includegraphics[clip=,width=\mapplotwidth]{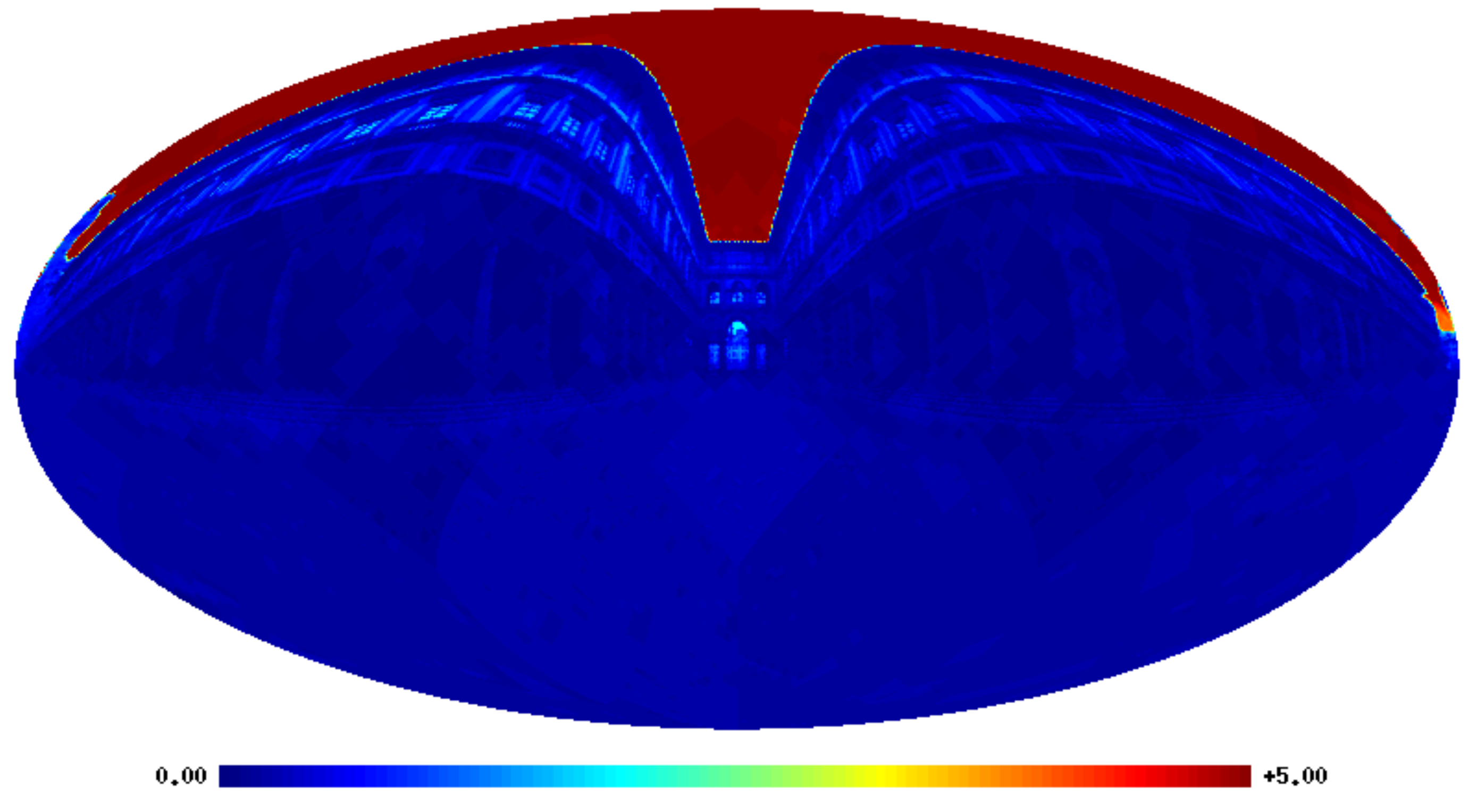}}
}
\caption{Compressed data for lossy compression applications \update{(data-spheres are
  displayed using the Mollweide projection)}.  Each row
  of panels shows the original, lossless and lossy compressed
  data-spheres.  File sizes for each data-sphere are also specified.  The
  lossless compressed data shown in the second
  column of panels is performed with a precision parameter of
  $\precision=3$.  The lossy compressed data 
  shown in the third column of panels is performed by retaining 5\% of
  detail coefficients only and including a RLE stage.  The full
  dynamic range of these images may not be visible in printed versions
  of this figure, hence this figure is best viewed online.  }
\label{fig:lossy_maps}
\end{figure*}

In \fig{\ref{fig:lossy_tradeoff}} we evaluate the performance of the
compression of these data more thoroughly.  Firstly, for the lossless
compression we examine the trade-off between compression ratio and
decompression error with respect to the precision parameter
$\precision$ (see the first column of panels of
\fig{\ref{fig:lossy_tradeoff}}).  For both the topographic and
illumination data it is apparent that we may reduce the precision
parameter to $\precision=3$, while introducing quantisation error
on the order of a few percent only.  If the precision parameter is reduced
to $\precision=2$, quantisation errors on the order of 10-20\% appear.
We therefore choose $\precision=3$ for the lossless compression since
this maximises the compression ratio while introducing an allowable
level of quantisation error.  We then examine the effect of increasing
the threshold level, by reducing the proportion of detail coefficients
retained, on the performance of the lossy compression (see the second
column of panels of \fig{\ref{fig:lossy_tradeoff}}).  Retaining 100\% of
the detail coefficients corresponds to the lossless compression case,
where no RLE is included.  RLE is included for all other lossy
compression cases.  Notice that when retaining 50\% of the detail
coefficients, the resulting improvement to the compression ratio is
offset by the additional encoding overhead of the RLE.  Consequently,
the compression ratio when retaining 50\% of coefficients (with RLE)
is often worse than when retaining 100\% of coefficients (without
RLE).  As the proportion of detail coefficients that are retained is
reduced, the improvement to compression ratio quickly exceeds the
additional overhead of RLE.  Decompression error remains at
approximately 5\% when retaining only 5\% of the detail coefficients,
but increases quickly as the proportion of coefficients retained is
reduced further.  Retaining 5\% of coefficients therefore appears to
give a good trade-off between compression ratio and fidelity of the
decompressed data, justifying this choice for the results presented in
\fig{\ref{fig:lossy_maps}}.  For this choice, the topographic and
illumination data are compressed to a ratio of approximately 40:1,
while introducing errors of approximately 5\%.  The images illustrated
in \fig{\ref{fig:lossy_maps}} show that errors of this order are not
significantly noticeable and are likely to be acceptable for many
applications.  Also notice that the curves shown in
\fig{\ref{fig:lossy_tradeoff}} for the environmental illumination data
are similar, indicating that the characteristics of natural
illumination are to some extent independent of the scene.  One
would therefore expect the compression performance observed for the
data-spheres considered here to be typical of environmental
illumination data in general.  Although we have made a number of
arbitrary choices here regarding acceptable levels of distortion in
the decompressed data, one may of course choose the number of detail
coefficients retained that provides a trade-off between compression
ratio and fidelity that is suitable for the application at hand.

\newlength{\tradeoffplotwidth}
\setlength{\tradeoffplotwidth}{48mm}

\begin{figure*}
\centering
\mbox{
\subfigure[\subfigcapsize Earth: lossless performance]
  {\includegraphics[clip=,width=\tradeoffplotwidth]{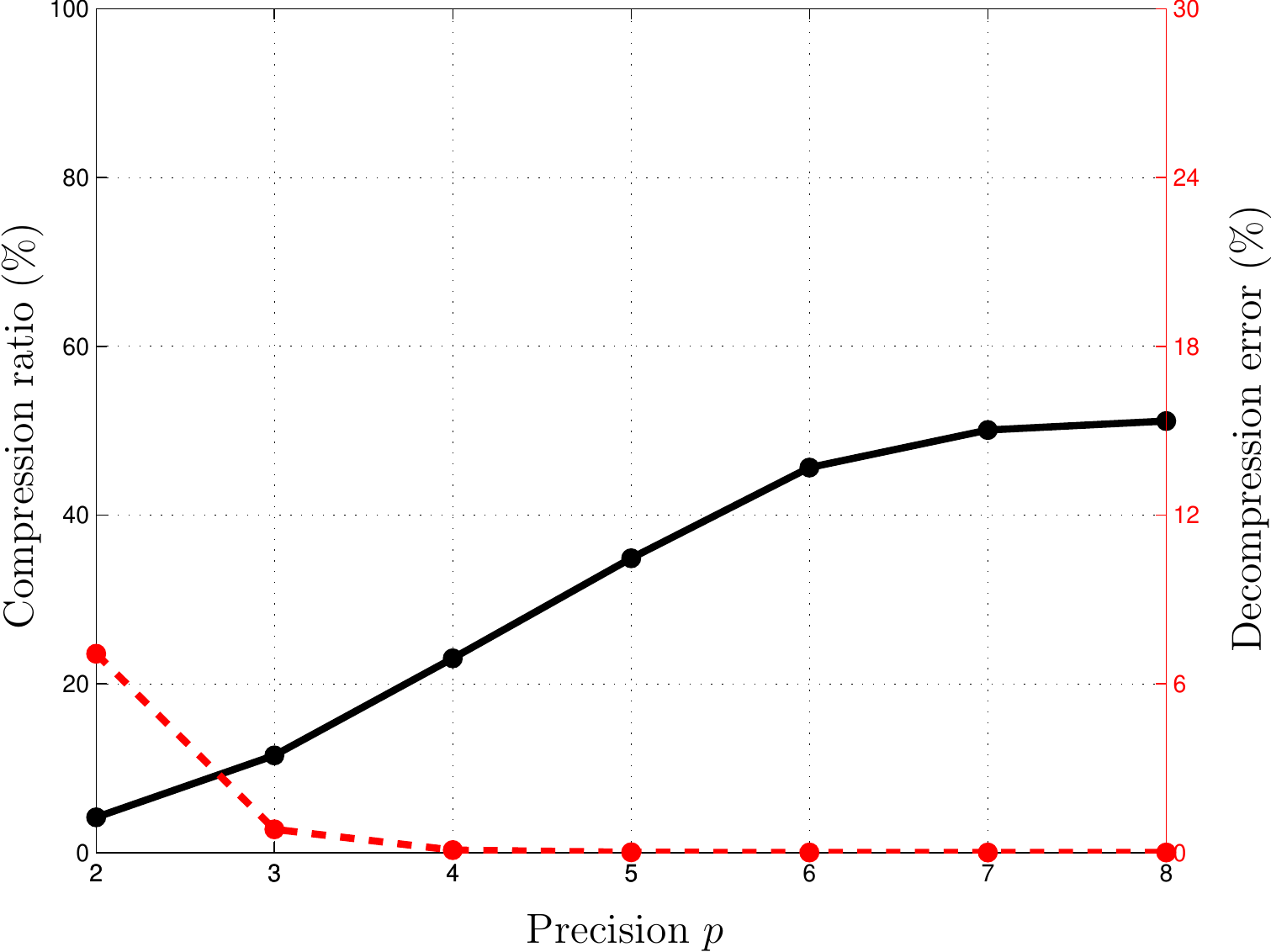}} \quad\quad
\subfigure[\subfigcapsize Earth: lossy performance]
  {\includegraphics[clip=,width=\tradeoffplotwidth]{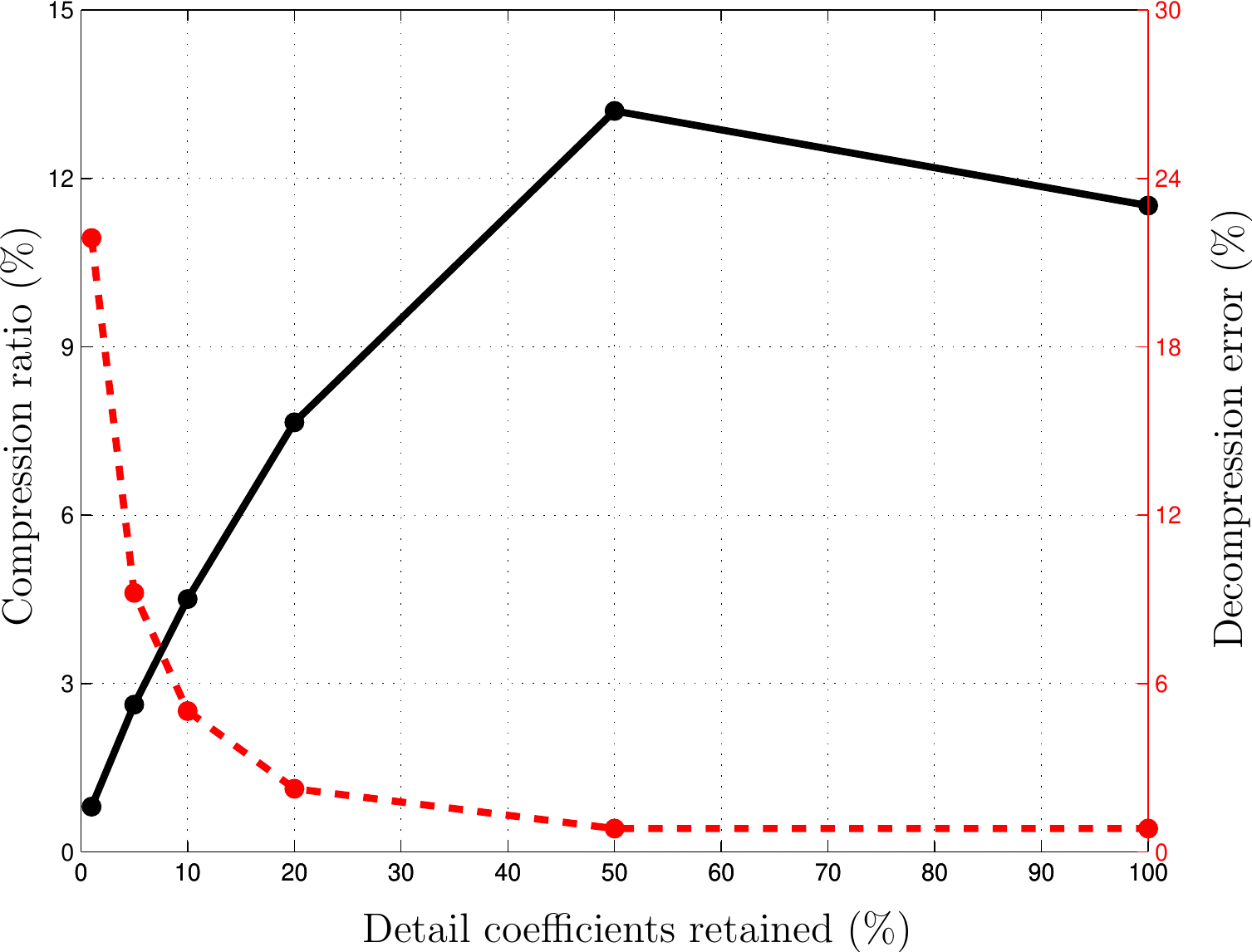}} 
} \\
\mbox{
\subfigure[\subfigcapsize Galileo: lossless performance]
  {\includegraphics[clip=,width=\tradeoffplotwidth]{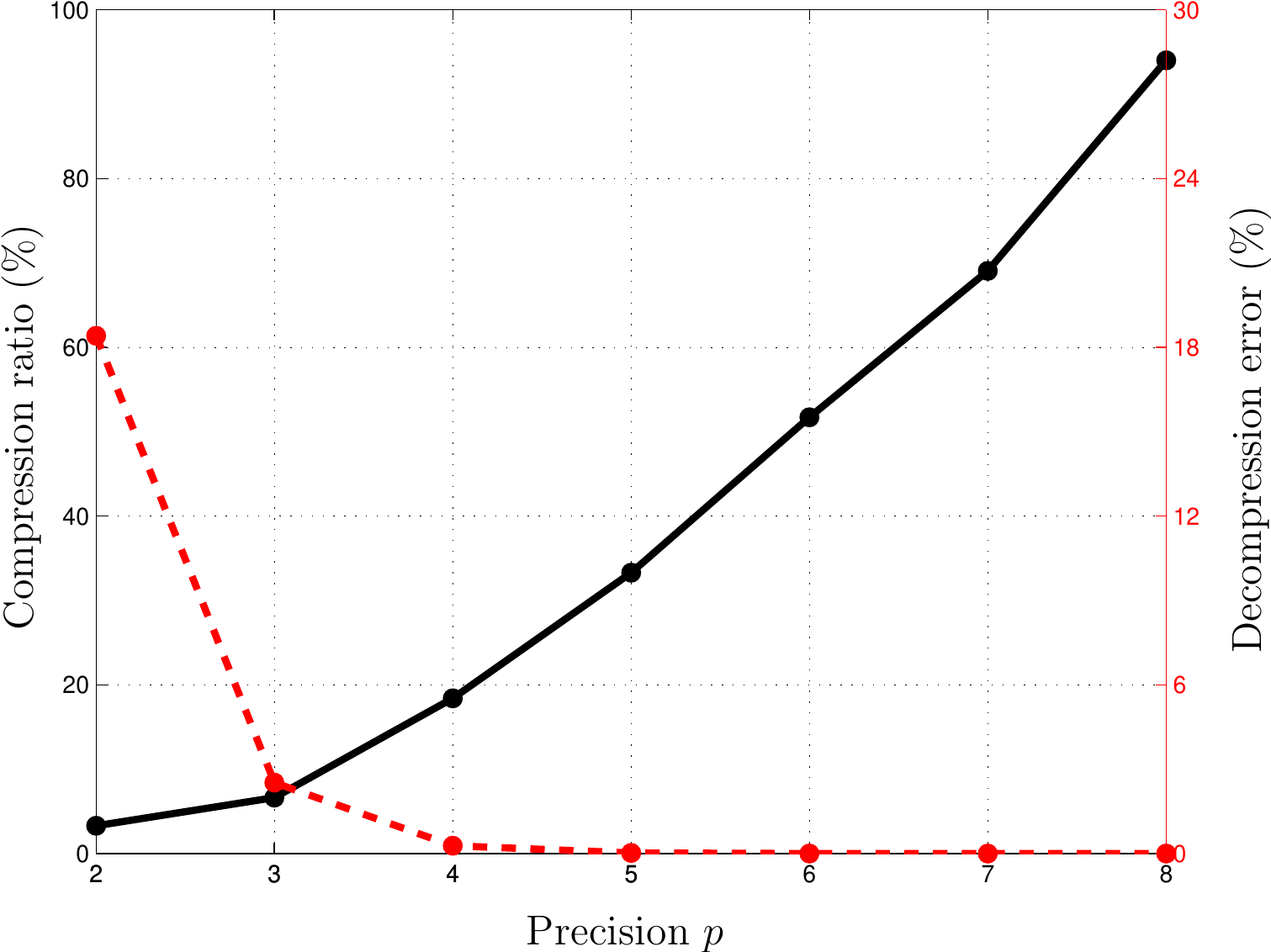}} \quad\quad
\subfigure[\subfigcapsize Galileo: lossy performance]
  {\includegraphics[clip=,width=\tradeoffplotwidth]{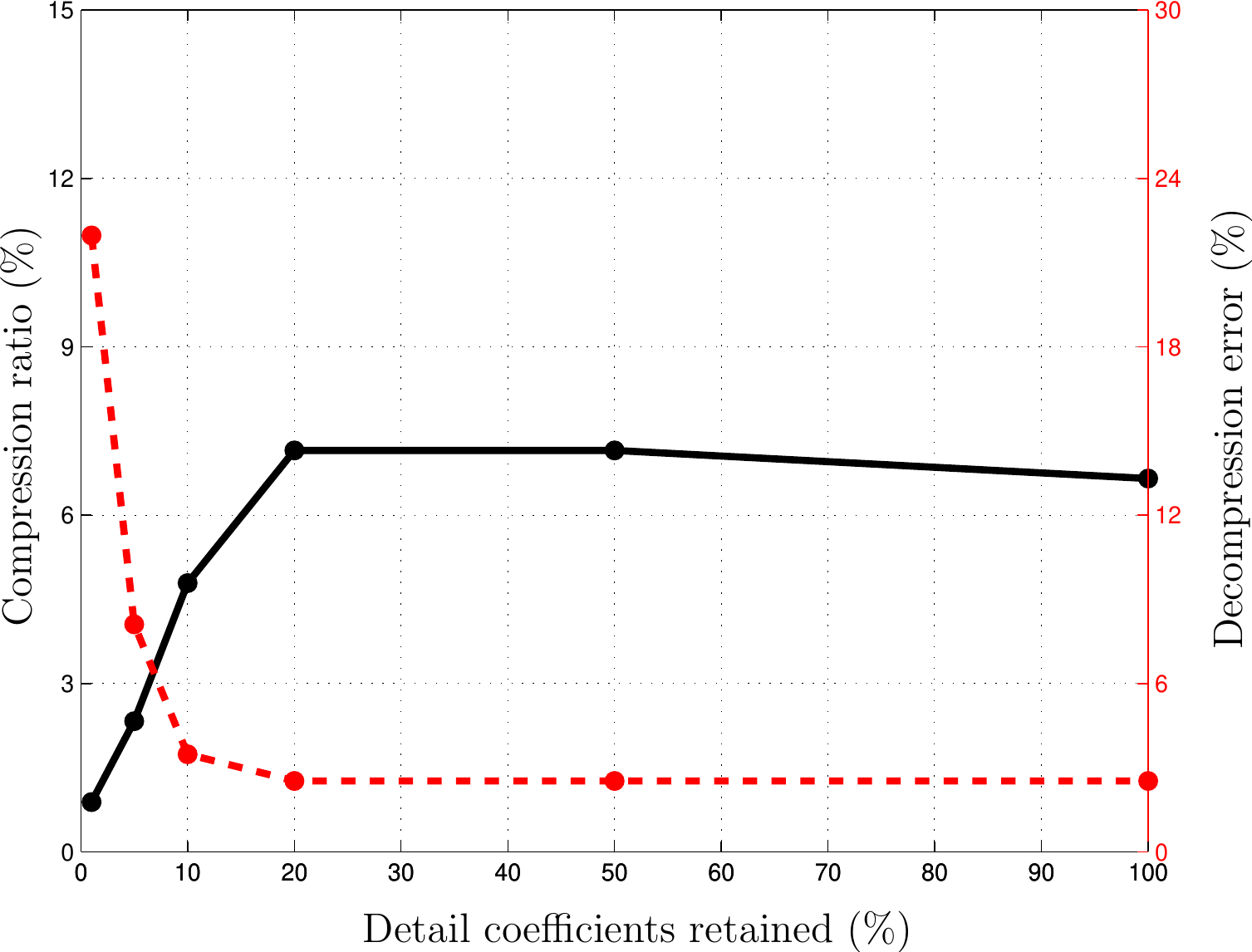}} 
} \\
\mbox{
\subfigure[\subfigcapsize St Peter's: lossless performance]
  {\includegraphics[clip=,width=\tradeoffplotwidth]{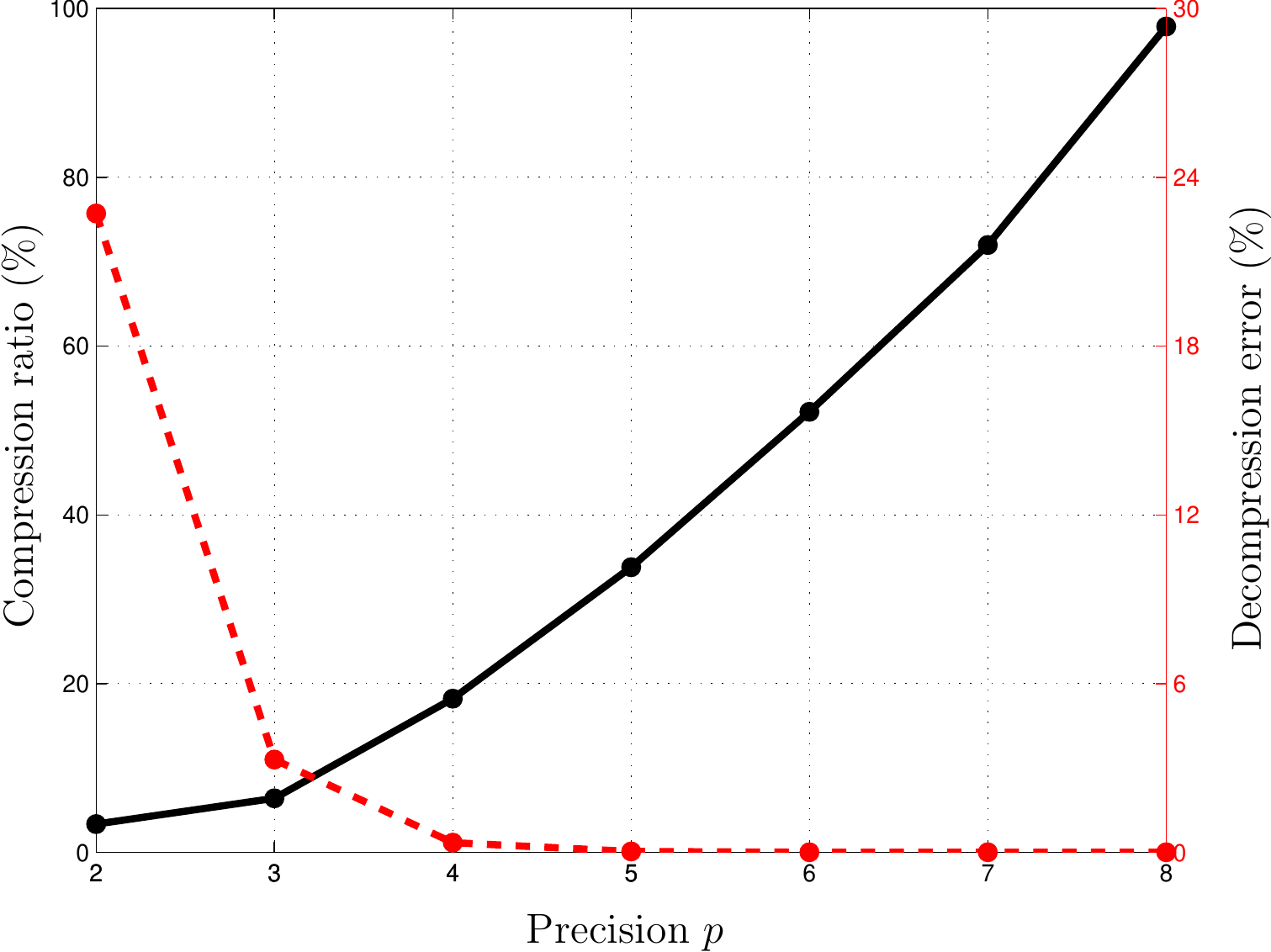}} \quad\quad
\subfigure[\subfigcapsize St Peter's: lossy performance]
  {\includegraphics[clip=,width=\tradeoffplotwidth]{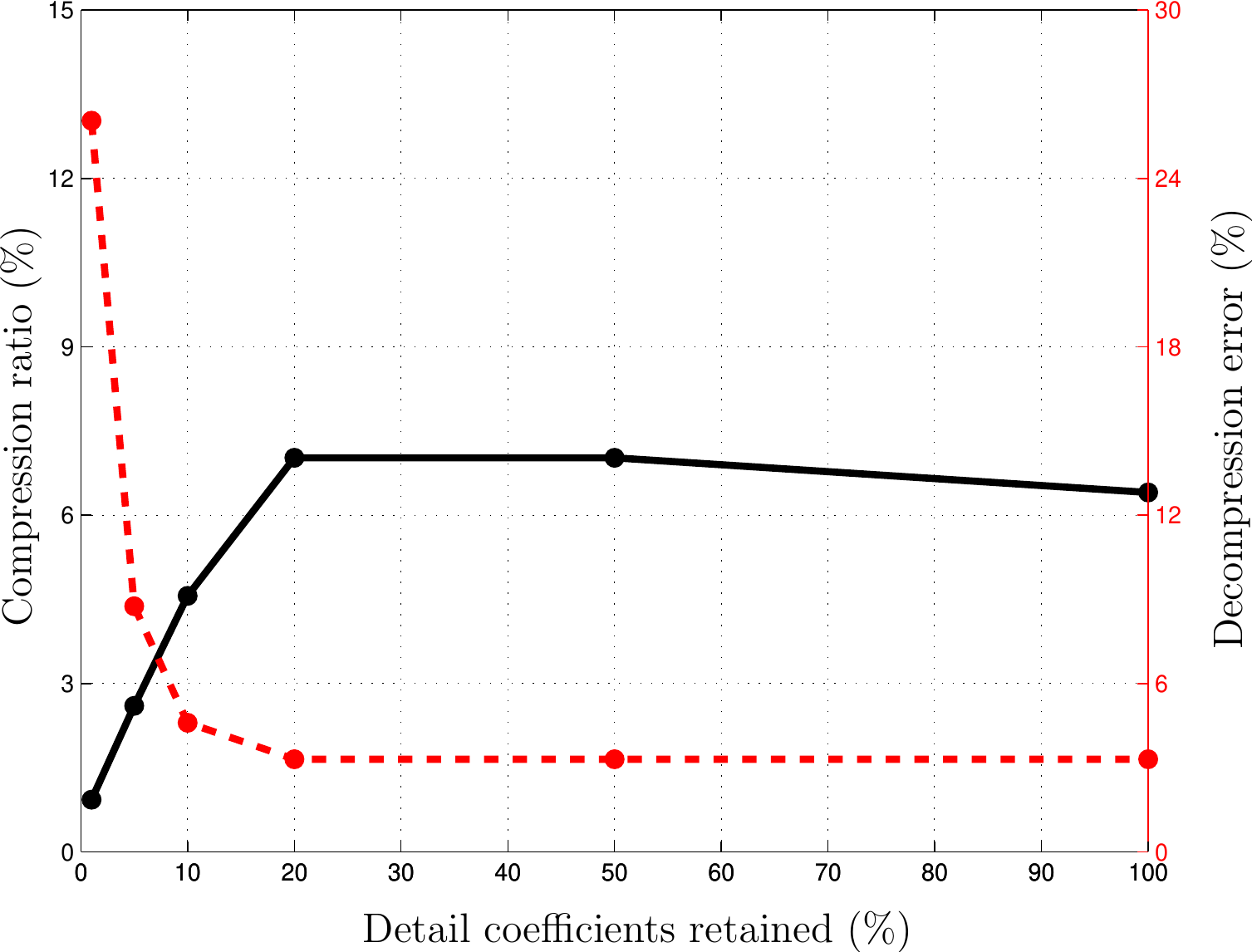}} 
} \\
\mbox{
\subfigure[\subfigcapsize Uffizi: lossless performance]
  {\includegraphics[clip=,width=\tradeoffplotwidth]{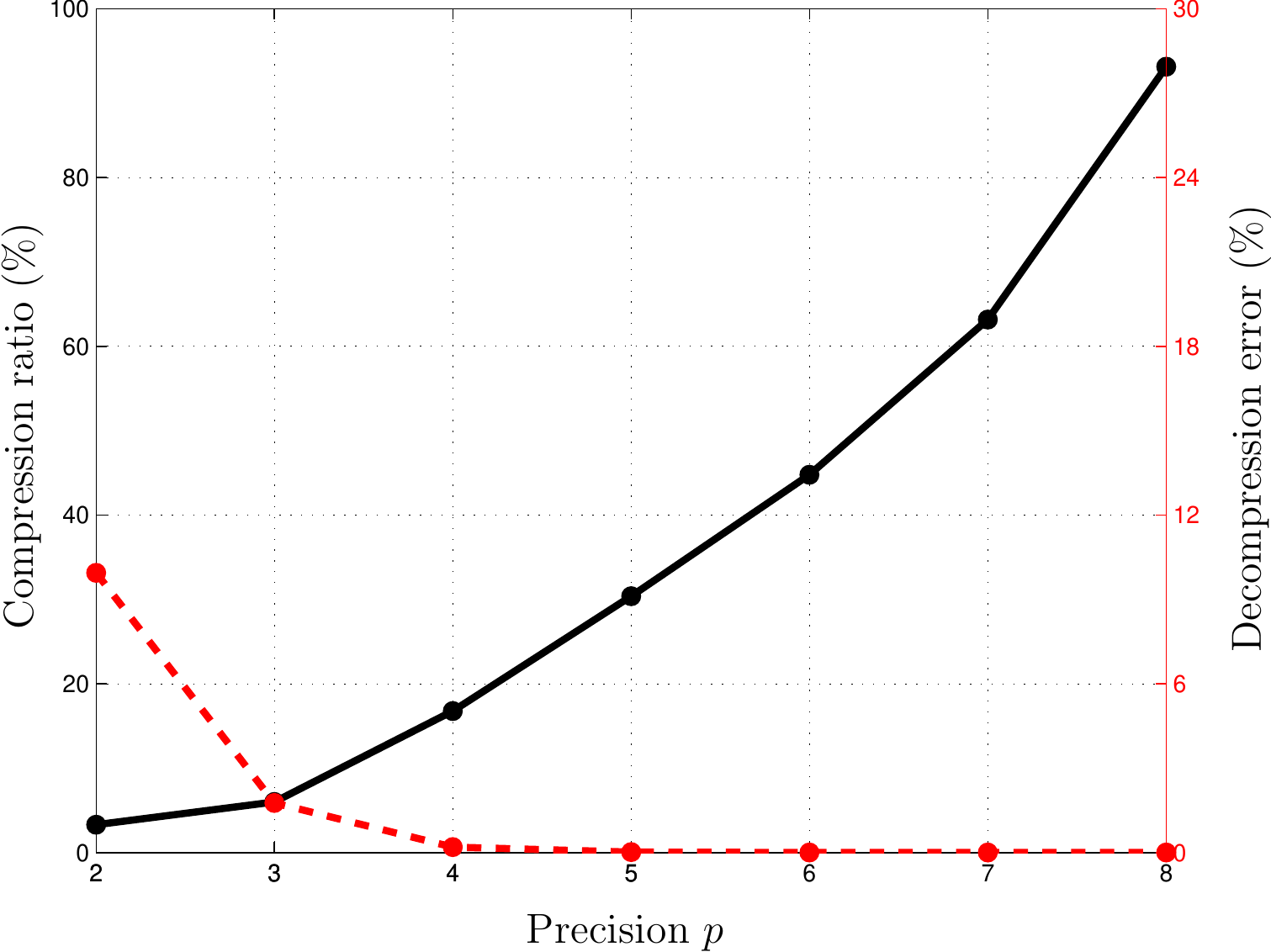}} \quad\quad
\subfigure[\subfigcapsize Uffizi: lossy performance]
  {\includegraphics[clip=,width=\tradeoffplotwidth]{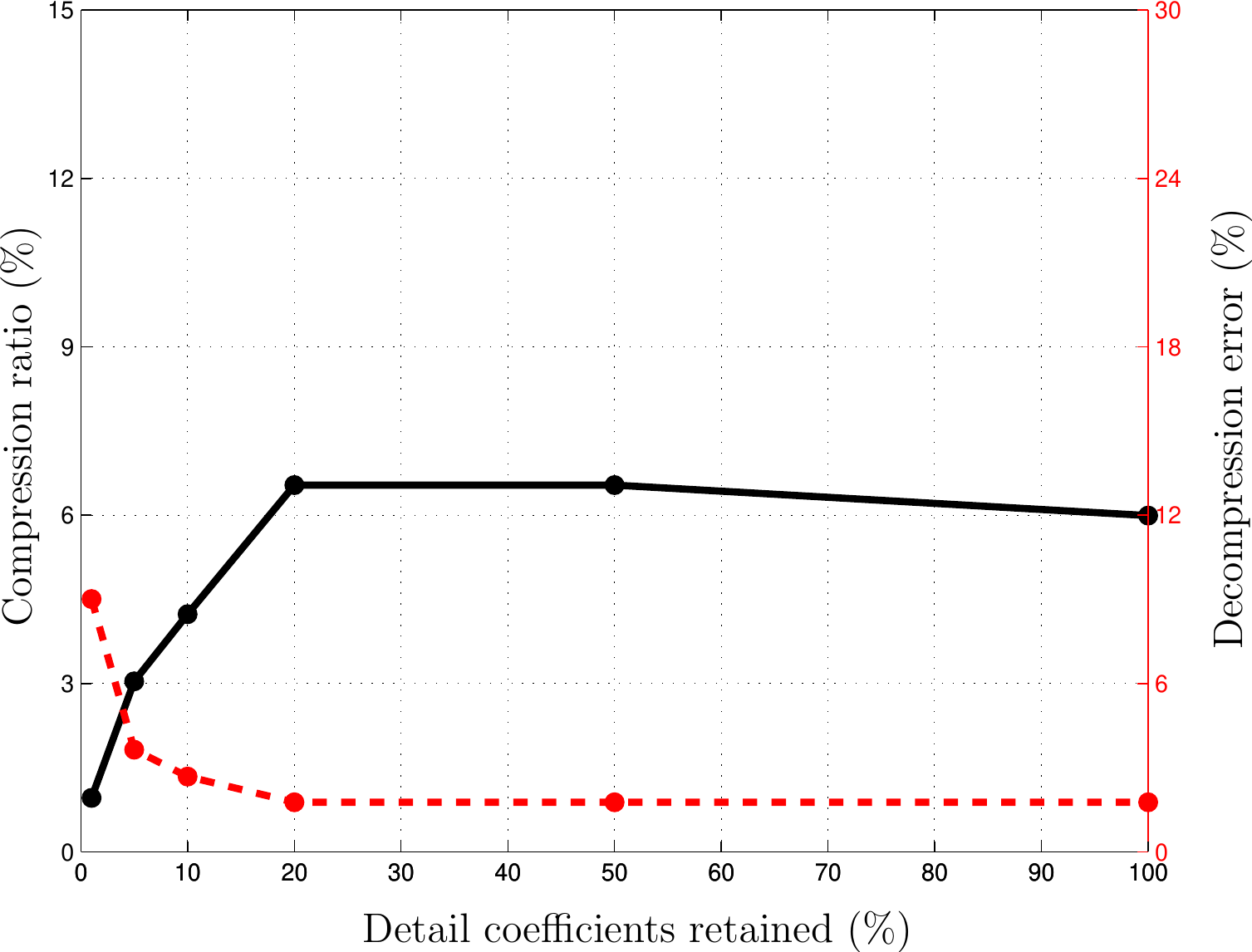}}
}
\caption{Compression performance for lossy
  compression applications.  Each row of panel shows performance
  plots for various data-spheres.  The first column of
  panels shows the trade-off between compression ratio and
  decompression fidelity with precision parameter $\precision$ for
  lossless compression.  The second column of panels shows the same
  trade-off, but with respect to the number of detail coefficients
  retained in the lossy compression.  A precision parameter of
  $\precision=3$ is used for all lossy compression results illustrated
  here.  Compression ratio (solid black line; left axis) and
  decompression error (dashed red line; right axis) are defined in the
  caption of \fig{\ref{fig:cmb}}.  
}
\label{fig:lossy_tradeoff}
\end{figure*}

\section{Concluding remarks}
\label{sec:conclusions}

We have developed algorithms to preform lossless and lossy compression
of data defined on the sphere.  These algorithms adopt a Haar wavelet
transform on the sphere to compress the energy content of the data,
prior to quantisation and Huffman encoding stages.  Note that the
resulting lossless compression algorithm is lossless to a user
specified numerical precision only. The lossy compression algorithm
incorporates, in addition, a thresholding stage so that only a user
specified proportion of detail coefficients are retained, and a RLE
stage.  By allowing a small degradation to the fidelity of the
compressed data in this manner, significantly greater compression
ratios can be attained.

The performance of these compression algorithms has been evaluated on
a number of data-spheres and the trade-off between compression ratio
and the fidelity of the decompressed data has been examined thoroughly.
Firstly, the lossless compression of \cmb\ data was performed and it
was demonstrated that the data can be compressed to 40\% of their
original size, while ensuring that essentially none of the
cosmological information content of the data is lost.  A compression
ratio of approximately 20\% can be achieved if a small loss of
cosmological information is tolerated.  Secondly, the lossy
compression of Earth topography and environmental illumination data
was performed.  For both of these data types compression ratios of
approximately 40:1 can be achieved, while introducing relative errors
of approximately 5\%.  By taking account of the geometry of the sphere
that these data live on, we achieve superior compression performance
than na\"{\i}vely applying standard compression algorithms to the data
(such as a JPEG compression of the six planes of a cross-cube
representation of a data-sphere, for example).  On inspection of the
decompressed data, it is possible to discern errors in the recovered
data by eye, nevertheless the overall structure and many of the
details of the data are well approximated.  The accuracy of the
compressed data remains suitable for many applications.

A number of avenues exist to improve the performance of the current
compression algorithms.  We choose Haar wavelets on the sphere for the
energy compression stage due to their simplicity and computational
efficiency.  However, the scale discretised wavelet methodology
developed by \citet{wiaux:2007:sdw} may yield better compression
performance due to the ability to represent directional structure in
the original data efficiently.  However, this wavelet transform is
computed in spherical harmonic space; forward and inverse spherical
harmonic transforms are not exact on a \healpix\ pixelisation.
Consequently, greater errors will be introduced in any compression
strategy based on this transform.  \update{For alternative constant
  latitude pixelisations of the sphere, however, exact (and fast)
  spherical harmonic transforms do exist and could be adopted
  \citep{mcewen:fssht}.  Nevertheless,} the implementation of the
scale discretised wavelet is also considerably more demanding
computationally.
Scope also remains to improve the lossy
compression algorithm by treating the detail coefficients at each
level differently, perhaps by using an annealing scheme to dynamically
specify the proportion of detail coefficients to retain at each level.
Nevertheless, the na\"{\i}ve thresholding strategy adopted currently has
been demonstrated to perform very well.  

The algorithms that we have developed to compress data defined on the
sphere have been demonstrated to perform well and we hope that our
publicly available implementation will now find practical use.
Obviously these compression algorithms may be used to reduce the
storage and dissemination costs of data defined on the sphere, but the
compressed representation of data-spheres may also find use in the
development of fast algorithms that exploit this representation (\eg\
\citealt{ng:2004}).  Furthermore, data defined on other
two-dimensional manifolds may be compressed by first mapping these
data to a sphere, before applying our data-sphere compression
algorithms.  We are currently pursuing this idea for the compression of
three-dimensional meshes used to represent computer graphics models.

\section*{Acknowledgements}

During the completion of this work JDM was supported by a Research
Fellowship from Clare College, Cambridge, and by the Swiss National
Science Foundation (SNSF) under grant 200021-130359.  YW is supported
in part by the Center for Biomedical Imaging (CIBM) of the Geneva and
Lausanne Universities, EPFL, and the Leenaards and Louis-Jeantet
foundations, and in part by the SNSF under grant PP00P2-123438.  We
acknowledge the use of the Legacy Archive for Microwave Background
Data Analysis (LAMBDA). Support for LAMBDA is provided by the NASA
Office of Space Science.


\bibliographystyle{mymnras_eprint}
\bibliography{bib}

\label{lastpage}
\end{document}